\documentclass[pdflatex,sn-mathphys]{sn-jnl}% Math and Physical Sciences Reference Style
\jyear{2021}%

\theoremstyle{thmstyleone}%
\theoremstyle{thmstyletwo}%

\theoremstyle{thmstylethree}%

\usepackage{graphicx} 
\usepackage{dcolumn}
\usepackage{bm}
\usepackage{amssymb}
\usepackage{amsmath}
\usepackage{wasysym}
\usepackage{color}
\usepackage{hyperref}
\usepackage{dcolumn}
\usepackage{float}
\usepackage{natbib}
\usepackage{flushend}
\usepackage{multirow}
\usepackage{cancel}
\hypersetup{
	colorlinks,
	citecolor=blue,
	filecolor=blue,
	linkcolor=blue,
	urlcolor=blue}
\newcommand{\la}{\left\langle}
\newcommand{\ra}{\right\rangle}
\newcommand{\be}{\begin{equation}}
	\newcommand{\ee}{\end{equation}}
\newcommand{\bse}{\begin{subequations}}
	\newcommand{\ese}{\end{subequations}}
\newcommand{\bea}{\begin{eqnarray}}
	\newcommand{\eea}{\end{eqnarray}}
\newcommand{\ba}{\begin{array}}
	\newcommand{\ea}{\end{array}}
	\newcommand{\etal}{{\em et al. }}

\begin{document}

\title[Turbulent Drag Reduction in MHD Turbulence]{Turbulent Drag Reduction in Magnetohydrodynamic Turbulence and Dynamo from Energy Flux Perspectives}

\author*[1]{\fnm{Mahendra K.} \sur{Verma}}\email{mkv.iitk.ac.in}
%\equalcont{MKV constructed the formalism, as well as wrote major part of the paper.}

\author[2]{\fnm{Manohar K.} \sur{Sharma}}\email{manohar-kumar.sharma@univ-grenoble-aples.fr}
%\equalcont{MKS performed numerical simulations, analyzed the data, and made the plots.}

\author[1]{\fnm{Soumyadeep } \sur{Chatterjee}}\email{soumyade@iitk.ac.in}
%\equalcont{SC performed numerical simulations, analyzed the data, and made the plots.}

\affil*[1]{\orgdiv{Department of physics}, \orgname{Indian institute of Technology Kanpur}, \orgaddress{\street{Kalyanpur}, \city{Kanpur}, \postcode{208016}, \state{Uttar Pradesh}, \country{India}}}

\affil[2]{\orgdiv{Department of Mathematics}, \orgname{University of Grenoble Aples }, \orgaddress{\street{Gires}, \city{Grenoble}, \postcode{38000}, \state{Grenoble}, \country{France}}}


\abstract{In this review, we describe turbulent drag reduction in a variety of flows using a universal framework of energy flux. In a turbulent flow with dilute polymers and magnetic field, the kinetic energy injected at large scales cascades to the velocity field at intermediate scales, as well as to the polymers and magnetic field at all scales. Consequently, the kinetic energy flux, $ \Pi_u(k) $, is suppressed in comparison to the pure hydrodynamic turbulence. We argue that the suppression of $ \Pi_u(k) $ is an important factor in the reduction of the inertial force $ \la {\bf u \cdot \nabla u} \ra$ and   \textit{turbulent drag}.  This feature of turbulent drag reduction is observed  in polymeric, magnetohydrodynamic, quasi-static magnetohydrodynamic, and stably-stratified turbulence, and in dynamos.  In addition, it is shown that turbulent drag reduction in thermal convection is due to the smooth thermal plates, similar to the turbulent drag reduction over bluff bodies.   In all these flows, turbulent drag reduction often leads to  a strong large-scale velocity in  the flow.  }

\keywords{Turbulent drag reduction, Magnetohydrodynamic turbulence, Energy flux, Dynamo, Quasi-static magnetohydrodynamics, Turbulent thernal convection}
%\keywords{keyword1, Keyword2, Keyword3, Keyword4}



\maketitle

\section{Introduction}
\label{sec:intro}

It has been observed that an introduction of polymers and magnetic field to a turbulent flow reduces turbulent  drag \cite{Lumley:ARFM1969,Tabor:EPL1986,deGennes:book:Intro,deGennes:book:Polymer,Sreenivasan:JFM2000,Lvov:PRL2004,Benzi:PRE2003,White:ARFM2008,Benzi:PD2010,Benzi:ARCMP2018,Verma:PP2020}. Turbulence drag is also suppressed over bluff bodies with particular shapes, e.g., aerofoils. This phenomena, known as \textit{turbulent drag reduction}, or \textit{TDR} in short, depends on many factors---properties of the boundaries and fluids, bulk turbulence, nature of polymers, etc.  In this review, using   energy flux, we describe a universal framework to  explain TDR in polymeric, magnetohydrodynamic (MHD), quasi-static MHD, and stably-stratified turbulence, and in dynamo.

\begin{figure}[tbhp]
	\centering
	\includegraphics[scale = 0.4]{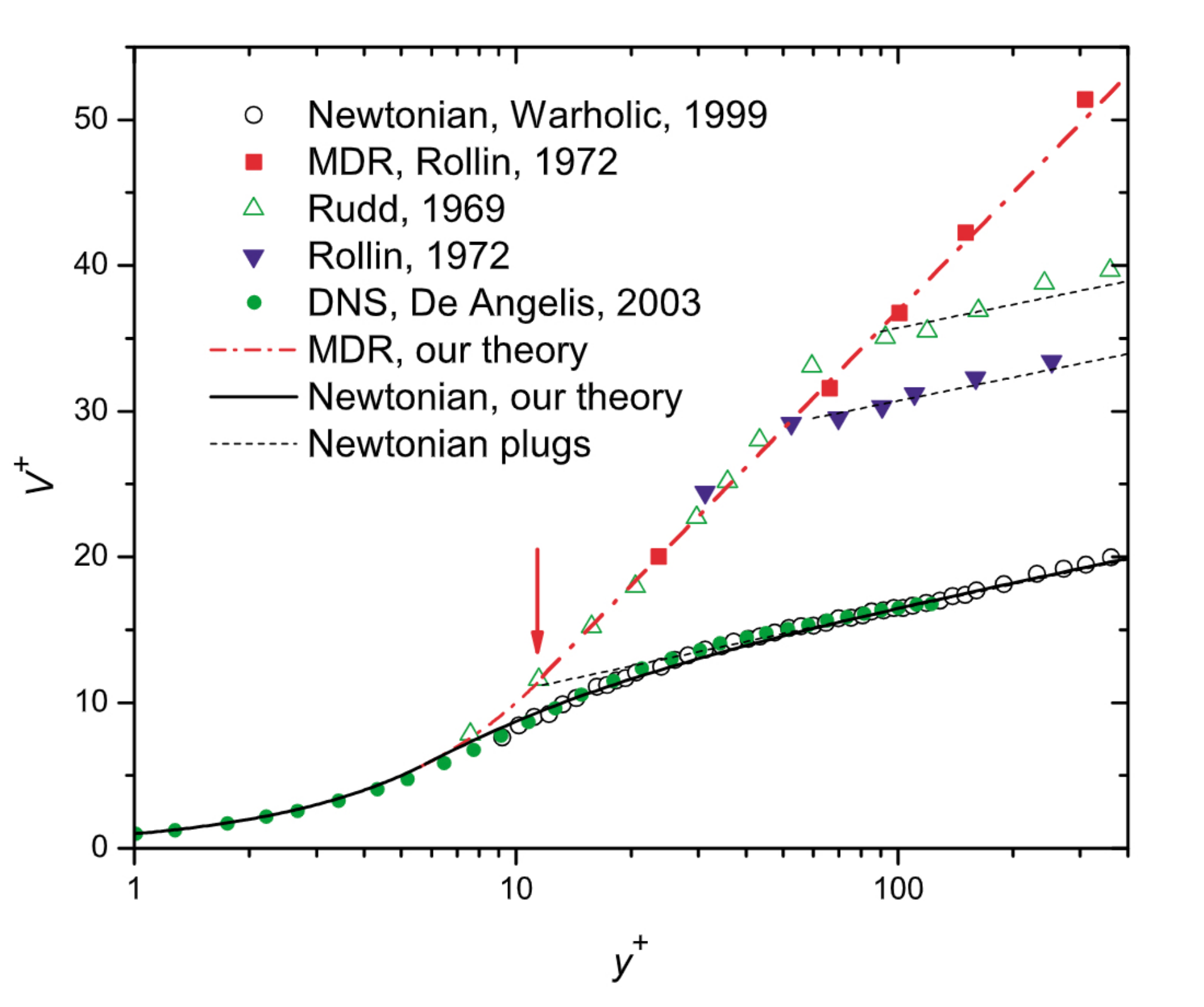}
	\caption{For a wall-bound flow, mean normalized velocity profiles ($ V^+ $) as a function of the normalized distance from the wall ($ y^+ $).  The bottom curve with green dots is for pure HD turbulence, whereas the chained-curve with red squares is for the polymeric turbulence.   From L'vov \etal~\cite{Lvov:PRL2004}. Reproduced with permission from APS. }
	\label{fig:Lvov}
\end{figure}

A pipe flow exhibits viscous drag at  small Reynolds numbers, but it experiences turbulent drag at large Reynolds numbers~\cite{Landau:book:Fluid,Kundu:book}. It has been observed that an introduction of small amount of polymers in the flow suppresses the turbulent drag up to 80\% \cite{Lumley:ARFM1969,Tabor:EPL1986,deGennes:book:Intro,deGennes:book:Polymer,Sreenivasan:JFM2000,Lvov:PRL2004,Benzi:PRE2003,White:ARFM2008,Benzi:PD2010,Benzi:ARCMP2018,Verma:PP2020}. In Fig.~\ref{fig:Lvov}, we illustrate the mean normalized velocity profiles ($ V^+ $) as a function of  normalized distance from the wall  ($ y^+ $) in a hydrodynamic (HD) flow with and without polymers.  The bottom curve with green dots represents  $ V^+ $ for pure HD turbulence and it exhibits K\'{a}rman's log layer, whereas the chained curve with red squares is for polymeric turbulence and it shows TDR. L'vov \etal \cite{Lvov:PRL2004} constructed a phenomenological model for the \textit{maximum drag reduction asymptote} (represented by the chained curve in the figure) that matches with numerical and experimental data quite well. Study of TDR is particularly important due to its wide-ranging practical applications. For example,  firefighters mix polymers in water to increase the range of  fire-hoses. Also, polymers are used to increase  the flow rates in oil pipe, etc.  

Bluff bodies too experience viscous and turbulent drag at small and large Reynolds numbers respectively. Turbulent drag over  bluff bodies depend on the surface properties, e.g., smoothness and curvature~\cite{Anderson:book:Aero,Anderson:book:History_aero}. Keeping these factors in mind, airplanes, automobiles, missiles, and ships are designed  to minimize  turbulent drag. 

%Years of scientific research has revealed that the turbulent drag in polymeric turbulence and on a bluff body depends  on many factors---Reynolds number, boundary layer, surface properties of the boundaries, bulk properties of the flow, etc. In this short review, we focus on turbulent drag in the bulk of the flow, and ignore the effects of the boundaries of the flow. We show that the turbulent drag reduction observed in polymeric turbulence is also present in magnetofluids. 

In a recent paper, Verma \etal \cite{Verma:PP2020} argued  that TDR occurs in MHD turbulence analogous to TDR in turbulent flows with dilute polymers. They showed that the kinetic energy (KE) flux ($ \Pi_u(k) $) is suppressed in  polymeric and MHD turbulence due to the transfer of energy from the velocity field to polymers and magnetic field respectively. The energy fluxes in polymeric and MHD turbulence  have been studied in a number of earlier works~\cite{Lumley:ARFM1969,Dar:PD2001,Mininni:ApJ2005,Valente:JFM2014,Valente:PF2016,Verma:PP2020}. It was argued that the turbulent drag and the nonlinearity $ \la {\bf u \cdot \nabla u} \ra $ are proportional to $ \Pi_u(k)/U $, where $ {\bf u} $ is the velocity field, $ U $ is the large-scale velocity, and $ \la . \ra $ represents averaging. Thus, Verma {\em et al.}'s \cite{Verma:PP2020} formalism provides a general framework for TDR in variety of flows, including polymeric and MHD turbulence.

An introduction of polymers or magnetic field in a turbulent flow enhances the mean flow, but suppresses $ \la {\bf u \cdot \nabla u} \ra $ \cite{Lumley:ARFM1969,Tabor:EPL1986,deGennes:book:Intro,deGennes:book:Polymer,Sreenivasan:JFM2000,Lvov:PRL2004,Benzi:PRE2003,White:ARFM2008,Benzi:PD2010,Benzi:ARCMP2018,Verma:PP2020}.  Verma \etal \cite{Verma:PP2020}  observed the above phenomena in a shell model of MHD turbulence. Note that $ \la {\bf u \cdot \nabla u} \ra $ and $ \Pi_u(k) $ depend critically on the phase relations between the Fourier modes.  Verma \etal \cite{Verma:PP2020} argued that the velocity correlations in polymeric and MHD turbulence are enhanced compared to pure  HD turbulence. These correlations lead to suppressed $ \la {\bf u \cdot \nabla u} \ra $ and $ \Pi_u(k) $ in spite of amplification of  $ U $. Thus, TDR, energy flux, and enhancement of $ U $ are  related to each other. 

Based on past results, Verma \etal \cite{Verma:PP2020}  argued for TDR in quasi-static MHD (QSMHD) turbulence~\cite{Moreau:book:MHD,Knaepen:ARFM2008}. The Joule dissipation suppresses $ \Pi_u(k) $ at all wavenumbers \cite{Moreau:book:MHD,Knaepen:ARFM2008,Reddy:PF2014,Reddy:PP2014}, and hence  $ \Pi_u(k) $ for QSMHD turbulence is lower than the corresponding flux for HD turbulence.  In addition,  large-scale $ U $ increases  with the increase of interaction parameter, thus indicating  TDR in QSMHD turbulence. 

Generation of magnetic field in astrophysical objects, such as planets, stars, and galaxies, are explained using dynamo  mechanism \cite{Moffatt:book,Roberts:RMP2000,Brandenburg:PR2005,Yadav:PRE2012}.  Here, magnetic field grows and saturates at some level due to the self-induced currents.   In the present review, we discuss TDR in dynamo using the energy flux. Based on earlier dynamo simulations (e,g., \cite{Olson:JGR1999,Yadav:PRE2012}), we show that the fluctuations in the velocity and magnetic fields are suppressed when a large-scale magnetic field emerges in the system.  This feature signals  TDR in dynamo.

Planetary and stellar atmospheres often exhibit stably stratified turbulence. In such flows, lighter fluid is above the heavier fluid with gravity acting downwards~\cite{Davidson:book:TurbulenceRotating,Verma:book:BDF}. The KE flux in stably stratified turbulence is suppressed, as in polymeric and MHD turbulence. Based on these observations, we argue for TDR in stably stratified turbulence.

Researchers have reported that  compared to HD turbulence, viscous dissipation rate ($ \epsilon_u $) and thermal dissipation rate ($ \epsilon_T $)  are suppressed in turbulent thermal convection. For example, Pandey  \etal \cite{Pandey:PF2016} and Bhattacharya \etal \cite{Bhattacharya:PF2021} showed that $ \epsilon_u \sim (U^3/d) \mathrm{Ra}^{-0.2}$ and $ \epsilon_T\sim (U (\Delta T)^2/d) \mathrm{Ra}^{-0.2}$, where $ \Delta T $ is the temperature difference between the top and bottom thermal plates separated by distance $ d $, and Ra is the Rayleigh number, which is the ratio of buoyancy and diffusion in thermal convection. In addition, Pandey  \etal \cite{Pandey:PF2016} observed that $ \la {\bf u \cdot \nabla u} \ra/(Ud/\nu) \approx \mathrm{Re Ra}^{-0.14}$, where Re is the Reynolds number. Thus, nonlinearity is suppressed in turbulent thermal convection. In this review, we relate the above suppression of nonlinearity and dissipation rates to  TDR over  bluff bodies.  It has been argued that TDR in turbulent convection arises due to  large-scale circulation (LSC) over  thermal plates, and  that the smooth thermal plates affect bulk turbulence.   

Thus, KE flux and $ \la {\bf u \cdot \nabla u} \ra$ provide valuable insights into the physics of TDR.  TDR is also related to the enhanced correlations in the velocity field. The present review  focusses on these aspects for a variety of flows---polymeric, MHD,  QSMHD, and stably-stratified turbulence; dynamo; and  turbulent thermal convection. Here, we focus on bulk turbulence, and avoid  discussion on boundary layers and smooth surfaces.  The latter aspects are covered in many books and reviews, e.g., \cite{deGennes:book:Intro,deGennes:book:Polymer,Sreenivasan:JFM2000,Benzi:ARCMP2018,	Anderson:book:Aero,Anderson:book:History_aero}. We remark that the energy flux is a well known quantity in turbulence literature \cite{Kolmogorov:DANS1941Dissipation,Kolmogorov:DANS1941Structure,Lesieur:book:Turbulence,Frisch:book,Verma:book:ET}. However,  the connection between the energy flux and TDR has been brought out only recently \cite{Verma:PP2020}, and the number of papers highlighting the above connection is relatively limited. 

The increase in the  mean velocity field during TDR is related to relaminarization. Narasimha and Sreenivasan \cite{Narasimha:AAM1979} studied relaminarization in stably stratified turbulence, rotating turbulence, and thermal convection, and related it  to the reduction in  $ \la {\bf u \cdot \nabla u} \ra $. Thus, the mechanism of relaminarization is intimately related to the TDR.

An outline of this  review  is as follows. In Section \ref{sec:Hydro} we briefly review viscous and turbulent drag in a pipe flow and over a bluff body.  In Section \ref{sec:drag} we describe a general framework for TDR using energy fluxes. In Section \ref{sec:polymer} we review the energy fluxes in a turbulent flow  with dilute polymers and relate it to  TDR in the bulk. Section \ref{sec:MHD} contains a framework of TDR in MHD turbulence via energy fluxes.  In Section \ref{sec:dns} we describe  signatures of TDR  in direct numerical simulations (DNS) and shell models of MHD turbulence. Sections \ref{sec:dynamo} and \ref{sec:QSMHD} deal with TDR in dynamos and in QSMHD turbulence respectively. In Section \ref{sec:misc} we describe TDR in stably stratified turbulence and in turbulent thermal convection.  We conclude in Section \ref{sec:conclusions}.

\section{Viscous and turbulent drag in hydrodynamic turbulence}
\label{sec:Hydro}
The equations for incompressible hydrodynamics are
\bea
\frac{\partial{\bf u}}{\partial t} + ({\bf u}\cdot\nabla){\bf u}
& = & -\nabla({p}/{\rho}) +  \nu\nabla^2 {\bf u}   +  {\bf F}_\mathrm{ext},  \label{eq:U}  \\
\nabla \cdot {\bf u}  & = & 0, \label{eq:incompress}
\eea
where ${\bf u}, p$ are respectively the velocity  and pressure fields; $\rho$ is the density which is assumed to be unity;   $\nu$ is  the kinematic viscosity;  and    $ {\bf F}_\mathrm{ext}$ is the external force employed at large scales that helps  maintain a steady state.  An important parameter for the fluid flows is Reynolds number, which is
 \be
 \mathrm{Re} = \frac{UL}{\nu},
 \ee
 where $ L $ and $ U $ are the large-scale length and velocity respectively. For homogeneous and isotropic turbulence, Re is  the ratio of the nonlinear term and the viscous term. However, in more complex flows like polymeric turbulence, MHD turbulence, and turbulent convection, 
  \be
\frac{ \mathrm{Nonlinear~term}}{ \mathrm{Viscous~term}} =  f \mathrm{Re},
 \ee
where the prefactor $ f  $ may differ from unity and may provide a signature for TDR.  For example, $ f \approx \mathrm{Ra}^{-0.2} $ for turbulent convection, where Ra is the Rayleigh number~\cite{Pandey:PF2016}. We expect complex $ f $ for MHD and polymeric turbulence.
\begin{figure}%[tbhp]
	\centering
	\includegraphics[width=1.0\linewidth]{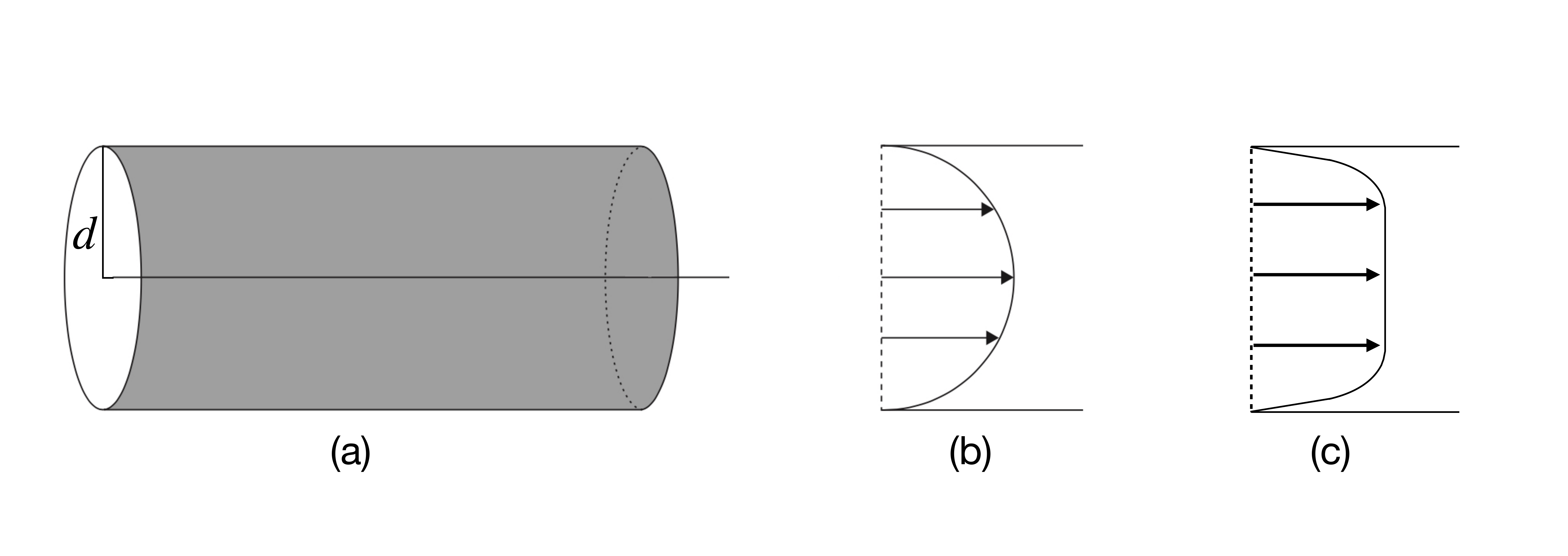}
	\caption{Schematic illustrations of (a) pipe flow and (b) its viscous flow profile. (c) The profile of the mean velocity  in a turbulent pipe flow.}
	\label{fig:pipe}
\end{figure}

A fluid moving in a pipe of radius $ d $ experiences drag (see Fig.~\ref{fig:pipe}).   At low Reynolds numbers, this drag is called \textit{viscous drag}. In this case, under steady state, the pressure gradient, $ -\nabla({p}/{\rho}) $, which can be treated as $ {\bf F}_\mathrm{ext} $, matches with the viscous term, $ \nu \nabla^2  {\bf u}$.  Hence, we estimate the viscous drag as~\cite{Kundu:book,Verma:book:Mechanics}
\be
F_\mathrm{drag} \approx \frac{ \nu U}{d^2}.
\ee
The proportionality constant is of the order of unity.  At large Reynolds number, the nonlinear term becomes significant, and hence~\cite{Kundu:book,Landau:book:Fluid,Anderson:book:Aero,Anderson:book:History_aero},
\be
F_\mathrm{drag} \approx \frac{  U^2}{d} + \frac{ \nu U}{d^2},
\ee
apart from the proportionality constants.  In the above formula, $ U^2/d $ is the turbulent drag that is larger than the viscous drag by a factor of Re. Clearly, the turbulent drag dominates the viscous drag at large Re. Note that the above drag force is in the units of force per unit mass; we will follow this convention throughout the paper.

A related problem is the frictional force experienced by a bluff body in a flow. Analogous to  a pipe flow, a bluff body experiences viscous drag at small Re, but turbulent drag at large Re. In literature, the drag coefficient is defined as~\cite{Kundu:book,Anderson:book:Aero}
\be
C_d = \frac{F_\mathrm{drag}}{\rho U^2 A},
\ee
where $ A $ is the area of the bluff body.

It is customary to describe fluid flows in Fourier space, where  Eqs.~(\ref{eq:U}, \ref{eq:incompress}) get transformed to \cite{Lesieur:book:Turbulence,Frisch:book,Verma:book:ET}
\bea
 \frac{d}{dt} {\bf u(k)} =- i \sum_{\bf p} \{ {\bf k \cdot u(q)}  \} {\bf u(p)}  - i{\bf k} p({\bf k} ) -\nu k^2 {\bf u(k)} + {\bf F}_\mathrm{ext}({\bf k}), 
\eea
where \textbf{k, p, q} are the wavenumbers with $ {\bf k = p+q} $; and $ {\bf u(k)},  {\bf u(p)}, {\bf u(q)}$ are the corresponding velocity Fourier modes. An equation for the modal  energy  $E_u({\bf k}) = \lvert {\bf u(k)} \rvert^2/2$ is \cite{Lesieur:book:Turbulence,Frisch:book,Verma:book:ET,Verma:JPA2022}
\bea
\frac{d}{dt} E_u(\mathbf{k})  & = & T_{u}({\bf k})  +   \mathcal{F}_\mathrm{ext}({\bf k})-D_u(\mathbf{k}),
\label{eq:Eu_dot_Fext_hydro} 
\eea
where 
\bea
T_u({\bf k}) & = & \sum_{\bf p} \Im \left[ {\bf  \{  k \cdot u(q) \} \{ u(p) \cdot u^*(k) \} } \right] ,  \label{eq:Tuk} \\
\mathcal{F}_\mathrm{ext}({\bf k}) & = &   \Re[ {\bf F}_\mathrm{ext}({\bf k}) \cdot {\bf u}^*({\bf k})  ], \\
D_u(\mathbf{k}) & = & 2 \nu k^2 E_u({\bf k}).
\eea
 Here, $\Re, \Im$ stand respectively for the real and imaginary parts of the argument; $ T_u({\bf k}) $ is the nonlinear energy transfer to the mode $ {\bf u(k)} $; $ D_u(\mathbf{k})  $ is the energy dissipation rate at wavenumber $ {\bf k} $; and $ \mathcal{F}_\mathrm{ext}({\bf k}) $  is the KE injection rate to  $ {\bf u(k)} $ by the external force $ {\bf F}_\mathrm{ext}({\bf k}) $.  
 
\begin{figure}%[tbhp]
	\centering
	\includegraphics[width=1\linewidth]{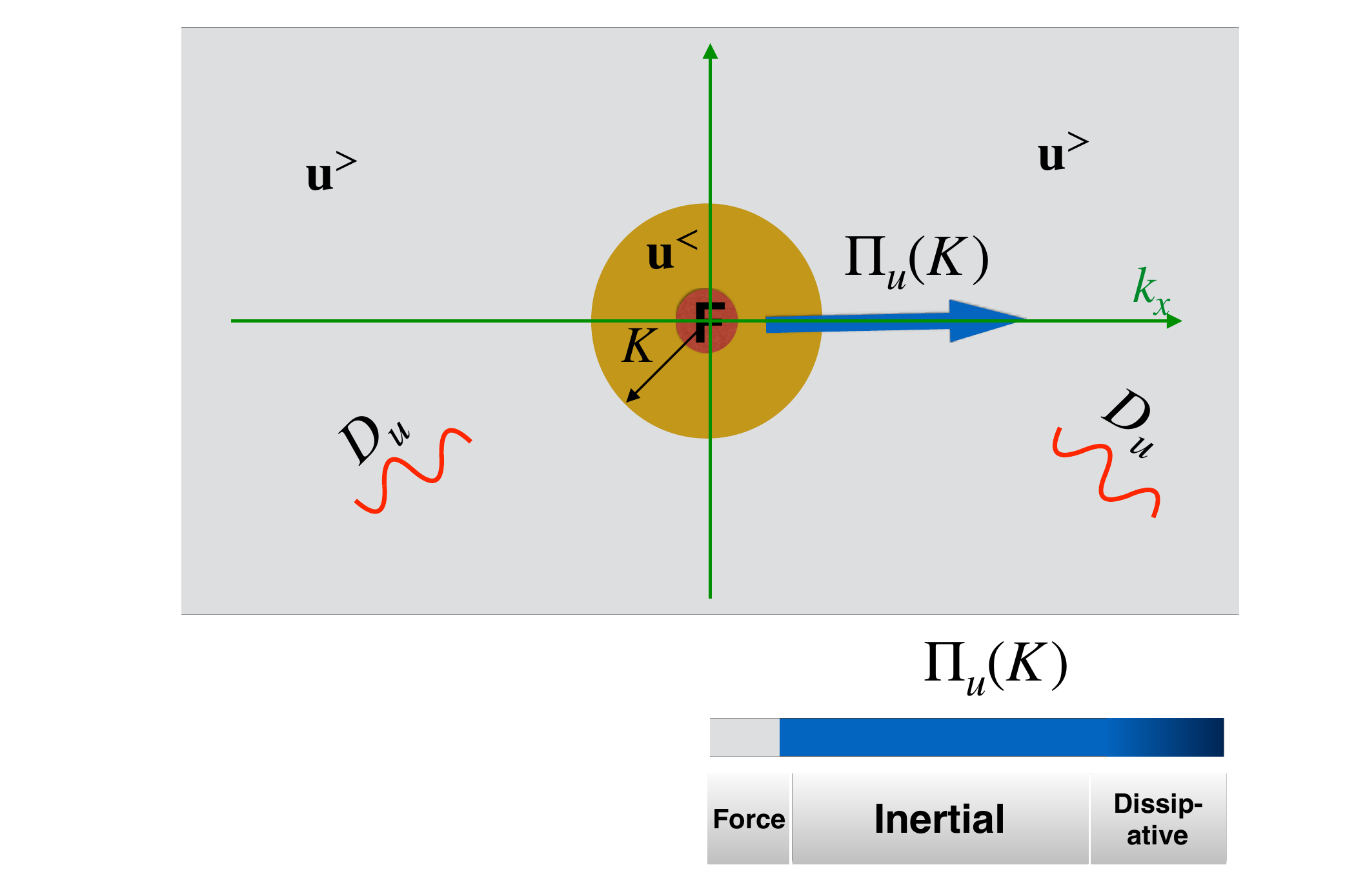}
	\caption{An illustration of KE flux $\Pi_u(K)$.   KE is injected into the small red sphere.  $\Pi_u(K)$  is constant in the inertial range, and it is dissipated at small scales with a  dissipation rate of $D_u$.  From Verma \etal~\cite{Verma:PP2020}. Reprinted with permission from AIP. }
	\label{fig:U_flux}
\end{figure}

 We assume that the external force   injects KE at large scales, e.g.,  in a wavenumber band $ (0, k_f) $ with small $ k_f $. Therefore, the total KE injection rate, $ \epsilon_\mathrm{inj}$, is 
\be
\int_0^{k_f}  d{\bf k} \mathcal{F}_\mathrm{ext}({\bf k}) \approx \epsilon_\mathrm{inj}.
\ee
This injected KE cascades to intermediate and small scales as KE flux, $ \Pi_u(K) $, which is defined as the cumulative KE transfer rate from the velocity modes inside the sphere of radius $ K $ to velocity  modes outside the sphere. In Fig.~\ref{fig:U_flux}, we illustrate the inner and outer modes as $ {\bf u}^< $ and $ {\bf u}^> $ respectively. In terms of Fourier modes, the above flux is~\cite{Kraichnan:JFM1959,Dar:PD2001,Verma:PR2004,Verma:book:ET}
\bea
\Pi_u(K) &= & - \sum_{k\le K} T_u({\bf k})  = \sum_{p\le K} \sum_{k>K} \Im \left[ {\bf  \{  k \cdot u(q) \} \{ u(p) \cdot u^*(k) \} } \right],
\label{eq:fluid_flux}
\eea 
where $ {\bf q=k-p} $.

The above energy flux is dissipated in the dissipative range, with the total viscous dissipation rate as
\be
\epsilon_u = \int d{\bf k} D_u({\bf k}) = \int d{\bf k}   2 \nu k^2 E_u({\bf k}).
\ee
 At large Reynolds numbers, it has been shown that in the inertial range \cite{Sreenivasan:PF1998, Kolmogorov:DANS1941Dissipation, Onsagar:Nouvo1949_SH,Frisch:book,Lesieur:book:Turbulence},
\be
\Pi_u(k) \approx     \epsilon_\mathrm{inj} \approx \epsilon_u \approx  \frac{U^3}{d} .
\ee
That is, the inertial-range energy flux, the viscous dissipation rate, and the  energy injection rate are all equal. 
Note that in the inertial range, $ \Pi_u(k) = \epsilon_\mathrm{inj}$ due to absence of external force and negligible viscous  dissipation~\cite{Kolmogorov:DANS1941Dissipation,Verma:book:ET,Verma:JPA2022}.  We show later that the magnetic field and polymers, as well as smooth walls, suppress the energy flux relative to $  \epsilon_\mathrm{inj}  $. We argue that this feature leads to TDR.

%\begin{figure}%[tbhp]
%	\centering
%	\includegraphics[width=1\linewidth]{fig/Figure1.pdf}
%	\caption{(color online)   In hydrodynamic turbulence, the energy supplied by the external force ${\bf F}_\mathrm{ext}$ at small wavenumbers (red sphere) cascades to the inertial range, leading to a constant KE flux $\Pi_u(k)$ in the inertial range. This flux is dissipated at small scales by viscosity. }
%	\label{fig:flux}
%\end{figure}

For a steady state, an integration of Eq.~(\ref{eq:U}) over a bluff body  yields the following formula for the drag force:
\be
{\bf F}_\mathrm{drag} = \int  d{\bf r} \left[  ({\bf u}\cdot\nabla){\bf u}
  + \nabla({p}/{\rho}) -   \nu\nabla^2 {\bf u}    \right]. \label{eq:drag_foce_tot}  
\ee
 The viscous force dominates the inertial term near the surface of a bluff body. Hence, for  bluff bodies, the inertial term of the above equation is ignored. Prandtl~\cite{Prandtl,Anderson:book:History_aero} was first to compute $ {\bf F}_\mathrm{drag}  $ for a bluff body as a sum of viscous drag and adverse pressure gradient.   The drag forces for a cylinder and aerofoil are computed in this manner~\cite{Anderson:book:Aero,Anderson:book:History_aero,Kundu:book}.

%In a pipe flow, pressure difference at the two ends provide the external force. Hence,
%\be
%{\bf F}_\mathrm{drag} = \frac{1}{\rho} (p_2 - p_1) A =  \int  d{\bf r} \left[  ({\bf u}\cdot\nabla){\bf u}  -   \nu\nabla^2 {\bf u}    \right], \label{eq:drag_pipe}  
%\ee
%where $ A  $ is the cross sectional area of the pipe.

Computation of $ {\bf F}_\mathrm{drag}  $ for a pipe flow  is also quite complex involving many factors---walls, fluid properties, bulk turbulence, Reynolds number, etc.    In the present review, we focus on the turbulent drag in bulk where we can ignore the effects of  walls.  The above simplification enables us to compute turbulent drag in many diverse flows---polymeric turbulence, MHD turbulence, dynamo, liquid metals---using a common framework.

We focus on a turbulent flow within a periodic box for which $ \int d{\bf r} \nabla (p/\rho) = 0$. By ignoring the viscous drag, we deduce the turbulent drag as (see Eqs.~(\ref{eq:U}, \ref{eq:drag_foce_tot}))
\be
{\bf F}_\mathrm{drag} = {\bf F}_\mathrm{ext}  = \int  d{\bf r} \left[  ({\bf u}\cdot\nabla){\bf u}   \right]. \label{eq:drag_box}  
\ee
Since the external force is active at large scales,  under  steady state,  
\be
\la {\bf F}_{\mathrm{drag}} \ra_\mathrm{LS} \approx \la  { \lvert \bf { (u \cdot \nabla) u } \rvert}  \ra_\mathrm{LS} \approx \la   {\bf F}_\mathrm{ext}   \ra ,
\ee
where $\la . \ra_\mathrm{LS}$ represents ensemble averaging over  large scales. To estimate $ \la {\bf F}_{\mathrm{drag}} \ra_\mathrm{LS}  $, we perform a dot product of  Eq.~(\ref{eq:U})  with $ {\bf u} $ and integrate it over a  wavenumber sphere of radius $ k_f $ (forcing wavenumber band) that leads to
\be
\int_\mathrm{LS} d{\bf r}  [{\bf F}_{\mathrm{ext}} \cdot {\bf u}]  = \int_\mathrm{LS} d{\bf r}  [{\bf F}_{\mathrm{drag}} \cdot {\bf u}] = f_1 U F_{\mathrm{drag}} ,
 \ee
 with $ f_1 \approx 1 $. Under steady state, using Eqs.~(\ref{eq:Eu_dot_Fext_hydro},\ref{eq:fluid_flux}) we deduce that 
 \be
\int_\mathrm{LS} d{\bf r}  [{\bf F}_{\mathrm{ext}} \cdot {\bf u}]  = \la  { \lvert \bf {[ (u \cdot \nabla) u] \cdot u } \rvert}  \ra_\mathrm{LS} = - \int_0^{k_f} T_u(k') dk' = \Pi_u(k).
 \ee
 Therefore, 
\be
U F_{\mathrm{drag}}   \approx \Pi_u \approx \frac{U^3}{d} \approx \epsilon_\mathrm{inj},
\label{eq:FD_HD_flux}
\ee
or
\be
 F_{\mathrm{drag}}  \approx \frac{\Pi_u}{U} \approx    \frac{U^2}{d}.
\label{eq:FD_HD}
\ee
Note that the viscous dissipation can be ignored at  large scales.     

It has been observed that  polymers and magnetic field suppress  turbulent drag. We detail these phenomena in the subsequent sections.

\section{General framework  for TDR using energy flux}
\label{sec:drag}

In this section, we describe a general framework for TDR in a turbulent flow with a secondary  field ${\bf B}$. At present, for convenience, we assume ${\bf B}$ to be a vector, however, it could  also be a scalar or a tensor. The present formalism is taken from Verma \etal \cite{Verma:PP2020}.

The equations for the velocity and secondary fields are ~\cite{Fouxon:PF2003,Verma:book:ET,Verma:PP2020,Davidson:book:TurbulenceRotating}:
\bea
\frac{\partial{\bf u}}{\partial t} + ({\bf u}\cdot\nabla){\bf u}
& = & -\nabla({p}/{\rho}) +  \nu\nabla^2 {\bf u} + {\bf F}_u({\bf u,B}) +  {\bf F}_\mathrm{ext},  \label{eq:U2} \\
\frac{\partial{\bf B}}{\partial t} + ({\bf u}\cdot\nabla){{\bf B}}
& = &   \eta \nabla^2 {{\bf B}} + {\bf F}_B({\bf u,B}),   \label{eq:B} \\
\nabla \cdot {\bf u}  & = & 0, \label{eq:incompress_B}
 \eea
 where ${\bf u}, p$ are  the velocity and pressure fields respectively; $\rho$ is the density which is assumed to be unity;   $\nu$ is  the kinematic viscosity;  $ \eta$ is the diffusion coefficient for \textbf{B}; and ${\bf F}_u$ and $ {\bf F}_B$ are the force fields acting on ${\bf u}$ and ${\bf B}$ respectively.   Note that ${\bf F}_u$ and $ {\bf F}_B$ typically represent interactions between $ {\bf u} $ and $ {\bf B} $. The external field $ {\bf F}_\mathrm{ext}$ is employed at large scales of the velocity field to maintain a steady state.  
 
\begin{figure}%[tbhp]
\centering
\includegraphics[width=0.8\linewidth]{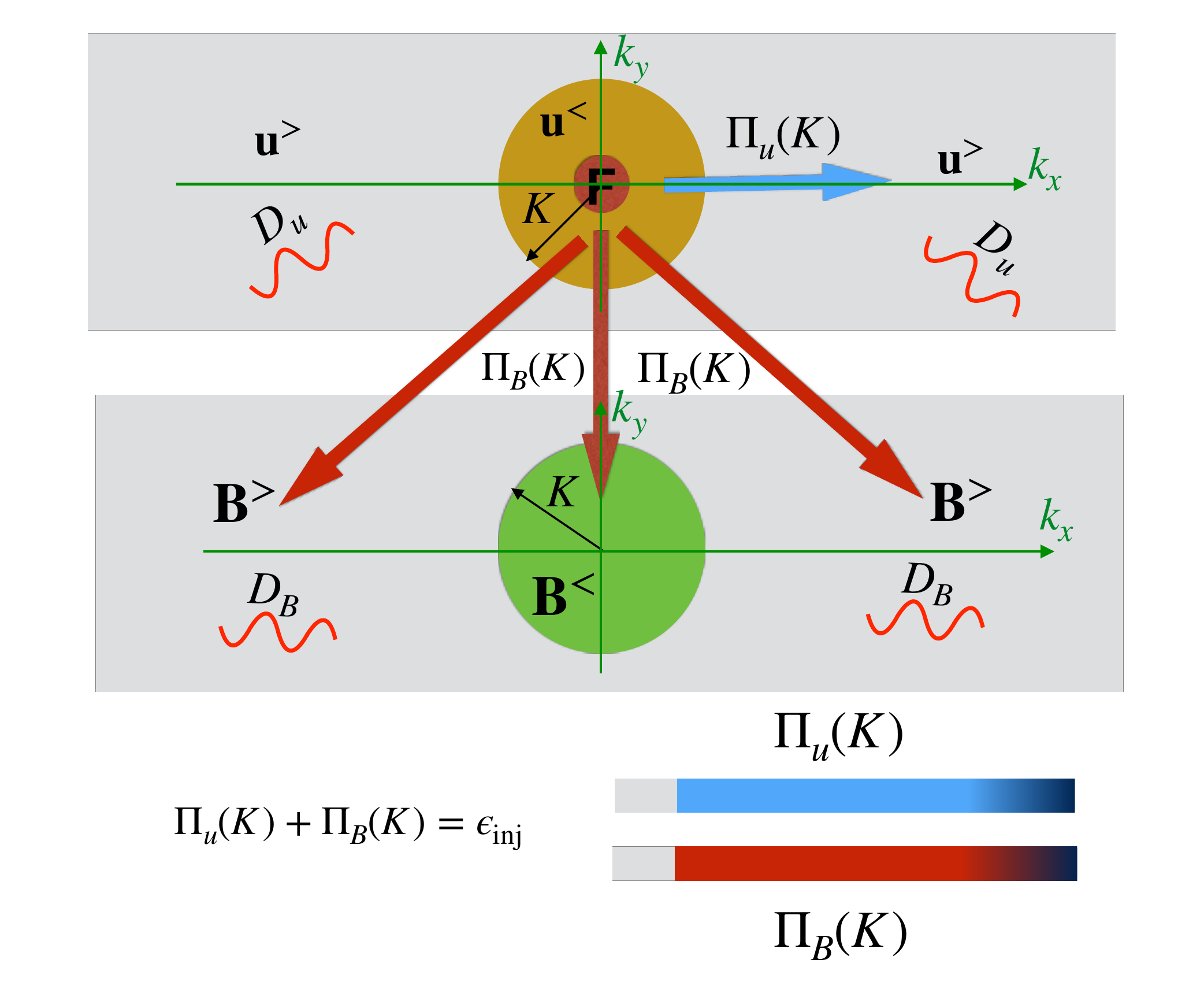}
\caption{The   external force injects KE  into the small red sphere with the rate of $\epsilon_\mathrm{int}$.     $\Pi_u(K)$ is the KE flux for the velocity wavenumber sphere of radius $K$ (yellow sphere), and $\Pi_B(K)$ is the net energy transfer from ${\bf u}$ modes inside the sphere to all the ${\bf B}$ modes.  The energy flux $\Pi_u(K)$  is dissipated with  dissipation rates $D_u$.  For small wavenumbers and inertial range, $\Pi_u(K) + \Pi_B(K) \approx \epsilon_\mathrm{int}$. From Verma \etal \cite{Verma:PP2020}. Reprinted with permission from AIP.}
\label{fig:flux_B}
\end{figure}

 Using Eq.~(\ref{eq:U2})  we derive the following equation for the KE density $u^2/2$ (with $ \rho =1 $):
 \be
 \frac{\partial}{\partial t} \frac{u^2}{2} +  \nabla \cdot \left[ \frac{u^2}{2} {\bf u} \right] = - \nabla \cdot ( p {\bf u}) + [{\bf F}_u + {\bf F}_\mathrm{ext} ] \cdot {\bf u} -\nu {\bf u} \cdot  \nabla^2 {\bf u}.
 \label{eq:basic_hd:framework:Eu_dynamics2}
 \ee
 In Fourier space, the equation for the modal KE,  $E_u({\bf k}) = \lvert {\bf u(k)} \rvert^2/2$, is 
 \bea
 \frac{d}{dt} E_u(\mathbf{k})  & = & T_{u}({\bf k})  +  \mathcal{F}_u({\bf k})  + \mathcal{F}_\mathrm{ext}({\bf k})-D_u(\mathbf{k}),
 \label{eq:Eu_dot_Fext} 
 \eea
 where 
 \bea
 T_u({\bf k}) & = & \sum_{\bf p} \Im \left[ {\bf  \{  k \cdot u(q) \} \{ u(p) \cdot u^*(k) \} } \right] , \\
 \mathcal{F}_u({\bf k}) & =  & \Re[ {\bf F}_u({\bf k}) \cdot {\bf u}^*({\bf k})  ], \\
 \mathcal{F}_\mathrm{ext}({\bf k}) & = &   \Re[ {\bf F}_\mathrm{ext}({\bf k}) \cdot {\bf u}^*({\bf k})  ], \\
 D_u(\mathbf{k}) & = & -2 \nu k^2 E_u({\bf k}),
 \eea
 with  ${\bf q=k-p}$.  We sum  Eq.~(\ref{eq:Eu_dot_Fext}) over  the $ {\bf u} $ modes of the wavenumber sphere of radius $K$ that yields~\cite{Verma:book:ET,Verma:JPA2022}:
 \bea
 \frac{d}{dt} \sum_{k \le K}  E_u({\bf k})   &= &  \sum_{k \le K} T_u({\bf k}) + \sum_{k \le K}\mathcal{F}_u({\bf k})   + \sum_{k \le K}\mathcal{F}_\mathrm{ext}({\bf k}) - \sum_{k \le K} D_u({\bf k}). 
 \label{eq:Pi_K_from_Ek}
 \eea  
A physical interpretation of the terms in the right-hand side of Eq.~(\ref{eq:Pi_K_from_Ek}) are as follows:
 \begin{enumerate}
 	\item $\sum_{k\le K} T_u({\bf k}) $  is the net KE transfer from the $ {\bf u} $ modes outside the sphere to the $ {\bf u} $ modes inside the sphere due to the nonlinearity $ ({\bf u}\cdot\nabla){\bf u}$. Equivalently, $\sum_{k\le K} T_u({\bf k}) = -\Pi_u(K)$ of Eq.~(\ref{eq:fluid_flux}). 
 	
 	\item $\sum_{k\le K} \mathcal{F}_u({\bf k}) $ is the total energy transfer rate by the interaction force ${\bf F}_u({\bf k})$  to $ {\bf u(k)} $ modes inside the  sphere.
 	\item $\sum_{k\le K} \mathcal{F}_\mathrm{ext}({\bf k}) $ is the net KE injected by the external force ${\bf F}_\mathrm{ext}$ (red sphere of Fig.~\ref{fig:flux_B}).  For   $K > k_f$,    $\sum_{k\le K} \mathcal{F}_\mathrm{ext}({\bf k}) = \epsilon_\mathrm{inj}$ because  ${\bf F}_\mathrm{ext} = 0$ beyond $ k=k_f $.
 \end{enumerate}

The $ {\bf u}^< $ modes lose energy to $ {\bf u}^> $ and   ${\bf B}$ modes via nonlinear interactions.  The term $-\sum_{k\le K} \mathcal{F}_u({\bf k}) $ of Eq.~(\ref{eq:Pi_K_from_Ek}) represents the net energy transfer from the ${\bf u}^<$ modes (those inside the sphere) to all the ${\bf B}$ modes ($ {\bf B}^< $ and ${\bf B}^> $) via the interaction force $\mathbf{F}_u({\bf k})$. We define the corresponding flux $\Pi_B(K)$ as
 \be
 \Pi_B(K)=-\sum_{k\le K} \mathcal{F}_u({\bf k}) .
 \ee
Thus,   $ {\bf u}^<$ modes   lose energy to $ {\bf u}^>$ modes, as well as to \textbf{B} modes, via nonlinear interactions.  In addition,  $ {\bf u}^<$ modes lose energy via viscous dissipation, which is the last term of Eq.~(\ref{eq:Pi_K_from_Ek}). Therefore, under steady state,   the kinetic energy injected  by ${\bf F}_\mathrm{ext}$  must match (statistically)  with the sum of   $\Pi_u(K)$, $ \Pi_B(K)$, and the viscous dissipation rate~\cite{Verma:book:ET,Verma:JPA2022}\footnote{In this paper we do not discuss the energetics of ${\bf B¯}$ field because TDR is related to the energy fluxes associated with the velocity field.}. That is, 
 \be
 \Pi_u(K) + \Pi_B(K) + \sum_{k \le K} D_u({\bf k}) = \epsilon_\mathrm{inj}.
 \ee
 In the inertial range where $D_u({\bf k}) \approx 0$, we obtain
 \be
 \Pi_u(K) + \Pi_B(K)  \approx \epsilon_\mathrm{inj}.
 \label{eq:Pi_tot}
 \ee

In later sections, we show that   $\Pi_B(k)>0$ in MHD, QSMHD,  polymeric, and stably-stratified turbulence.   Therefore, using Eq.~(\ref{eq:Pi_tot}) we deduce that for the same injection rate $\epsilon_\mathrm{inj}$,  $\Pi_u(k) $ in the mixture (with field ${\bf B}$) is lower than that in HD turbulence, that is,
 \be
 \Pi_{u,\mathrm{mix}} < \Pi_{u,\mathrm{HD}}.
 \ee

Now we estimate the drag force in the presence of $ {\bf B} $.  As discussed below, there are several ways to estimate this drag force.
\begin{enumerate}
	\item As  discussed in Section~\ref{sec:Hydro}, we average  Eq.~(\ref{eq:U2}) over small wavenumbers. Using 
\be
		\int_\mathrm{LS} d{\bf r}  [{\bf F}_{\mathrm{ext}} \cdot {\bf u}]  = \int_\mathrm{LS}  d{\bf r}  [{\bf F}_{\mathrm{drag}} \cdot {\bf u}] = f_2 U F_{\mathrm{drag,mix}} .
\ee
Under steady state, using Eqs.~(\ref{eq:Eu_dot_Fext_hydro},\ref{eq:fluid_flux}) we deduce that 
\be
\int_\mathrm{LS} d{\bf r}  [{\bf F}_{\mathrm{ext}} \cdot {\bf u}]  =  - \int_0^{k_f} [T_u(k') + \mathcal{F}_u(k')]dk' = \Pi_u(k) + \Pi_B(k).
\ee
Hence,
\be
F_{\mathrm{drag,mix}}  \approx \frac{\Pi_u+ \Pi_B}{ f_2 U}  \approx \frac{\epsilon_\mathrm{inj}}{ f_2 U} .  
\label{eq:FD_secondary}
\ee
It is observed that in a mixture,  $ U $  is typically larger than that in HD turbulence~\cite{Sreenivasan:JFM2000,Verma:PP2020}. Computation of $ f_2 $ may be quite complex, and it is difficult to compare $ f_1 $ and $ f_2 $. Still, considering $ U_\mathrm{mix} > U_\mathrm{HD} $, we expect  $ F_{\mathrm{drag,mix}}   $  to be weaker than the corresponding drag in HD turbulence. This is the origin of TDR  in the bulk when $ {\bf B} $ field (polymers or magnetic field) is present.

\item Considering the uncertainties  in $ f_2 $, it is proposed that turbulent drag is proportional to $ ({\bf u}\cdot\nabla){\bf u} $ \cite{Verma:PP2020}. For MHD turbulence, the force $ {\bf F}_u $, which is the Lorentz force, may be treated separately, and $ ({\bf u}\cdot\nabla){\bf u} $ may be considered as the drag force. This assumption simplifies the calculation with 
\be 
F_\mathrm{drag,mix} \approx \frac{ \Pi_u }{U}.
\ee
In a typical scenario,  $ \Pi_{u,\mathrm{mix}}< \Pi_{u,\mathrm{HD}} $, and $ U_\mathrm{mix} > U_\mathrm{HD} $ \cite{Sreenivasan:JFM2000,Verma:PP2020}. Therefore, we expect that
\be
F_{\mathrm{drag,mix}} <  F_{\mathrm{drag,HD}}.
\ee 
Thus, turbulent drag is reduced in the presence of a secondary fields, such as magnetic field and polymers.   Verma \etal \cite{Verma:PP2020} adopted this scheme for the computation of  turbulent drag.  We will use this scheme throughout the paper.
\end{enumerate}

\begin{figure}[tbhp]
	\centering
	\includegraphics[scale = 0.5]{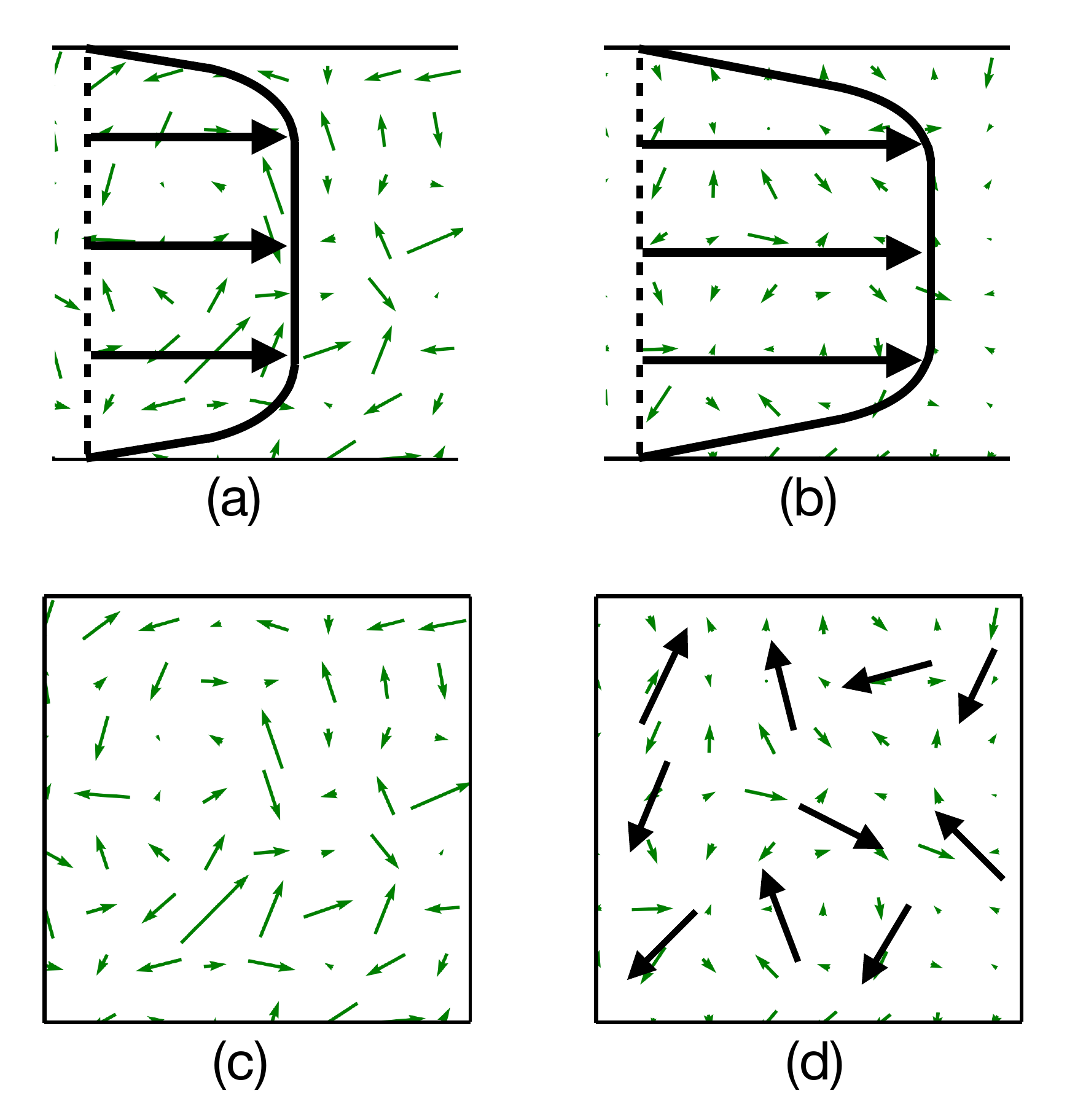}
	\caption{(a) Mean velocity profile (D profile) and fluctuations (green arrows)  in  a pipe flow without polymers. (b) With dilute polymers, the mean flow is enhanced, but the fluctuations are suppressed. (c) Velocity fluctuations in HD turbulence. (d)  With polymers and magnetic field, the fluctuations (green arrows) are suppressed, but  the large-scale $ U $ (black arrows) is enhanced.}
	\label{fig:drag_red_schematic}
\end{figure}

 In Fig.~\ref{fig:drag_red_schematic}, we present a schematic diagram illustrating TDR in a pipe flow and in bulk turbulence. An introduction of polymers in a pipe flow weakens the fluctuations and enhances the mean flow (see Fig.~\ref{fig:drag_red_schematic}(a,b)).  Similarly, in  bulk turbulence, polymers and magnetic field can induce strong large-scale $ U $ and weaken the  fluctuations in comparison to HD turbulence (see Fig.~\ref{fig:drag_red_schematic}(c,d)).

 We propose the following  drag coefficients to quantify TDR in the bulk:
\bea
\bar{C}_{d1} &= & \frac{ \la \Pi_u \ra }{U^3/L},
\label{eq:c1} \\
\bar{C}_{d2} & = & \frac{ \la \lvert ({\bf u \cdot \nabla}) {\bf u} \rvert \ra }{U^2/L},
\label{eq:c2}
\eea
where $ L $ is  the integral length scale, and $ U $  is the large-scale velocity. We obtain $ \bar{C}_{d1}  \approx 1 $ and $ \bar{C}_{d2}\approx 1 $ for HD turbulence. However,  $ \bar{C}_{d1} $ and $ \bar{C}_{d2} $  for a mixture are smaller than those for HD turbulence.  In subsequent sections, we will compute the above drag coefficients for a variety of flows, but with an emphasis on MHD and QSMHD turbulence, and dynamo.

In the next section, we provide a brief introduction to TDR in a turbulent flow with dilute polymers.

\section{TDR in flows with dilute polymers via energy flux}
\label{sec:polymer}

An introduction of small amount of polymers in a turbulent flow suppresses  turbulent drag \cite {Lumley:ARFM1969,Tabor:EPL1986,deGennes:book:Intro,deGennes:book:Polymer,Sreenivasan:JFM2000,Lvov:PRL2004,Benzi:PRE2003,White:ARFM2008,Benzi:PD2010,Benzi:ARCMP2018,Verma:PP2020}.  As discussed in Section \ref{sec:intro}, TDR in polymeric turbulence depends on the boundaries, bulk turbulence, properties of fluids and polymers, anisotropy, etc.  However, in this paper we focus on the TDR due to suppression of KE  flux in the presence of polymers. For detailed discussions on  TDR due to polymers, refer to the  references \cite {Lumley:ARFM1969,Tabor:EPL1986,deGennes:book:Intro,deGennes:book:Polymer,Sreenivasan:JFM2000,Lvov:PRL2004,Benzi:PRE2003,White:ARFM2008,Benzi:PD2010,Benzi:ARCMP2018,Verma:PP2020}.

One of the popular models for polymers is  \textit{finitely extensible nonlinear elastic-Peterlin model} (\textit{FENE-P})~\citep{Benzi:PD2010,Perlekar:PRL2006}.  In this model, the  governing equations for the velocity field $ {\bf u} $ and configuration tensor $ \mathcal{C}$ are~\citep{Sagaut:book, Benzi:PD2010, Fouxon:PF2003}
\bea
\frac{\partial{u_i}}{\partial t} +u_j \partial_j u_i
& = & -\partial_i p /\rho +  \nu \partial_{jj} u_i + \frac{\mu}{\tau_p} \partial_j (f   \mathcal{C}_{ij})
+ F_{\mathrm{ext}, i}, \label{eq:tensor:FENE-u} \\
\frac{\partial{\mathcal{C}_{ij}}}{\partial t} +  u_l \partial_l \mathcal{C}_{ij}
& = &     \mathcal{C}_{il} \partial_l u_j +  \mathcal{C}_{jl} \partial_l u_i +  \frac{1}{\tau_p}   [f \mathcal{C}_{ij} -\delta_{ij} ],
\label{eq:tensor:FENE-C} \\
\partial_i u_i  & = & 0, 
\eea
where  $\rho$ is the mean density of the solvent, $\nu$ is the kinematic viscosity, $\mu$ is an additional viscosity parameter, $\tau_p$ is the polymer relaxation time, and $f$ is the renormalized Peterlin's function.    In the above equations, the following forces are associated with ${\bf u}$ and $\mathcal{C}$ (apart from constants)~\cite{deGennes:book:Intro,Perlekar:PRL2006,Benzi:ARCMP2018,Verma:book:ET,Verma:JPA2022}:
\bea
F_{u,i} & = &   \partial_j (f \mathcal{C}_{ij}), \\
F_{u,i}({\bf k}) & = &  \sum_{\bf p}  \left[  i k_j f({\bf q}) \mathcal{C}_{ij}({\bf p}) \right], \\
\mathcal{F}_u({\bf k}) & = & \Re[F_{u,i}({\bf k}) u_i^*({\bf k})] =  -c_1 \sum_{\bf p} \Im \left[  k_j f({\bf q}) \mathcal{C}_{ij}({\bf p})u^*_i({\bf k}) \right],
\eea
where  ${\bf q=k-p}$, and $ c_1 $ is a constant.  Note that the field   $\mathcal{C}$  replaces  ${\bf B}$ of Eqs. (\ref{eq:U2}-\ref{eq:incompress_B}). Using the above equations, we derive the energy flux $\Pi_\mathcal{C}(K)$, which is the net energy transfer rate from $ {\bf u}^<$ to $ \mathcal{C} $, as~\cite{Verma:book:ET,Verma:JPA2022} 
\bea
\Pi_\mathcal{C}(K) &= & \sum_{k \le K}  \sum_{\bf p}  -c_1 \Im \left[  k_j f({\bf q}) \mathcal{C}_{ij}({\bf p})u^*_i({\bf k}) \right] 
\eea
with $ {\bf q=k-p} $.
 \begin{figure}%[tbhp]
	\centering
	\includegraphics[width=0.8\linewidth]{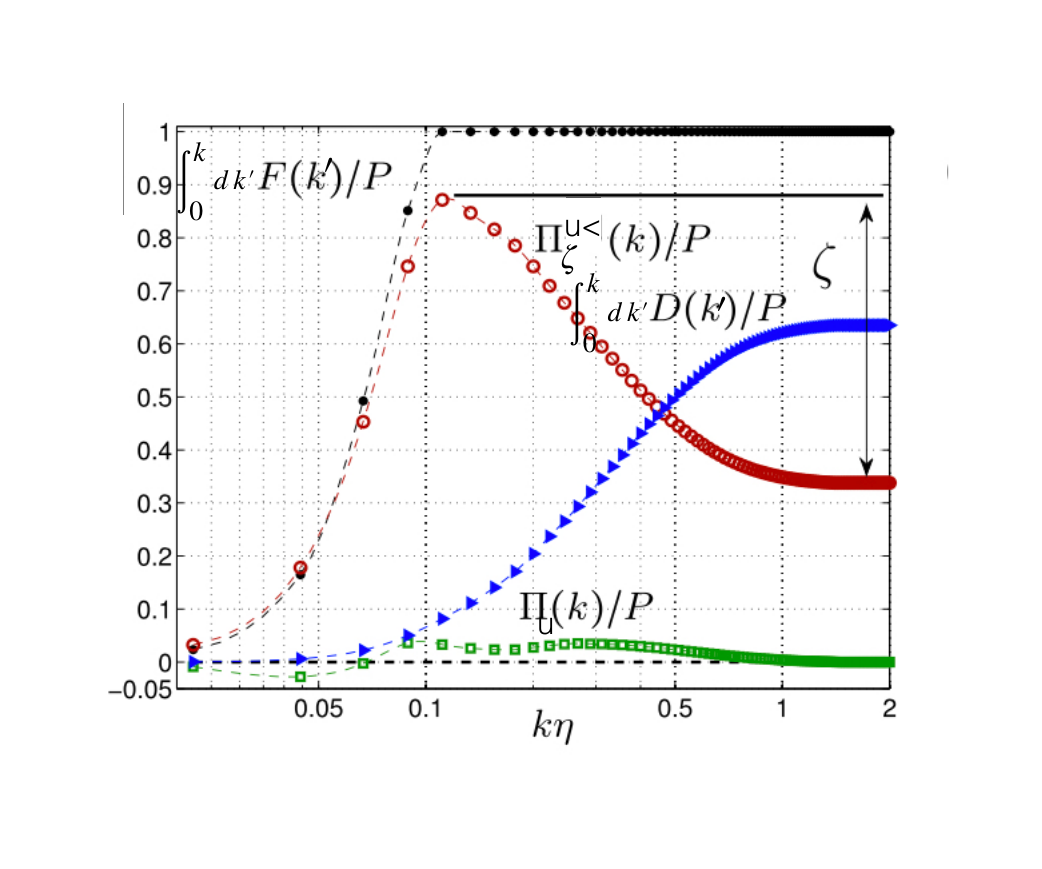}
	\caption{For a polymeric flow with De = 16.2, the energy fluxes $ \Pi_u(k) $ and $ \Pi_\mathcal{C}(k)$ normalized with the KE injection rate $ P $, and dissipation rate $ D_u(k) $~\cite{Valente:PF2016}.  The injected KE, $ P $, is transferred to $ {\bf u}^> $ and $ \mathcal{C} $  as $ \Pi_u(k)$ and $ \Pi_\mathcal{C}(k) $ respectively. The rest of the injected energy is dissipated. Adapted from a figure from Valente {\em et al.} \cite{Valente:PF2016}. Reprinted with the permission of AIP.}
	\label{fig:polymer}
\end{figure}

Valente \etal \cite{Valente:JFM2014,Valente:PF2016} analysed the energy fluxes $\Pi_u(k)$ and $\Pi_\mathcal{C}(k)$ in a turbulent flow with dilute polymers and observed that  $\Pi_\mathcal{C}(k) >0$.  One of their figures illustrating $\Pi_u(k)$ and $\Pi_\mathcal{C}(k)$  is reproduced in Fig.~\ref{fig:polymer} \cite{Valente:PF2016}.  As shown in the figure, for $\mathrm{De} = 16.2$, $\Pi_\mathcal{C}(k)/P$ ($ P $ = total injected power) peaks at approximately 0.9 when $ k\eta \approx 0.1 $, where $\eta$ is Kolmogorov's wavenumber. However, $\Pi_u(k)/P$ remains less than 0.1 for all $ k\eta $.  Valente \etal \cite{Valente:JFM2014,Valente:PF2016} also reported that $\Pi_u(k)$ and $\Pi_\mathcal{C}(k)$ depend on the Deborah number, $\mathrm{De}$,  which is the ratio of the relaxation  time scale of the polymer and the characteristic time scale for the energy cascade.   Notably, $\Pi_\mathcal{C}(k)$ is maximum when $\mathrm{De} \sim 1$. Thus,  Valente \etal \cite{Valente:JFM2014,Valente:PF2016}  showed that $\Pi_u(k)$ is  reduced  significantly from $ \epsilon_\mathrm{inj}  $ due to the energy transfer from the velocity field to polymers. That is, $ \Pi_u(k) < \epsilon_\mathrm{inj}  $. 

%By simulating various shell models, Benzi \etal \cite{Benzi:PRE2003} and  Ray and Vincenzi~\cite{Ray:EPL2016} arrived at similar conclusions. 
 
\begin{figure}%[tbhp]
	\centering
	\includegraphics[width=0.6\linewidth]{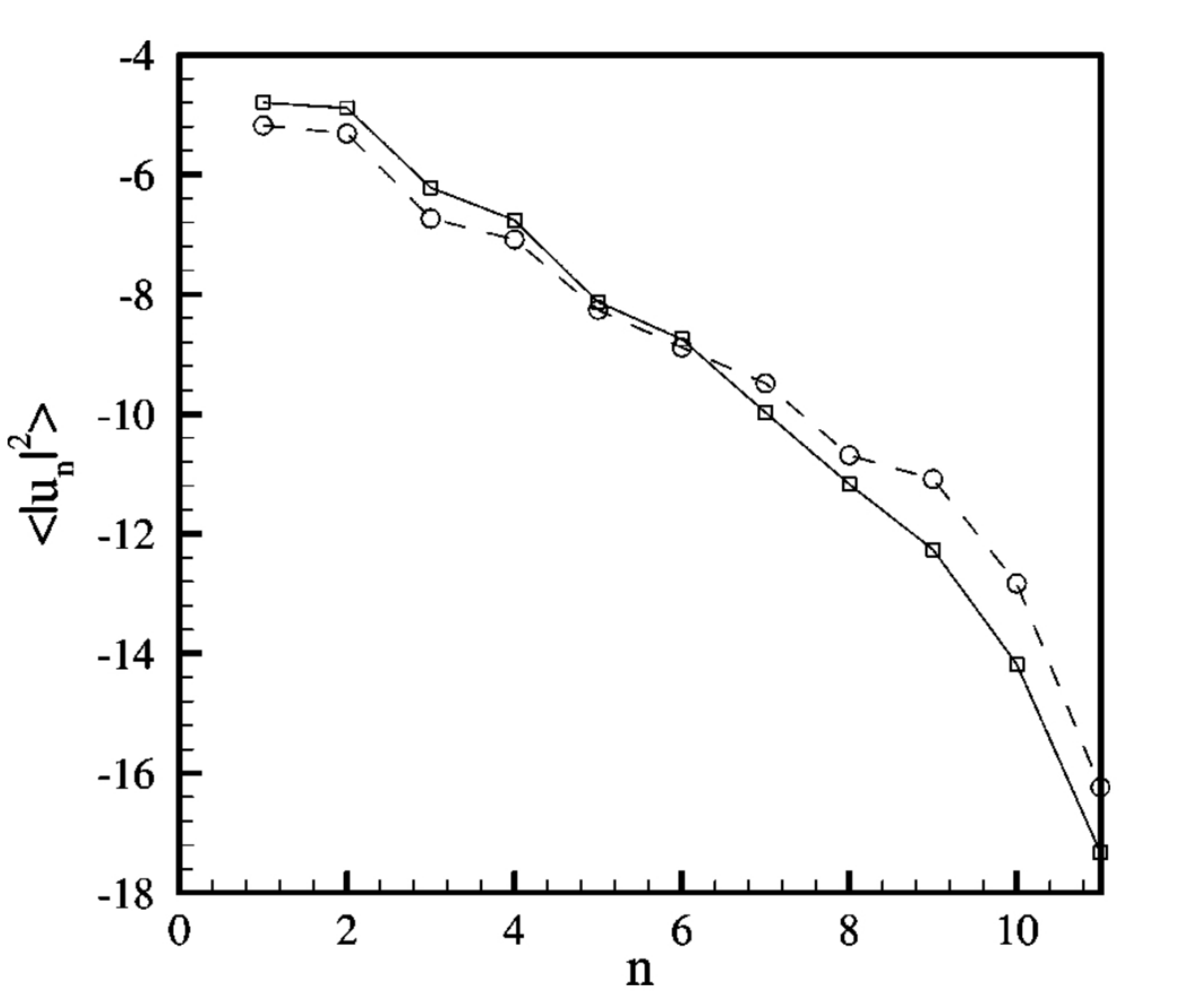}
	\caption{KE spectra for pure HD turbulence (dashed line with circle)  and polymeric turbulence (solid line with squares). At small wavenumbers, $ E_u(k) $ with polymers is larger than that without polymers. 	From Benzi \etal \cite{Benzi:PRE2003}. Reprinted with permission from APS. }
	\label{fig:Benzi}
\end{figure}

Benzi \etal \cite{Benzi:PRE2003} and 
Ray and Vincenzi \cite{Ray:EPL2016}  showed that during TDR, the large-scale KE is enhanced compared to HD turbulence.  Figure~\ref{fig:Benzi}  illustrates the energy spectra of  Benzi \etal for pure HD and polymeric turbulence. In the figure we observe that at small wavenumbers,  $ E_u(k) $ is larger for  polymeric turbulence than that for HD turbulence.  Hence, we deduce that large-scale $ U $ is enhanced in the presence of polymers.  Thais et al.~\cite{Thais:IJHFF2013} and Nguyen et al.~\cite{Nguyen:PRF2016} arrived at  similar conclusions using direct numerical simulation of polymeric turbulence.  Based on these observations, we deduce that
\be
\Pi_{u,\mathrm{Polymeric}} <  \Pi_{u,\mathrm{HD}}~~~\mathrm{and}~~~U_\mathrm{Polymeric} >   U_\mathrm{HD}.
\ee   
Therefore, using  $ F_\mathrm{drag} =  \Pi_{u}/U$, we deduce that
\be
F_{\mathrm{drag,Polymeric}} <  F_{\mathrm{drag,HD}}.
\ee  
Thus, reduction in KE flux leads to a decrease in nonlinearity, and hence, TDR in polymeric turbulence.

L'vov et al.~\cite{Lvov:PRL2005} and others have observed TDR in flows with bubbles.   In a bubbly flow, the KE is transferred to the elastic energy of the bubbles that leads to TDR.   We also remark  that in the laminar regime,  the polymers induce additional drag via the term $\mu\partial_j (f   \mathcal{C}_{ij})/ \tau_p $ of Eq.~(\ref{eq:tensor:FENE-u}). Hence, polymers enhance the drag in the viscous limit  \cite{Sreenivasan:JFM2000}.  Also note that in the present review, we focus on  TDR in bulk turbulence and have avoided discussions on boundary layers, anisotropy, effects of polymer concentration, etc.   

Earlier, Fouxon and Lebedev~\citep{Fouxon:PF2003} had related the equations of a turbulent flow with dilute polymers to those of MHD turbulence.  In the next section, we will show that the  energy transfers in MHD turbulence are similar to those in polymeric turbulence.

\section{TDR in MHD turbulence via  energy flux}
\label{sec:MHD}

\textit{Magnetofluid} is quasi-neutral and highly conducting charged fluid, and its dynamics is described by magnetohydrodynamics (MHD). Our universe is filled with magnetofluids, with prime examples being solar wind, solar corona, stellar convection zone, interstellar medium, and intergalactic medium \cite{Cowling:book,Priest:book,Goldstein:ARAA1995} .  
 
The equations for incompressible MHD  are~\cite{Cowling:book,Priest:book}  
\bea
\frac{\partial{\bf u}}{\partial t} + ({\bf u}\cdot\nabla){\bf u}
& = & -\nabla({p}/{\rho}) +  \nu\nabla^2 {\bf u} + {\bf F}_u({\bf B,B}) +  {\bf F}_\mathrm{ext},  \label{eq:U_MHD} \\ 
\frac{\partial{\bf B}}{\partial t} + ({\bf u}\cdot\nabla){{\bf B}}
& = &   \eta \nabla^2 {{\bf B}} + {\bf F}_B({\bf B,u}),   \label{eq:W} \\
\nabla \cdot {\bf u}  & = & 0,\\
\nabla \cdot {\bf B}  & = & 0, \label{eq:b_incompress}
\eea
where ${\bf u,B}$ are  the velocity and magnetic fields respectively; $p$ is the total (thermal + magnetic) pressure; $\rho$ is the density which is assumed to be unity;   $\nu$ is  the kinematic viscosity;  $ \eta$ is the magnetic diffusivity;   $ {\bf F}_\mathrm{ext}$ is the external force employed  at large scales; and 
\bea
{\bf F}_u & = &  {\bf (B \cdot \nabla) B}, \\
{\bf F}_B & = & {\bf (B \cdot \nabla) u}
\eea 
represent respectively the Lorentz force and the stretching of the magnetic field by the velocity field. Note that  $ {\bf F}_u $ and $ {\bf F}_B $ induce  energy exchange among $ {\bf u} $ and $ {\bf B} $ modes. In the above equations, the magnetic field ${\bf B}$ is in  velocity units, which is achieved by ${\bf B}_\mathrm{cgs} \rightarrow {\bf B}_\mathrm{cgs}/\sqrt{4\pi \rho}$.

The evolution equation for the modal kinetic energy $E_u({\bf k}) = \lvert{\bf u(k)}\rvert^2/2$ is~\citep{Kraichnan:JFM1959,Frisch:book,Dar:PD2001, Verma:PR2004,Davidson:book:Turbulence,Verma:book:ET,Verma:JPA2022}   
\bea
\frac{d}{dt} E_u(\mathbf{k})  & = & T_{u}({\bf k})  +  \mathcal{F}_u({\bf k})  + \mathcal{F}_\mathrm{ext}({\bf k})-D_u(\mathbf{k}),
\label{eq:MHD_ET:Eu_dot_Fext} 
\eea
where 
\bea
T_u({\bf k}) & = & \sum_{\bf p} \Im \left[ {\bf  \{  k \cdot u(q) \} \{ u(p) \cdot u^*(k) \} } \right] , \\
\mathcal{F}_u({\bf k}) & =  & \Re[ {\bf F}_u({\bf k}) \cdot {\bf u}^*({\bf k})  ] = \sum_{\bf p} - \Im \left[ {\bf  \{  k \cdot B(q) \} \{ B(p) \cdot u^*(k) \} } \right],  \label{eq:Fu_mhd} \\
\mathcal{F}_\mathrm{ext}({\bf k}) & = &   \Re[ {\bf F}_\mathrm{ext}({\bf k}) \cdot {\bf u}^*({\bf k})  ], \\
D_u(\mathbf{k}) & = & 2 \nu k^2 E_u({\bf k}),
\eea
with  ${\bf q=k-p}$.  Summing  Eq.~(\ref{eq:MHD_ET:Eu_dot_Fext}) over  the modes of the wavenumber sphere of radius $K$ yields~\cite{Davidson:book:Turbulence,Sagaut:book,Verma:book:BDF}:
\bea
-\frac{d}{dt} \sum_{k \le K}  E_u({\bf k})   &= & - \sum_{k \le K} T_u({\bf k}) - \sum_{k \le K}\mathcal{F}_u({\bf k})    - \sum_{k \le K}\mathcal{F}_\mathrm{ext}({\bf k}) + \sum_{k \le K} D_u({\bf k}) \nonumber \\
 & = & \Pi_u(K) + \Pi_B(K)  - \epsilon_\mathrm{inj} + \mathrm{total~viscous~dissipation}.
\label{eq:ET:Pi_k0_from_Ek}
\eea  
Note that
\be
\Pi_B(K) = - \sum_{k \le K}  \mathcal{F}_u({\bf k}) = \sum_{k \le K} \sum_{\bf p}  \Im \left[ {\bf  \{  k \cdot B(q) \} \{ B(p) \cdot u^*(k) \} } \right].
\ee
In Fig.~\ref{fig:flux_B}, we illustrate $ \Pi_B(K) $ using the red arrows. 

%Note that $ \Pi_B(K) $ represents energy transfers from $ {\bf u}^> $ to all the magnetic modes.  In Fig.~\ref{fig:Mininnii}, we illustrate the positivity of  $ \mathcal{F}_u({\bf k})$, and hence show that $ \Pi_B(K) >0$. In the next section, we will show these results using direct numerical simulations and shell models.

\begin{figure}%[tbhp]
	\centering
	\includegraphics[scale=0.7]{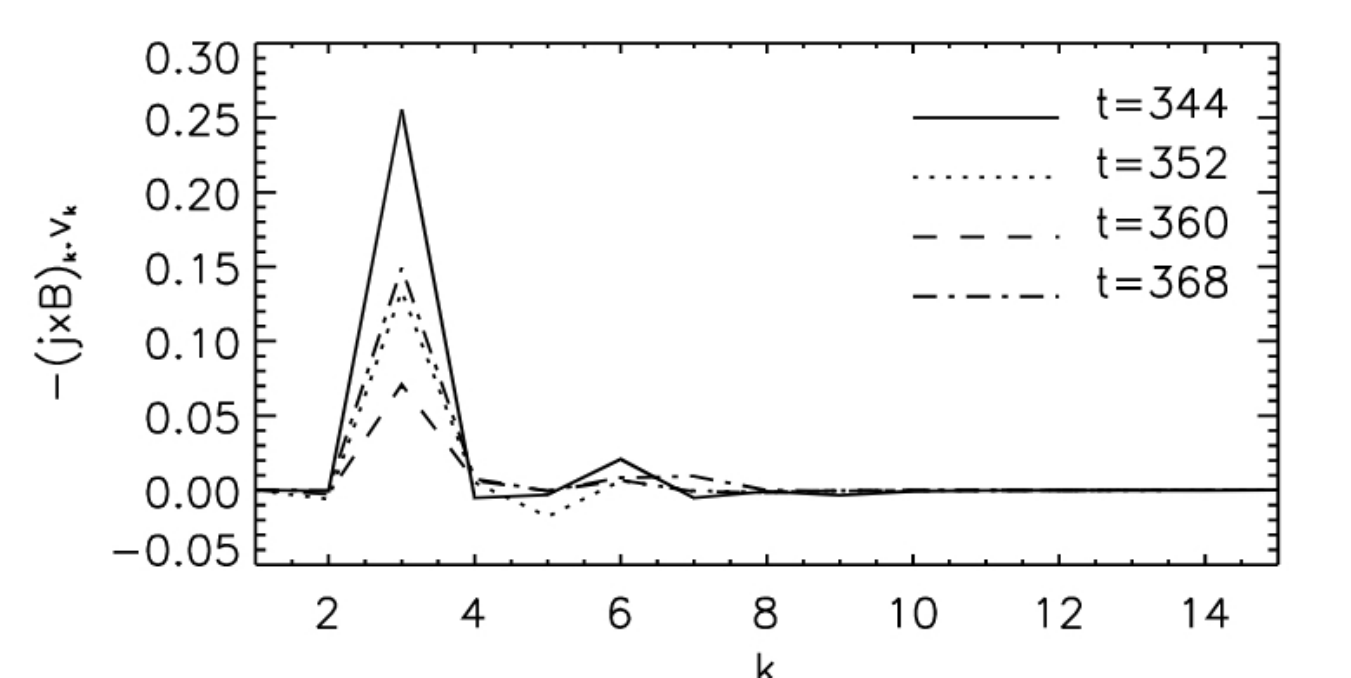}
	\caption{  Numerically computed $ \mathcal{F}_u({\bf k}) = \Re[{\bf [J \times B](k)} \cdot {\bf u^*(k)}] $ by Mininni \etal \cite{Mininni:ApJ2005}. Clearly, 	$ \mathcal{F}_u({\bf k}) > 0$, and hence $ \Pi_B(K) > 0$. 	From Mininni \etal \cite{Mininni:ApJ2005}.  Reproduced with permission from ApJ.}
	\label{fig:Mininni}
\end{figure}

Under a steady state ($d E_u({\bf k}) /dt =0$),   
\be
\Pi_u(K) + \Pi_B(K) + \sum_{k \le K} D_u({\bf k}) = \epsilon_\mathrm{inj}.
\ee
In the inertial range where $D_u({\bf k}) \approx 0$, we obtain
\be
\Pi_u(K) + \Pi_B(K)  \approx \epsilon_\mathrm{inj}.
\label{eq:Pi_sum}
\ee
Following similar lines of arguments as in Section~\ref{sec:drag}, we estimate the turbulent drag in MHD turbulence as
\be
\la F_{\mathrm{drag,MHD}} \ra \approx \la  { \lvert \bf (u \cdot \nabla) u \rvert}  \ra_\mathrm{LS} \approx    \frac{\Pi_u}{U} \approx  \frac{ \epsilon_\mathrm{inj}- \Pi_B}{U} .
\label{eq:Pi_B_MHD}
\ee
Researchers have studied the energy fluxes $\Pi_u$ and $\Pi_B$ in detail for various combinations of parameters---forcing functions, boundary condition, $\nu$ and $\eta$ (or their ratio $\mathrm{Pm} = \nu/\eta$, which is called the {\em magnetic Prandtl number}).   For example, Dar \etal \cite{Dar:PD2001}, Debliquy \etal \cite{Debliquy:PP2005},   Mininni \etal \citep{Mininni:ApJ2005}, and Kumar \etal \cite{Kumar:EPL2014, Kumar:JoT2015} computed the fluxes  $\Pi_u$ and $\Pi_B$ using numerical simulations and observed that   $\Pi_B > 0$ on most occasions.  Using numerical simulations, Mininni \etal \cite{Mininni:ApJ2005} showed that $ \mathcal{F}_u({\bf k}) < 0$, and hence $ \Pi_B({\bf k}) > 0$ (see Fig.~\ref{fig:Mininni}).

Hence, using Eq.~(\ref{eq:Pi_B_MHD}) we deduce that
\be
\Pi_{u,\mathrm{MHD}} <  \Pi_{u,\mathrm{HD}}.
\label{eq:Pi_MHD_reduced}
\ee   
That is, the KE flux in MHD turbulence is lower than the corresponding flux in HD turbulence (without magnetic field).   In addition, the speed $ U  $ may increase under the inclusion of magnetic field. Therefore,  using  $ F_\mathrm{drag} =  \Pi_{u}/U$, we deduce that
\be
F_{\mathrm{drag,MHD}} <  F_{\mathrm{drag,HD}}.
\ee 

In this next section, we will explore whether the above inequality holds in  numerical simulations of MHD turbulence.

\section{Numerical verification of TDR  in MHD turbulence }
\label{sec:dns}

Many researchers have simulated MHD turbulence, but TDR in MHD turbulence has not been explored in detail.  In this section, we will present numerical results on TDR from direct numerical simulations (DNS) and shell models.

\begin{figure}%[tbhp]
	\centering
	\includegraphics[scale=0.7]{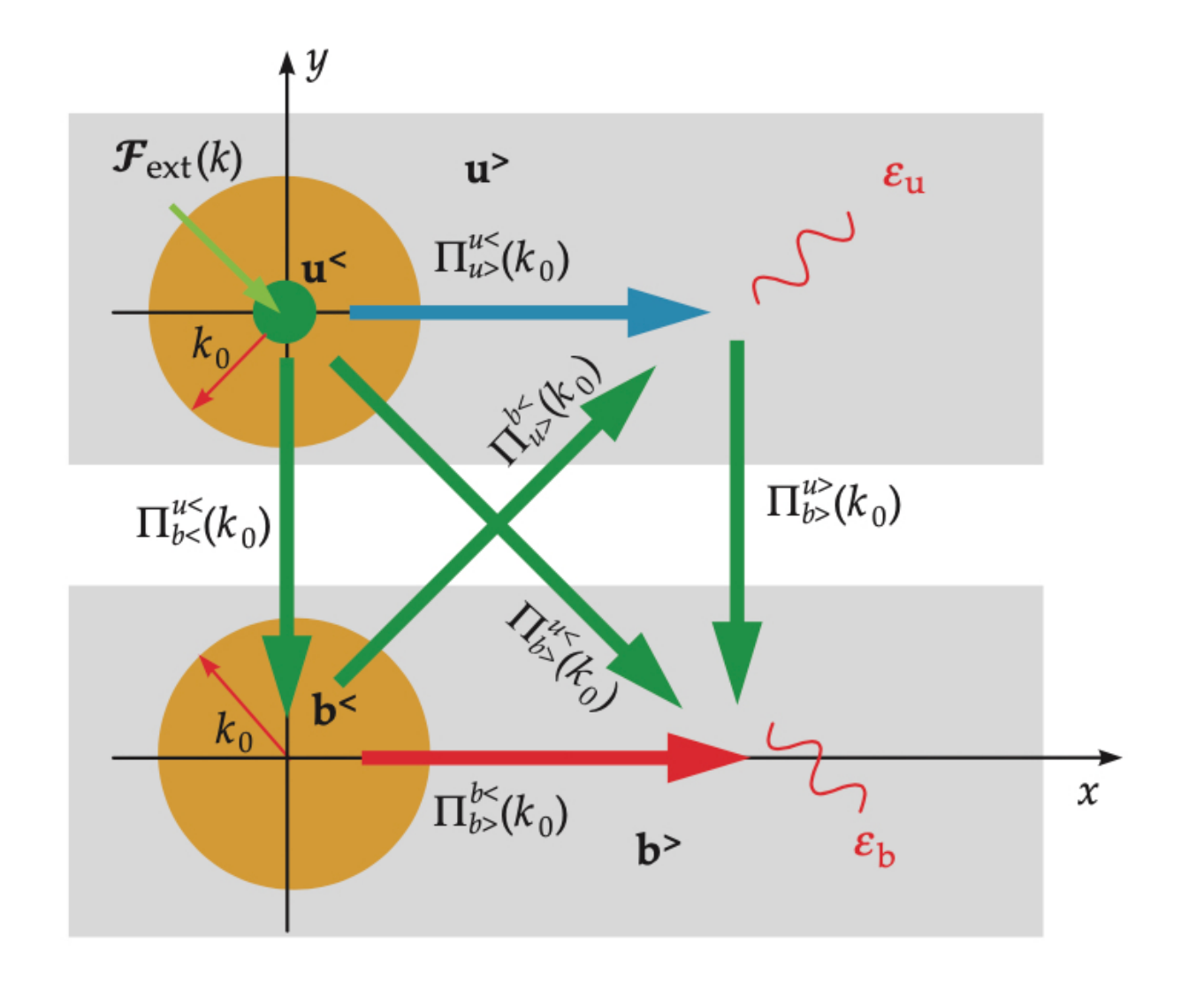}
	\caption{Six energy fluxes of MHD turbulence: $ \Pi^{u<}_{u>} $, $ \Pi^{u<}_{b<} $, $ \Pi^{u<}_{b>} $, $ \Pi^{b<}_{b>} $, $ \Pi^{b<}_{u>} $, $ \Pi^{u>}_{b>} $. From Verma~\cite{Verma:book:ET}.  Reproduced with permission from Verma.}
	\label{fig:mhd_fluxes}
\end{figure}

MHD turbulence exhibits six energy fluxes that are shown in Fig.~\ref{fig:mhd_fluxes}.  These fluxes represent energy transfers from $ u^< $ and $ u^> $ to $ b^< $ and $ b^> $ \cite{Dar:PD2001,Verma:PR2004,Verma:book:ET}. 
However, as we discussed in Section~\ref{sec:drag}, the relevant fluxes for TDR are $ \Pi_u $ and $ \Pi_B $. Also, TDR takes place at large scales, hence, we consider energy fluxes  from small wavenumber spheres.  In terms of the  fluxes of Fig.~\ref{fig:mhd_fluxes},
\bea
\Pi_u(K) & = & \Pi^{u<}_{u>}(K), \\
\Pi_B(K) & =  & \Pi^{u<}_{b<}(K) +  \Pi^{u<}_{b>}(K).
\eea
As discussed in Section \ref{sec:MHD}, $ \Pi_B > 0 $  \cite{Dar:PD2001,Verma:PRE2001,Debliquy:PP2005,Mininni:ApJ2005,Kumar:EPL2014, Kumar:JoT2015}.  Hence,  $ \Pi_u  < \epsilon_\mathrm{inj}$ that leads to TDR in MHD turbulence. In this section, we will report the energy fluxes and $\la \lvert ({\bf u}\cdot \nabla) {\bf u} \rvert  \ra$ for  HD and MHD turbulence from DNS and shell models, and compare them to quantify TDR in MHD turbulence. 

It is important to note that the velocity field receives parts of  $ \Pi_B $   via the energy fluxes $ \Pi^{b<}_{u>} $ and $ \Pi^{b>}_{u>}  $. However, these transfers are effective at intermediate and  large wavenumbers. In this review we focus on small wavenumbers, hence we can ignore these energy transfers.  In the following subsection, we discuss TDR in DNS  of MHD turbulence.

%%%%

\subsection{TDR in direct numerical simulation of MHD turbulence}
We solve the nondimensional MHD equations~(\ref{eq:U_MHD}-\ref{eq:b_incompress}) using pseudo-spectral code TARANG~\cite{Boyd:book:Plasma,Verma:Pramana2013tarang,Chatterjee:JPDC2018} in a cubic periodic box of size  $(2\pi)^3$.   We nondimensionalize  velocity, length, and  time  using the initial rms speed ($U_0$), box size $ (2\pi) $, and the initial eddy turnover time ($2\pi / U_0)$ respectively. We  employ the fourth-order Runge-Kutta (RK4) scheme for time marching; Courant-Friedrich-Lewis (CFL) condition for computing the time  step $\Delta t$; and  $2/3$ rule for dealising. We  perform our simulations on a $256^3$ grid  for  $ \mathrm{Pm}=1/3, 1, 10/3 $ (the details in the following discussion).  The mean magnetic field  ${\bf B}_0 = 0$.   Note that the $ 256^3 $ grid resolution is sufficient for computing the large-scale $ \Pi_u, \Pi_B $, and   $ \la {\bf u \cdot \nabla u} \ra $.  In addition, the low grid resolution helps us carry out  simulations for many eddy turnover times.

For the initial condition, we employ random velocity and magnetic fields at all wavenumbers.  For creating such fields, it is convenient to employ Craya-Herring basis~\cite{Craya:thesis, Herring:PF1974}, whose  basis vectors for wavenumber $ {\bf k} $ are
\begin{equation}
	\hat{\bf e}_3(\mathbf{k}) = \hat{\bf k};~~~\hat{\bf e}_1(\mathbf{k}) =  ( \hat{\bf k} \times \hat{\bf n})/ \lvert \hat{\bf k} \times \hat{\bf n}\rvert;~~~ \hat{\bf e}_2(\mathbf{k}) = \hat{\bf k} \times \hat{\bf e}_1(\mathbf{k})
\end{equation}
with $\hat{\bf n}$ along any arbitrary direction, and $\hat{\bf k}$ as the unit vector along  $\mathbf{k}$. We choose 3D incompressible flow, hence,
\begin{equation}
	\mathbf{u}(\mathbf{k}) = u_1(\mathbf{k}) \hat{\bf e}_1(\mathbf{k}) + u_2(\mathbf{k}) \hat{\bf e}_2(\mathbf{k}).
\end{equation}
For random initial velocity with the total kinetic energy as $ E_u $, we employ
\begin{eqnarray}
	u_1(\mathbf{k}) & = & \sqrt{(E_u/2N^3)} \ i \left(\exp(i \phi_1(\mathbf{k})) - \exp(i \phi_2(\mathbf{k})) \right), \\
	u_2(\mathbf{k}) & = & \sqrt{(E_u/2N^3)} \ \left(\exp(i \phi_1(\mathbf{k})) + \exp(i \phi_2(\mathbf{k}))\right),
\end{eqnarray}
where $N^3$ is the total number of modes, and the phases $\phi_1(\mathbf{k})$ and $\phi_2(\mathbf{k})$ are chosen randomly from uniform distribution in the band $[0, 2\pi]$. The above formulas ensure that the kinetic helicity remains zero. We employ $ E_u=0.5 $ for our simulation. A similar scheme is adopted for the random magnetic field with the   initial magnetic energy as 0.25.  We carry out the above run for $ \nu =  \eta = 0.01 $, or $ \mathrm{Pm}=1$.

We  employ random force to the velocity modes in a wavenumber shell $(2,3)$, denoted by $k_f = 2$,  so as to  achieve a steady state~\cite{Sadhukhan:PRF2019}.  The kinetic-energy injection rate $ \epsilon_\mathrm{inj}  = 0.4$. We carry out the  simulation till 29 eddy turnover times. Note, however, that the flow reaches a steady state in approximately $15$ eddy turnover times. 

At  the end of the above simulation, we perform four independent simulations given below.  We take the final state of the above run as the initial state  ($ t=0 $) for the following simulations.
\begin{enumerate}
	\item MHD1:  $ \nu = 0.01 $, $ \eta = 0.03$, and hence $ \mathrm{Pm}=1/3 $.

\item MHD2:   $ \nu = 0.01 $, $ \eta = 0.01$, and hence $ \mathrm{Pm}=1$. This is continuation of the  run described above.

\item MHD3:  $ \nu = 0.01 $, $ \eta = 0.003$, and hence $ \mathrm{Pm}=10/3 $.

\item HD: $ \nu = 0.01$ with magnetic field turned off.
\end{enumerate}
We carry out the HD and MHD2 simulations  till 40 eddy turnover times, whereas MHD1 and MHD3 runs till 5 eddy turnover times.  Subsequently, we  compare the energy fluxes and $\la \lvert ({\bf u}\cdot \nabla) {\bf u} \rvert  \ra$ of  the four runs after they have reached their respective steady states that occur in several eddy turnover times.   The Reynolds number ($ \mathrm{Re} = UL/\nu $) for the steady state of the HD run is 457. For the steady state of the MHD runs with $ \mathrm{Pm}=1/3, 1, $ and $10/3$,  $ \mathrm{Re} = $ 413, 347, and $338$ respectively, while $ \mathrm{Rm} = $ 137, 347  and $1127$ respectively. 
%%%%%%%%%%%%%%%%%%%%%%%%%%%%%%%%%%%
\begin{figure}%[tbhp]
	\centering
	\includegraphics[scale=0.5]{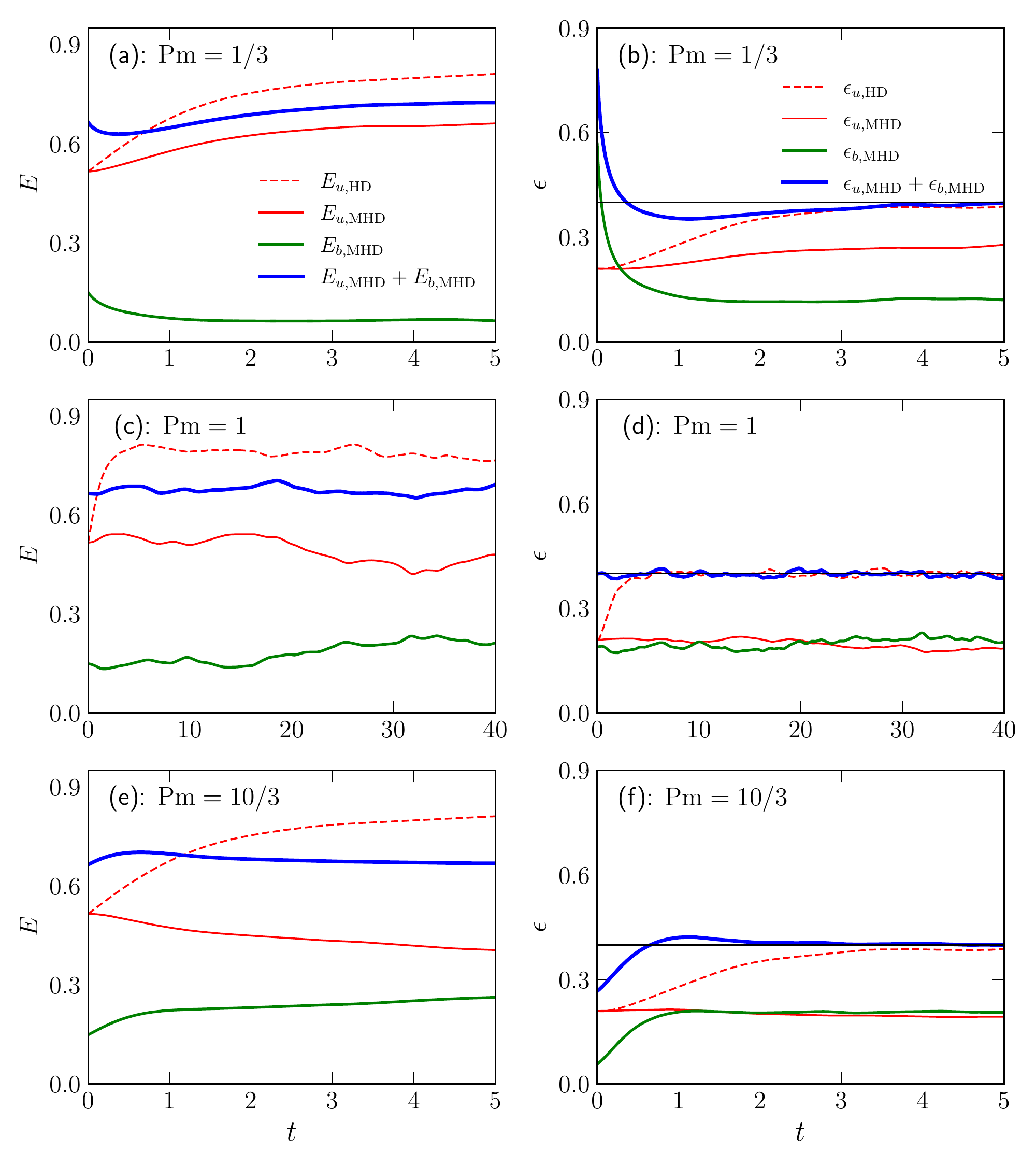}
	\caption{Left column: (a,c,e) Time series of KE of the HD run (dashed red curve); and KE  (solid red curve), magnetic energies (solid green curve), and total energies (solid blue curve) of the MHD runs for $ \mathrm{Pm}=1/3, 1, 10/3 $. Right column: (b,d,f) Corresponding energy dissipation rates with the same notation.}
	\label{fig:dns_energy_evolution}
\end{figure}

%%%%%%%%%%%%%%%%%%%%%%%%%%%%%%
%
%\begin{table}[h]
%		\begin{minipage}{\textwidth}
%			\caption{Simulation parameters: Grid resolution ($N$), energy injection rate ($\epsilon_{\rm{inj}}$), kinematic viscosity ($\nu$), magnetic diffusivity ($\eta$), Magnetic Prandtl number ($\mathrm {Pm}$), kinetic Reynolds number for hydrodynamic ($\mathrm {Re}_{\rm {HD}} = U L_b/ \nu$), kinetic Reynolds number for MHD ($\mathrm {Re}_{\rm {MHD}} = U L_b / \nu$) and magnetic Reynolds number ($\mathrm {Rm} = U L_b / \eta$), where $L_b = 2\pi$ is the box size. The parameters are averaged in time domain $t \in [20,40]$.} \label{tab:table1_dns}%
%			\vspace{10pt}%
%			\begin{center}
%			\begin{tabular}{@{}llllllll@{}}
%				\toprule
%				{$N$}	& $\epsilon_{\rm {inj}}$ & $\nu$ & $\eta$ & $\rm{Pm}$ & $\rm{Re}_{\rm {HD}}$ & {$\rm{Re}_{\rm {MHD}}$} & {${\rm {Rm}}$}\\
%				\midrule
%				$256^3$		& $0.4$	 & $0.01$& $0.01$ & $1$       & $457 $               &$ 348$ & $348$\\
%				\botrule
%			\end{tabular}
%		\end{center}
%		\end{minipage}
%\end{table}

%%%%%%%%%%%%%%%%%%%%%%%%%%%%%%%%%%%%%%%%%%%%%%%%%%%%%%%%%%%%%%%%%%%%%%%%%%%%%%%

In Fig.~\ref{fig:dns_energy_evolution} (left column), we exhibit the time  series of KE  of the HD run, and as well as KE, magnetic energies (ME), and the total energies of the three MHD runs. The corresponding dissipation rates are exhibited in the right column of Fig. \ref{fig:dns_energy_evolution}.  As shown in the figures, all the runs reach steady states after several eddy turnover times. The KE dissipation rate for the HD run increases rapidly to 0.4, which is the KE injection rate ($\epsilon_{\rm {inj}}$). The KE for the MHD runs  with  $ \mathrm{Pm} = 1/3, 1 $, and $10/3$ saturate respectively to approximate values of $0.65, 0.47$ and $0.41$, but the  respective magnetic energies  saturate at approximately  $0.07, 0.2$ and $0.26$. Note that energies for the MHD runs  exhibit significant fluctuations, however, the dissipation rates of the total energy remain at 0.4. 

%%%%%%%%%%%%%%%%%%%%%%%%%%%%%%%%%%%%%%%%%%%%%

\begin{figure}[tbhp]
	\centering
	\includegraphics[scale = 0.7]{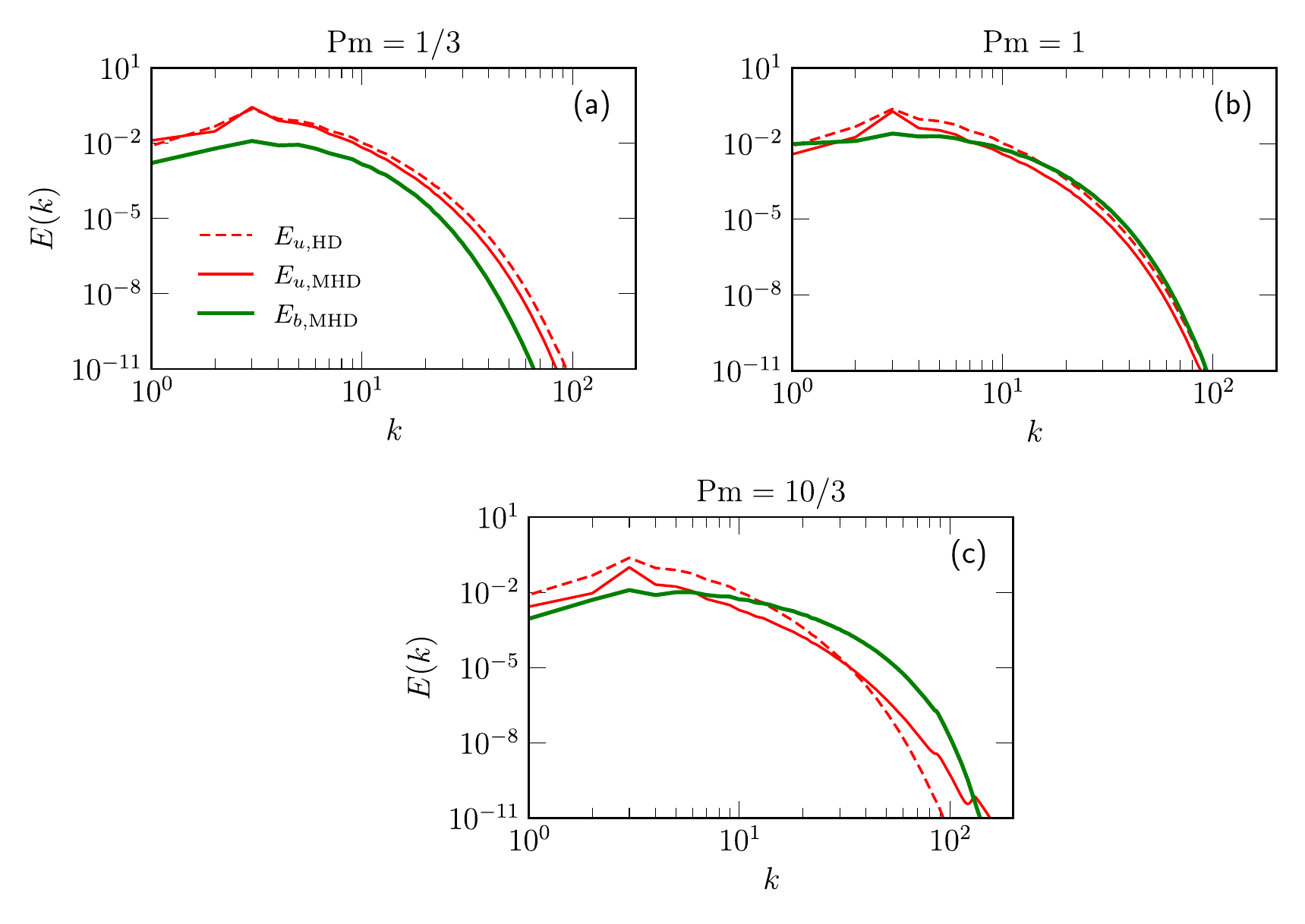}
	\caption{(a,b,c) For MHD runs with $ \mathrm{Pm} = 1/3, 1, 10/3 $,   the KE spectra (solid red curve) and the magnetic energy spectra (solid green curve). We also exhibit the plots of the KE spectra of the HD run (dashed red curve). }
	\label{fig:spectrum_comp_dns}
\end{figure}

Now, we report the  energy spectra for the velocity and magnetic fields for a wavenumber $ k $. Numerically, we compute them using
\begin{eqnarray}
	E_u(k) &=& \frac{1}{2} \sum_{k-1 < \lvert \mathbf {k'} \rvert \leq k} \lvert \bf{u}(\bf {k'}) \rvert^2, \label{eq:dns_ke} \\
	E_b(k) &=& \frac{1}{2} \sum_{k-1 < \lvert \mathbf {k'} \rvert \leq k} \lvert \bf{b}(\bf {k'}) \rvert^2 \label{eq:dns_me}. 
\end{eqnarray}
In Fig.~\ref{fig:spectrum_comp_dns}, we exhibit $ E_u(k)$ and $E_b(k) $ for the MHD runs, along with $ E_u(k)$ for the HD run. These quantities are averaged  over several time frames in the  steady state.   We observe that  $ E_u(k) $ for  the HD run is larger than those for the MHD runs, except at several small wavenumbers for $ \mathrm{Pm}=1/3 $ where $ E_b(k) > E_u(k) $.

Further, for the HD and MHD runs, we report the large-scale velocity $U$, integral length scales $L$, and Reynolds numbers based on  Taylor microscale, $\mathrm {Re}_{\lambda} = U \lambda/\nu$, where Taylor microscale $\lambda = (15 \nu U^2/\epsilon)^{1/2}$~\cite{Lesieur:book:Turbulence,Verma:book:ET}. Following Sreenivasan \cite{Sreenivasan:PF1998}, we compute $ U $  as the rms value for each component of the velocity field, or
\be
U = \left[ \frac{2}{3} \int dk E(k) \right]^{1/2},
\ee
whereas  the integral length $ L $ is computed using
\be
L = \frac{\int dk k^{-1} E(k)}{\int dk E(k)}.
\ee
\begin{figure}[tbhp]
	\centering
	\includegraphics[width=0.9\linewidth]{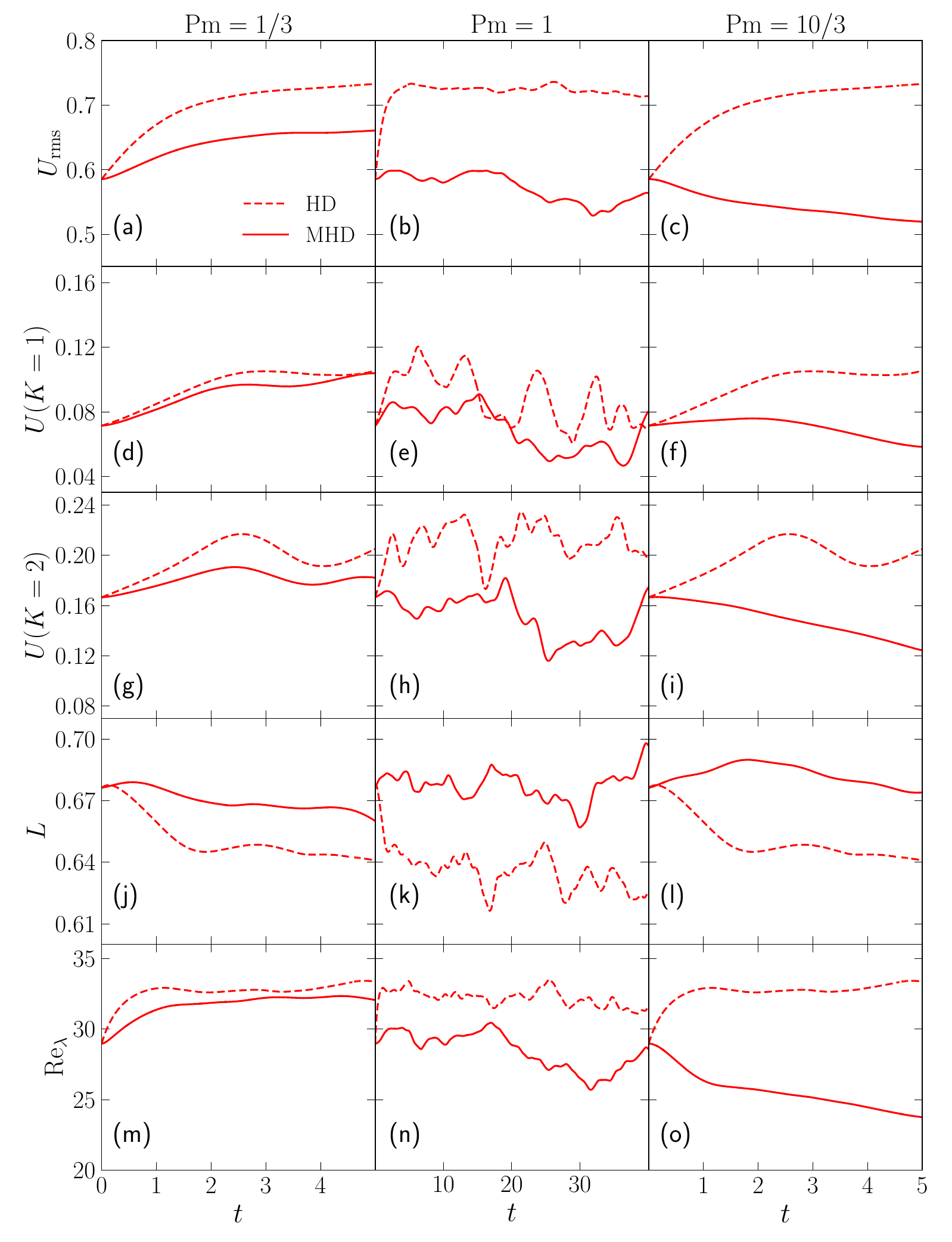}
	\caption{Time evolution of rms velocity ($U_\mathrm{rms}$), $ U(K=1) $, $ U(K=2) $, integral length scale ($L$), and $\mathrm{Re}_{\lambda}$ for the HD run (dashed red curve) and the MHD runs (solid red curve) for $ \mathrm{Pm}=1/3, 1, 10/3 $. $ U(K=1) $ and $ U(K=2) $ are computed using the KE contained in the  waveumber spheres of radii 1 and 2 respectively. }
	\label{fig:time_evol_dns}
\end{figure}

\begin{table}[h]
	\begin{minipage}{\textwidth}
		\caption{For MHD runs with  Pm = 1/3, 1, 10/3, numerical values of average KE flux ($\la \Pi_{u} \ra $) in the inertial range,  rms velocity ($U_\mathrm{rms}$), and $ \la \bar{C}_{d1} \ra$.  We also list $\langle \lvert (\mathbf{u}\cdot\nabla)\mathbf{u} \rvert \rangle$   and $\la \bar{C}_{d2} \ra$ for the wavenumber spheres of  radii $K=1$ and $K=2$.  The table contains the corresponding quantities for the HD run. For all the runs, $\epsilon_\mathrm{inj}=0.4$ }
		\label{tab:table2_dns}
		\vspace{10pt}
		\begin{tabular*}{\textwidth}{@{\extracolsep{\fill}}lcccccccccc@{\extracolsep{\fill}}}
			\toprule%
			& & & & & \multicolumn{2}{@{}c@{}}{$K=1$} & \multicolumn{2}{@{}c@{}}{$K=2$} \\\cmidrule{6-7}\cmidrule{8-9}%
			& $ \mathrm{Pm} $ & $ \la \Pi_{u} \ra $ & $U_\mathrm{rms}$ & $\la \bar{C}_{d1} \ra $ &$\la \lvert ({\bf u}\cdot \nabla) {\bf u} \rvert  \ra$ & $\la \bar{C}_{d2} \ra$ & $\la \lvert ({\bf u}\cdot \nabla) {\bf u} \rvert \ra $ & $\la \bar{C}_{d2} \ra$ \\ \\
			\midrule
			\rm{HD} & -&  $0.35$  & $0.72$ & $0.58$  & $0.1$  & $0.13$ & $0.3$  & $0.37$ \\
			\rm{MHD1} & $1/3$ & $0.28$ & $0.66$ & $0.65$ & $0.07$ & $0.11$	 & $0.3$ & $0.46$  \\
			\rm{MHD2} & $1$ & $0.25$ & $0.55$ & $0.98$ & $0.06$ & $0.13$ & $0.22$ & $0.49$ \\
			\rm{MHD3} & $10/3$ & $0.17$ & $0.53$ & $0.8$ & $0.04$ & $0.09$ & $0.17$ & $0.41$\\
			\botrule
		\end{tabular*}
	\end{minipage}
\end{table}
 We quantify $ U $ in three ways: $ U_\mathrm{rms} $; and $ U(K=1) $ and $ U(K=2) $, which are computed using the KE in the  wavenumber spheres of radii 1 and 2 respectively. We list $ U_\mathrm{rms}$ in Table~\ref{tab:table2_dns}.   In Fig.~\ref{fig:time_evol_dns}, we exhibit  the time series of $U_\mathrm{rms} $, $ U(K=1) $,  $ U(K=2) $, $L$, and $\mathrm {Re}_{\lambda}  $  for  the four runs. We observe that  $ U_\mathrm{rms} $, $ U(K=1) $,  and $ U(K=2) $ for the MHD runs are smaller than the corresponding quantities for the HD run, except  for MHD1 ($ \mathrm{Pm} =1/3 $) where $ U(K=1) $ is comparable to  that for the HD run.   Consequently,  $\mathrm {Re}_{\lambda} $ for MHD1    is close to that for the HD run, but $\mathrm {Re}_{\lambda} $  for the other  two MHD runs are smaller than those for the HD run. The integral lengths $ L $ for the three MHD runs are larger than the corresponding $ L $ for the HD run. Hence, the velocity fields are more ordered in the MHD runs compared to the HD run.

\begin{figure}[tbhp]
	\centering
	\includegraphics[scale = 0.6]{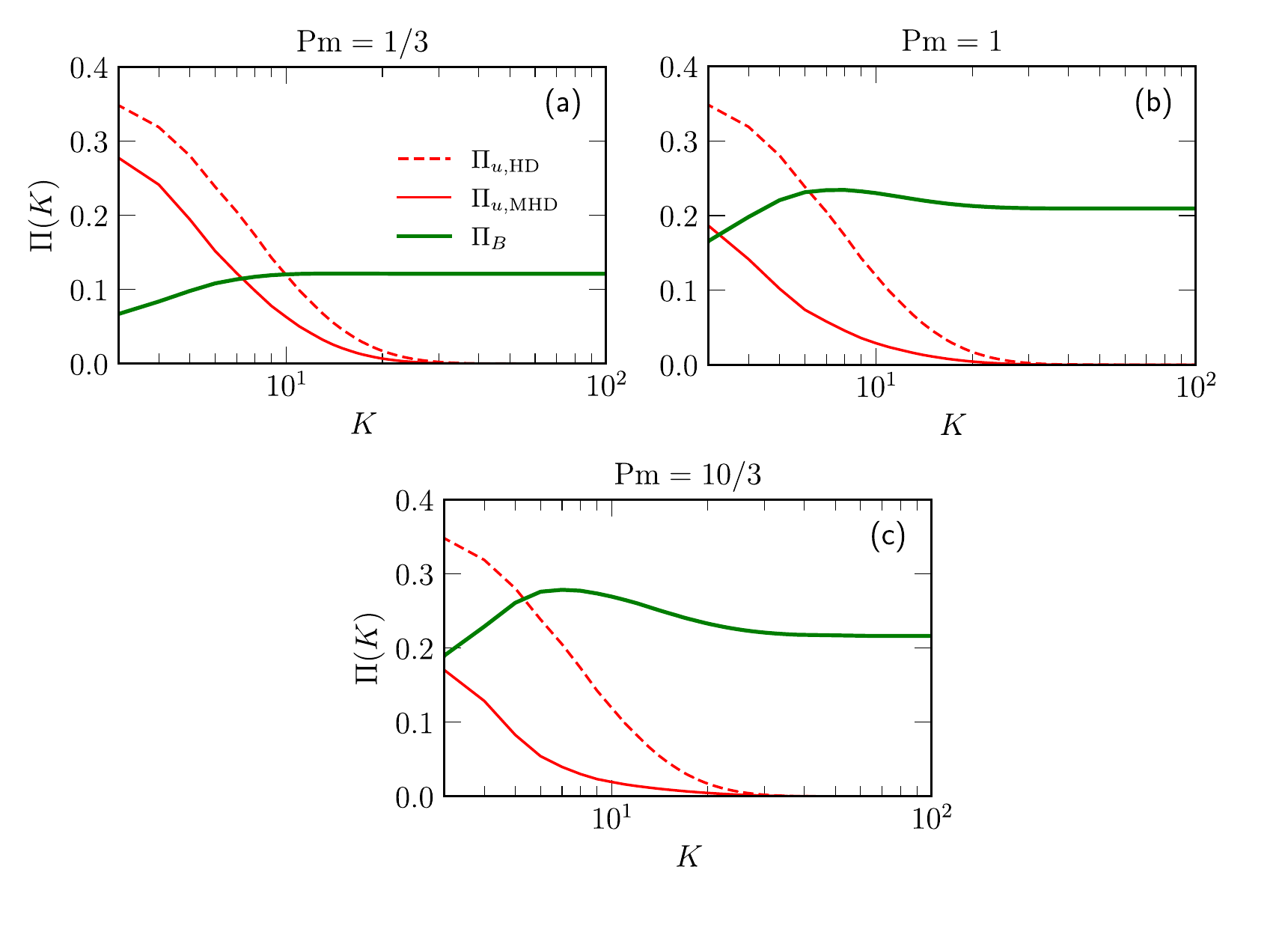}
	\caption{(a,b,c)  Plots   $ \Pi_u(K) $  (solid red curve) and  $ \Pi_B(K) $  (solid green curve)  for the MHD runs with $ \mathrm{Pm} = 1/3, 1, 10/3 $. Plots also illustrate $ \Pi_u(K) $  (dashed red curve) for the HD run. }
	\label{fig:flux_dns}
\end{figure}

Next, we compute $\Pi_u(K)$ for the HD and MHD runs, as well as $ \Pi_B(K) $ for the MHD runs.  These fluxes exhibit significant fluctuations, hence we average over  several time frames in the steady state.  The fluxes, shown in Fig~\ref{fig:flux_dns},  clearly show that $ \Pi_B > 0 $, indicating energy transfers from the velocity field to magnetic field at all scales, and that
\begin{eqnarray}
	\Pi_{u, \mathrm{MHD}} < \Pi_{u, \mathrm{HD}}.
\end{eqnarray}

\begin{figure}%[tbhp]
	\centering
	\includegraphics[scale=0.5]{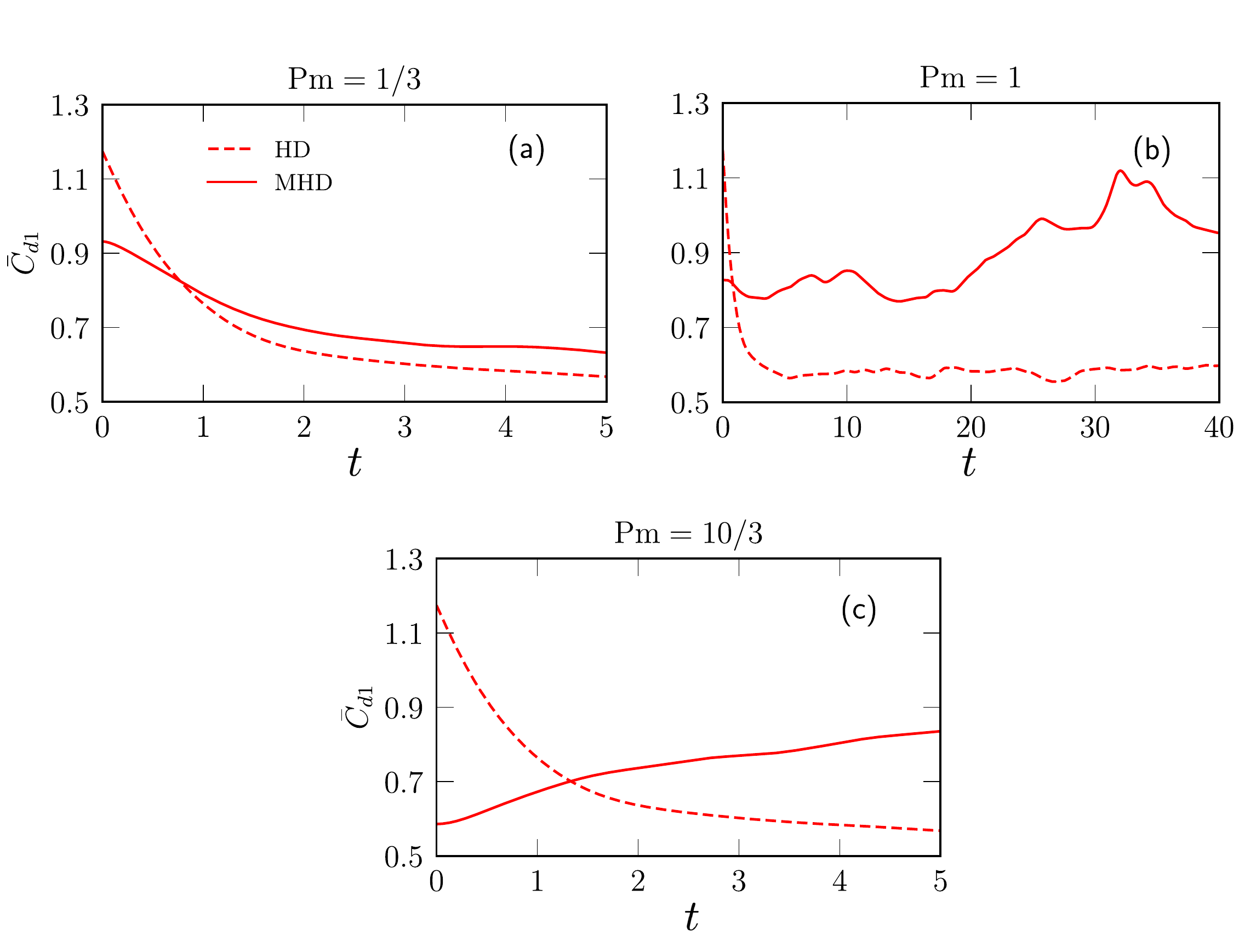}
	\caption{(a,b,c)  Time evolution of the drag reduction coefficient $\bar{C}_{d1}$ for the HD run (dashed red curve) and the MHD runs (solid red curve) with  $ \mathrm{Pm} = 1/3, 1, 10/3 $.}
	\label{fig:dns_drag_c1}
\end{figure}

We compute the drag coefficient $\bar{C}_{d1}$, which is defined in Eq.~(\ref{eq:c1}) as $ \la \Pi_u \ra / (U_\mathrm{rms}^3/L) $, and exhibit its time series  in Fig.~\ref{fig:dns_drag_c1}. In  Table~\ref{tab:table2_dns}, we list the average values of $\bar{C}_{d1}$ for the steady state.  We observe that $\bar{C}_{d1}$ for the steady state of the HD run is consistent with the results of Sreenivasan~\cite{Sreenivasan:PF1998}, thus validating our code and diagnostics. However,  $\bar{C}_{d1}$ for the steady states of the three MHD runs are larger than that for the HD run.  This is because the decrease in $ U_\mathrm{rms}^3 $ for the MHD runs overcompensates  the decrease in $ \Pi_u(K) $.  

%%%%%%%%%%%%%%%%%%%%%%%%%%%%%%%%%%%%%%%%%%%%

\begin{figure}[tbhp]
	\centering
	\includegraphics[width=\linewidth]{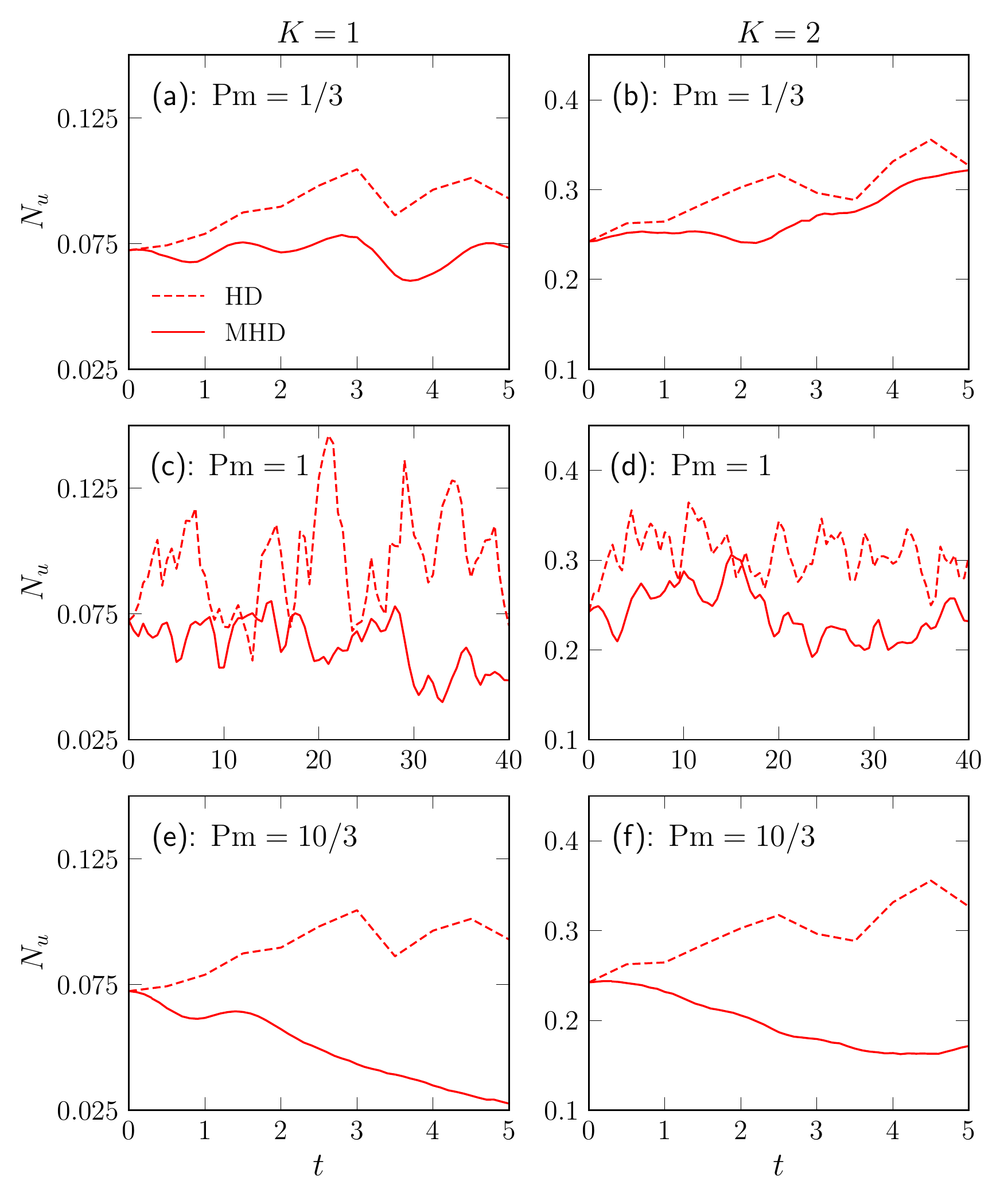}
	\caption{(a,b,c) Plots of the time series of nonlinear term $(N_u)$ for spheres of radii (a) $K=1$ and (b) $K=2$ for the HD run (dashed red curve) and the MHD runs (solid red curve) with $ \mathrm{Pm} = 1/3, 1, 10/3 $.}
	\label{fig:dns_nonlin}
\end{figure}

Now, we examine the   nonlinear term $N_u$ for the HD  and MHD runs. Since the drag force is effective at large scales, we  estimate $N_u$  by  its  rms value for a small  wavenumber sphere  of radius $ K $, that is,
\begin{eqnarray}
	\langle \lvert \left(\mathbf{u} \cdot \boldsymbol{\nabla} \right) \mathbf{u} \rvert \rangle_\mathrm{LS}  = N_{u}(K)  =  \sqrt{ \sum_{k\leq {K}} \lvert {\bf N}_u(\mathbf{k}) \rvert^2 }.   \label{eq:dns_nonlin} 
\end{eqnarray}
In particular, we choose $K=1$ and $K=2$. In Fig.~\ref{fig:dns_nonlin}(a,b), we illustrate the time series of $N_u(K)$ for the HD run (dashed red curve) and the MHD runs (solid red curve) for $K=1$ and $K=2$.  In Table~\ref{tab:table2_dns}, we list the average values of $N_u(K)$ for all the runs. We observe that $ N_u (K)$ for the three MHD runs are smaller than  $ N_u (K)$  for the HD counterpart. Hence, there is a reduction in $ \langle \lvert \left(\mathbf{u} \cdot \boldsymbol{\nabla} \right) \mathbf{u} \rvert \rangle_\mathrm{LS} $ for MHD turbulence compared to HD turbulence, signalling TDR in MHD turbulence.
%%%%%%%%%%%%%%%%%%%%%%%%%

\begin{figure}%[tbhp]
	\centering
	\includegraphics[width=\linewidth]{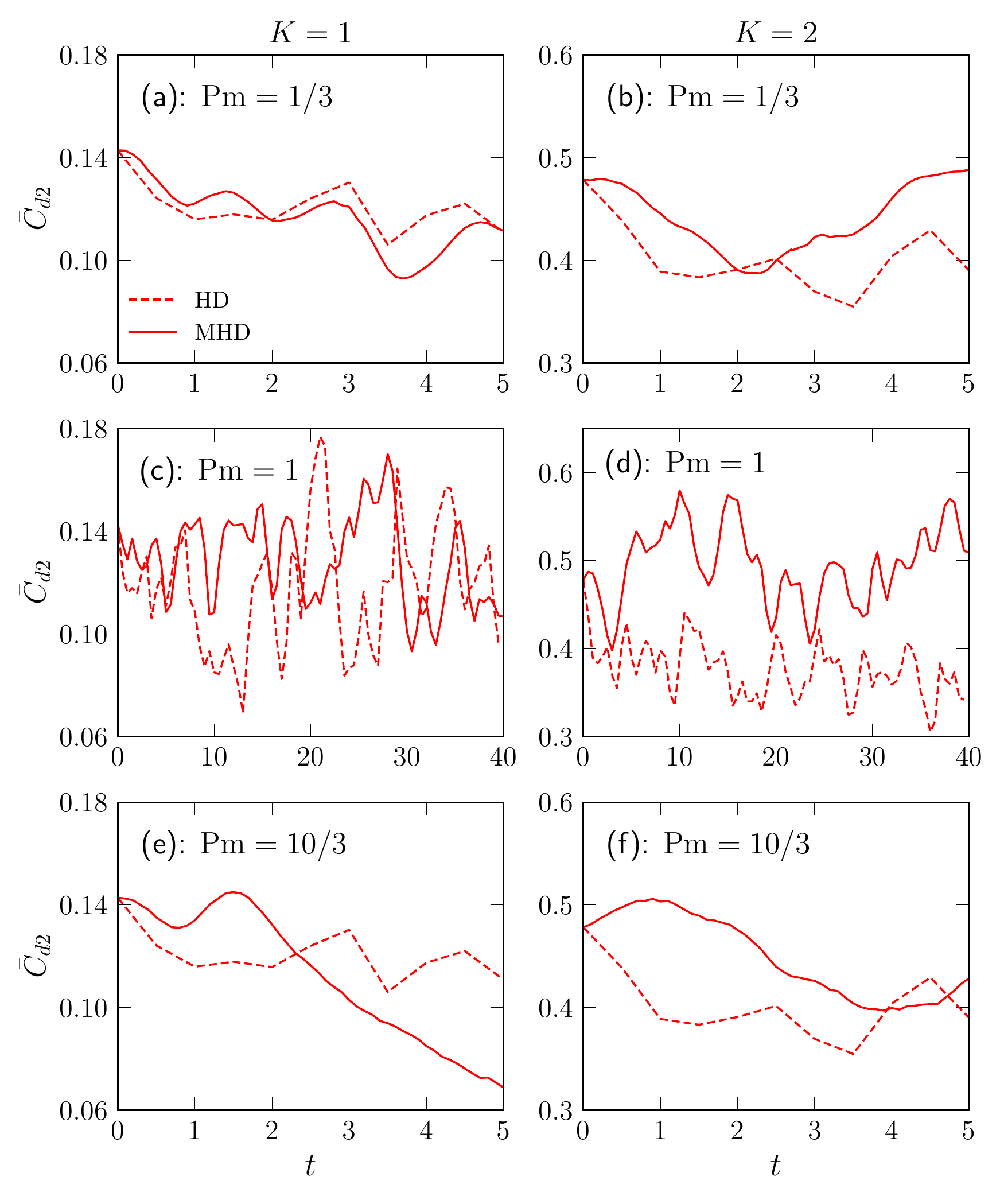}
	\caption{(a,b,c) Time evolution of drag reduction coefficient $\bar{C}_{d2}$ for sphere of radii (a) $K=1$, and (b) $K=2$ for HD turbulence (dashed red curve) and MHD turbulence (solid red curve) with $ \mathrm{Pm} = 1/3, 1, 10/3 $.}
	\label{fig:dns_drag_c2}
\end{figure}

After this, we compute the drag reduction coefficient  $\bar{C}_{d2}$, which is defined in Eq.~(\ref{eq:c2}) as $ \la \lvert ({\bf u \cdot \nabla}) {\bf u} \rvert \ra_\mathrm{LS}/(U_\mathrm{rms}^2/L) $.  The time series of $\bar{C}_{d2}$ for $K=1$ and $K=2$ are plotted in  Figure~\ref{fig:dns_drag_c2}, and their average values for their steady states are listed in Table~\ref{tab:table2_dns}.  We observe that $\bar{C}_{d2}(K=1)$ for the MHD runs with $ \mathrm{Pm} =1/3 $ and 10/3  are smaller than that for the HD run for $ t \gtrapprox 2 $.  For the other cases, $\bar{C}_{d2}$ for MHD runs are  larger than those for the HD run. 

Thus, for $ 1/3 \le \mathrm{Pm} \le 10/3 $, $ \Pi_u $ and $\la \lvert ({\bf u}\cdot \nabla) {\bf u} \rvert  \ra$ for the MHD runs are smaller than  the corresponding values for the HD run. For $ K=1 $, the drag coefficient $\bar{C}_{d2}$ exhibits similar behaviour for  Pm = 1/3 and 10/3, but not for Pm = 1.  This is in contrast to $\bar{C}_{d1}$, which is  typically larger  for MHD runs than that for the corresponding HD runs.  

We will show in Section \ref{sec:QSMHD} that QSMHD turbulence, which corresponds to $ \mathrm{Pm} = 0 $, exhibits larger $ U $  than the respective HD turbulence. Hence, we expect that MHD runs with very small $ \mathrm{Pm} $ will yield larger $ U $ than the corresponding HD runs.  This conjecture needs to be verified in future.  In addition, dynamo simulations exhibit enhancement in $ U $ on the emergence of a large-scale magnetic field (see Section \ref{sec:dynamo}). We will discuss these issues in later sections.

In summary, DNS of MHD turbulence exhibits reduction in $ \Pi_u(k) $ and $ 	\langle \lvert \left(\mathbf{u} \cdot \boldsymbol{\nabla} \right) \mathbf{u} \rvert \rangle_\mathrm{LS}  $ in comparison to HD turbulence. However, we do not observe enhancement in $ U $ in the MHD runs, at least for $ 1/3 \le \mathrm{Pm} \le 10/3 $.  We conjecture that MHD runs with very small Pm may exhibit  enhancement in $U $.

After the above discussion on DNS results on TDR in MHD turbulence, in the next subsection, we will discuss  TDR  in  the shell model of MHD turbulence.

\subsection{Numerical verification of TDR in shell models of MHD turbulence}
\label{subsec:shell}
In comparison to  DNS, shell models have much fewer  variables,  hence they are computationally faster than DNS. Therefore, shell models are often used to study turbulence, especially for extreme parameters. Beginning with Gledzer-Ohkitani-Yamada (GOY) shell model for HD turbulence \cite{Gledzer:DANS1973,Yamada:PRE1998,Ditlevsen:book}, researchers have developed several shell models for MHD turbulence \cite{Frick:PRE1998,Stepanov:ApJ2008,Plunian:PR2012,Verma:JoT2016}.    In this subsection, we report TDR in  a shell model of MHD turbulence  \cite{Verma:PP2020}. Verma \etal employed a revised version of GOY shell model  and computed  the drag forces and  nonlinear terms for the HD and MHD runs.  They  showed that the turbulent drag  in MHD turbulence is indeed reduced compared to HD turbulence.

In a  shell model of turbulence, all the Fourier modes in a wavenumber shell are represented by a single variable. A MHD shell model with $ N $ shells has $ N $ velocity and $ N $ magnetic shell variables that are coupled nonlinearly. The corresponding HD shell model has $ N $ velocity shell variables. In this subsection, we present the results of the shell model of Verma \etal \cite{Verma:PP2020}.

Verma \etal \cite{Verma:PP2020}  employed a shell model with $36$  shells, with random forcing  employed at shells $n = 1$ and $2$ such that the KE injection rate is maintained at a constant value~\cite{Stepanov:JoT2006}.     They performed three sets of HD and MHD  simulations with  KE injection rates $\epsilon_\mathrm{inj}=0.1, 1.0$ and $10.0$, and  $\nu =   \eta=10^{-6}$.  For time integration, they used Runge-Kutta fourth order (RK4) scheme with a fixed $\Delta t$. For  $\epsilon_\mathrm{inj}=0.1$ and $1.0$, they chose $\Delta t =5\times10^{-5}$,  but  for $\epsilon_\mathrm{inj}=10.0$, they took $\Delta t =1\times10^{-5}$.  The numerical results are summarized in Table~\ref{tab:HD_MHD_shell}. They carried out the HD and MHD simulations up to  1000 eddy turnover time.  For further details on the model and the numerical method, refer to Verma \etal \cite{Verma:PP2020}. 

%%%%%%%%%%%%%%%%%%%%%%%%%%%%%%%%%%%
%\begin{figure}%[tbhp]
%	\centering
%	\includegraphics[scale=0.3]{fig/Figure9.pdf}
%	\caption{Time evolution plots of kinetic energy for hydrodynamics; and of kinetic, magnetic, and total energy for MHD turbulence for different injection rates $\epsilon_\mathrm{inj}=0.1$(a),$1.0$(b) and $10.0$(c).  The solid red curve denotes the kinetic energy for hydrodynamic turbulence, and the dashed red, forest green and dark blue curves, respectively, show the kinetic, magnetic and total energy for MHD turbulence.}
%	\label{fig:energy_comp}
%\end{figure}

Both HD and MHD simulations reached their respective steady states after approximately $200$ eddy turnover time.   Interestingly, Verma \etal \cite{Verma:PP2020}  observed  that for the same $\epsilon_\mathrm{inj}$,  the KE and $ U $ for MHD turbulence are larger than those for  HD turbulence (see Table~\ref{tab:HD_MHD_shell}).  These observations clearly demonstrate an enhancement of $ U $ in MHD turbulence compared to HD turbulence, as is the case for turbulent flows with dilute polymers.

%%%%%%%%%%%%%%%%%%%%%%%%%%%%%%%%%%%%%%
\begin{table}[h]
	\begin{center}
		\begin{minipage}{\textwidth}
			\caption{For the shell model runs of HD and MHD turbulence with  $\epsilon_\mathrm{inj}=0.1,1.0,10.0$, numerical values of inertial-range KE flux $\Pi_{u}$, rms velocity  $U$,   $\la  \lvert ({\bf u}\cdot \nabla) {\bf u}\rvert  \ra=(\sum_{n} \lvert N_{n}[u,u]\rvert^{2} )^{1/2}$,  $ \bar{C}_{d1} $, and $ \bar{C}_{d2} $. \cite{Verma:PP2020}. }	\label{tab:HD_MHD_shell}
			\vspace{10pt}
			\begin{tabular*}{\textwidth}{@{\extracolsep{\fill}}lcccccccc@{\extracolsep{\fill}}}
				\toprule%
				%& \multicolumn{3}{@{}c@{}}{Hydrodynamics} & \multicolumn{3}{@{}c@{}}{MHD} \\\cmidrule{2-4}\cmidrule{5-7}%
				& $\epsilon_\mathrm{inj}$ & $\Pi_{u}$ & $U$ & $\la  \lvert ({\bf u}\cdot \nabla) {\bf u}\rvert  \ra$  & $ \bar{C}_{d1} $ & $ \bar{C}_{d2} $ \\
				\midrule
				HD & $0.1$ & $0.1$ & $0.87$ & $8.77$ & $0.15$ & $11.6$  \\
				MHD & $0.1$ & $0.02$ & $0.92$ & $4.17$ & $0.026$ & $4.93$  \\
				\midrule
				HD & $1.0$ & $1.0$ & $1.88$ & $47.48$ & $0.15$ & $13.4$  \\
				MHD & $1.0$ & $0.21$ & $2.02$ & $23.79$ & $0.026$ & $5.83$  \\
				\midrule
				HD & $10.0$ & $10.0$ & $3.95$ & $271.88$ & $0.16$ & $17.4$  \\
				MHD & $10.0$ & $2.06$ & $4.33$ & $136.44$ & $0.025$ & $7.28$  \\
				\botrule
			\end{tabular*}
		\end{minipage}
	\end{center}
\end{table}

\begin{figure}%[tbhp]
	\centering
	\includegraphics[scale=0.3]{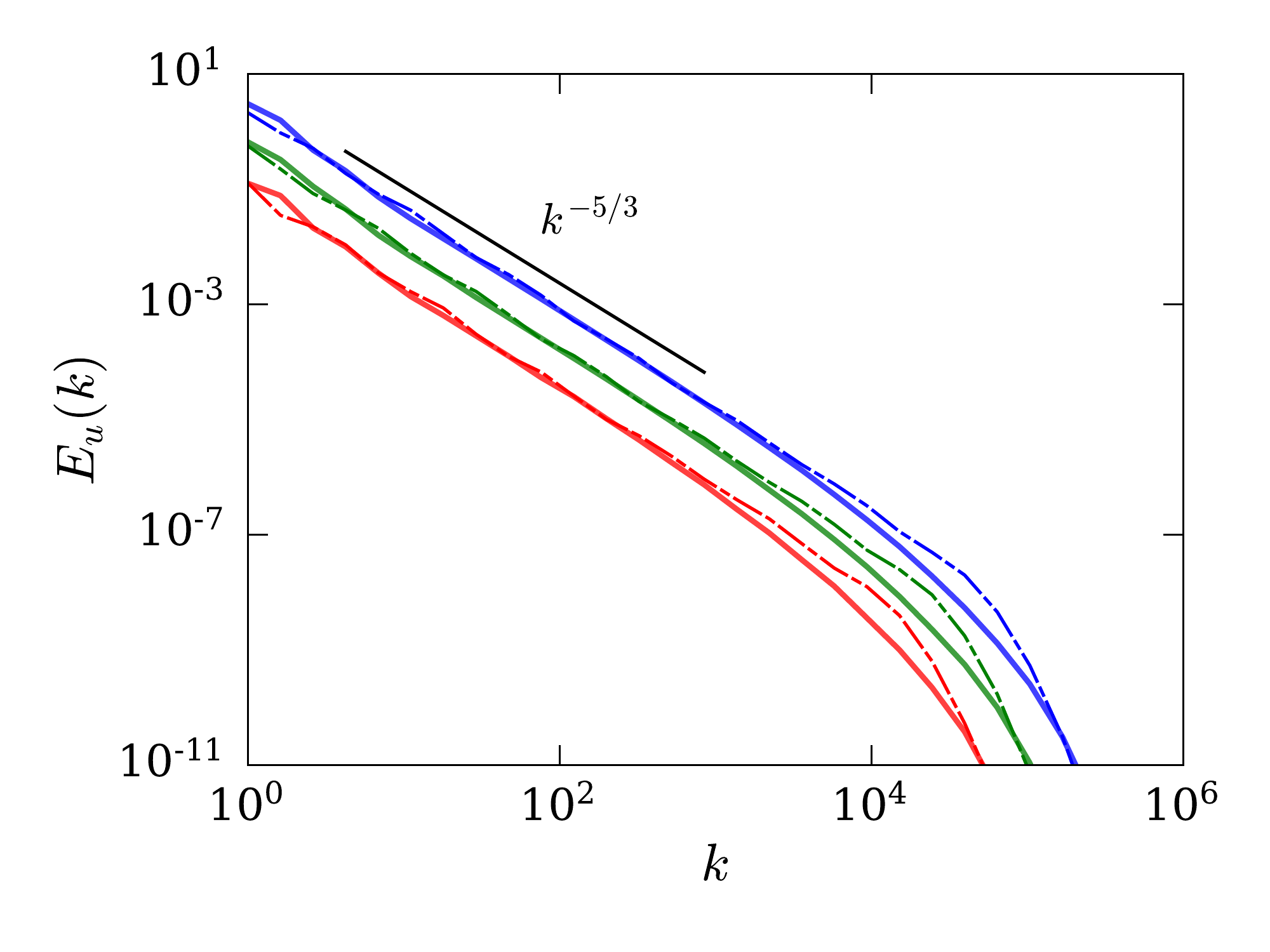}
	\caption{Plots of KE spectra $E_u(k)$ for the shell model runs with $\epsilon_\mathrm{inj}=0.1$ (red), $\epsilon_\mathrm{inj}= 1.0$ (green) and $\epsilon_\mathrm{inj}= 10.0$ (blue).  The dashed and solid curves represent the $E_u(k)$ for the MHD and HD runs respectively. Kolmogorov's $-5/3$ scaling (black) fits well in the inertial range for all the runs. From Verma \etal \cite{Verma:PP2020}.  Reproduced with permission from AIP.}
	\label{fig:spectrum_comp}
\end{figure}

The increase in $ U $ for the MHD runs compared to the HD runs has its origin in the energy spectra. Verma \etal \cite{Verma:PP2020}  computed the average KE spectra $  E_u(k) $ for the HD and MHD runs. These spectra, shown in Fig.~\ref{fig:spectrum_comp}, exhibit Kolmogorov's $k^{-5/3}$ spectrum. For a given $\epsilon_\mathrm{inj}$,  $ E_u(k) $ plots for the HD and MHD runs almost overlap with each other, except  for small wavenumbers where $E_u(k)$ for the MHD runs are larger than the HD counterpart. Since the energy is concentrated at small wavenumbers,  we observe that $U_\mathrm{MHD} > U_\mathrm{HD}$.   This is in sharp contrast to DNS results of Section \ref{sec:dns} where $ U $ and $ E_u(k) $ of the MHD runs with moderate Pm  are  smaller than the corresponding values for the HD runs.  However, in dynamo simulations, we do observe that $ U $ of MHD turbulence could be larger than that for HD turbulence; this topic will be discussed in the next section.

Next, using the numerical data of the shell model, Verma \etal \cite{Verma:PP2020}  estimated the rms values of $({\bf u}\cdot \nabla) {\bf u}$ for the HD and MHD runs using 
\be
\la  \lvert({ \bf u}\cdot \nabla) {\bf u} \rvert  \ra 
= \left(\sum_n \lvert N_n[u,u] \rvert^2 \right)^{1/2}.
\ee
To suppress the fluctuations, averaging was performed over a large number of states.   As  listed in Table~\ref{tab:HD_MHD_shell},  $ \la  \lvert({ \bf u}\cdot \nabla) {\bf u} \rvert  \ra  $ for  the MHD runs are suppressed compared to the corresponding HD runs.  These results reinforce the fact that the nonlinearity $  \la  \lvert({ \bf u}\cdot \nabla) {\bf u} \rvert  \ra  $ depends critically on the phases of the Fourier modes; larger $U$ does not necessarily imply larger $\la  \lvert({ \bf u}\cdot \nabla) {\bf u} \rvert  \ra$. We remark that  averaging over the small $ n $ would have been more appropriate for the estimation of $ \la  \lvert({ \bf u}\cdot \nabla) {\bf u} \rvert  \ra  $, as was done for the DNS.

\begin{figure}%[tbhp]
	\centering
	\includegraphics[scale=0.3]{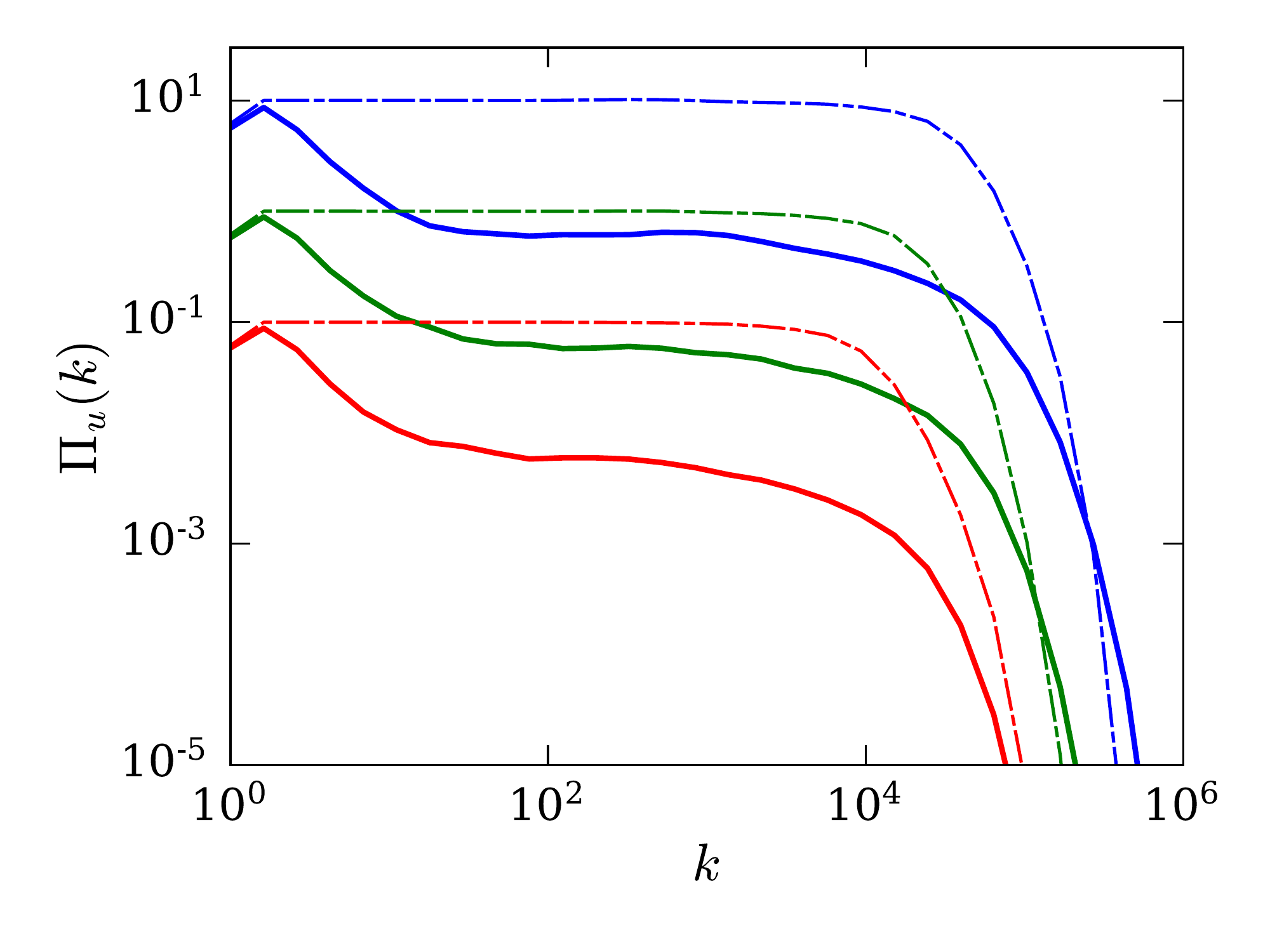}
	\caption{Plots of   $\Pi_u(k)$ for   $\epsilon_\mathrm{inj}=0.1$ (red), $\epsilon_\mathrm{inj}= 1.0$ (green) and $\epsilon_\mathrm{inj}= 10.0$ (blue). The dashed curves represent  $\Pi_u(k)$ for the HD runs, whereas the solid curves indicate the same for the MHD runs. From Verma \etal\cite{Verma:PP2020}.  Reproduced with permission from AIP.}
	\label{fig:flux_comp}
\end{figure}

Verma \etal \cite{Verma:PP2020} also computed the  average KE fluxes for the HD and MHD runs \cite{Verma:JoT2016, Verma:book:ET}. These fluxes are illustrated in Fig.~\ref{fig:flux_comp}, and their average values in the steady state are listed in Table~\ref{tab:HD_MHD_shell}. 
The figure illustrates  that for a given $ \epsilon_\mathrm{inj} $, the MHD run has a lower KE flux than corresponding HD run.  This is  consistent with the suppression of $ \la  \lvert({ \bf u}\cdot \nabla) {\bf u} \rvert  \ra  $;  lower $ \la  \lvert({ \bf u}\cdot \nabla) {\bf u} \rvert  \ra  $ leads to lower KE flux. In addition, we  compute $ \bar{C}_{d1} $ and $ \bar{C}_{d2} $ using the values of Table~\ref{tab:HD_MHD_shell} and $ L=1 $. Clearly, $ \bar{C}_{d1} $ and $ \bar{C}_{d2} $  for the MHD runs are lower than those for the corresponding HD runs, thus indicating TDR in MHD turbulence.

Thus, DNS  and the shell model results illustrate that MHD turbulence has lower $ \la  \lvert({ \bf u}\cdot \nabla) {\bf u} \rvert  \ra  $ and lower $ \Pi_u(k) $ compared to HD turbulence.  These results demonstrate TDR in MHD turbulence.   Note, however, that in DNS, $ U $ for the MHD runs with $ 1/3 \le \mathrm{Pm} \le 10/3 $ are smaller than  the corresponding  $ U $ for the HD runs, but it is other way round in the shell model.  As argued in Section \ref{sec:dns}, we expect that $ U $ for MHD runs with very small Pm would be larger than $ U $ for the HD runs. 

In the next section we will describe TDR in dynamos.

\section{TDR  in Dynamos}
\label{sec:dynamo}

Magnetic field generation, or \textit{dynamo process}, in astrophysical objects is an important subfield of MHD. In dynamo process, the velocity field is forced mechanically, or by convection induced via temperature and/or concentration gradients. Rotation  too plays an important role in dynamo.  There are many books and papers written on dynamo, see e.g.~\cite{Moffatt:book,Roberts:RMP2000}. In this section, we will discuss only a handful of dynamo studies that are related to TDR. 

\begin{figure}%[tbhp]
	\centering
	\includegraphics[width=0.6\linewidth]{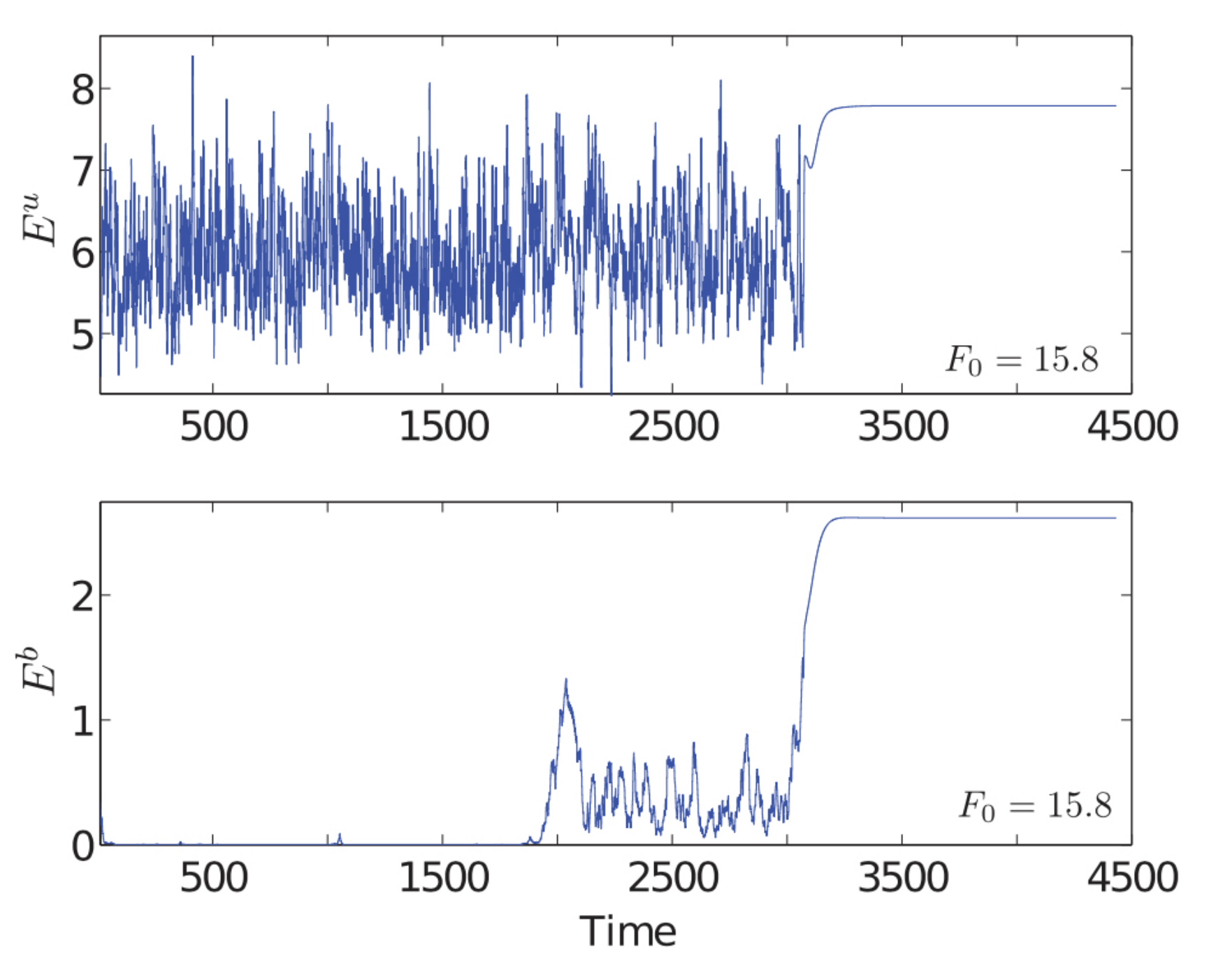}
	\caption{Plots of the total KE (top panel) and the total ME (bottom panel) for Taylor-Green dynamo with $F_0 =  15.8 $. We observe ordered velocity and magnetic fields after the onset of dynamo (time $ > $ 3000 units).  From Yadav \etal \cite{Yadav:PRE2012}. Reprinted with the permission of APS.}
	\label{fig:dynamo_yadav2}
\end{figure}

Yadav \etal \cite{Yadav:PRE2012} simulated Taylor-Green dynamo for magnetic Prandtl number Pm = 0.5. They reported many interesting properties, including subcritical dynamo transition, as well as steady, periodic, quasi-periodic, and chaotic dynamo states. Let us focus on an interesting feature of this dynamo that is related to TDR. 
 \begin{figure}%[tbhp]
	\centering
	\includegraphics[width=0.6\linewidth]{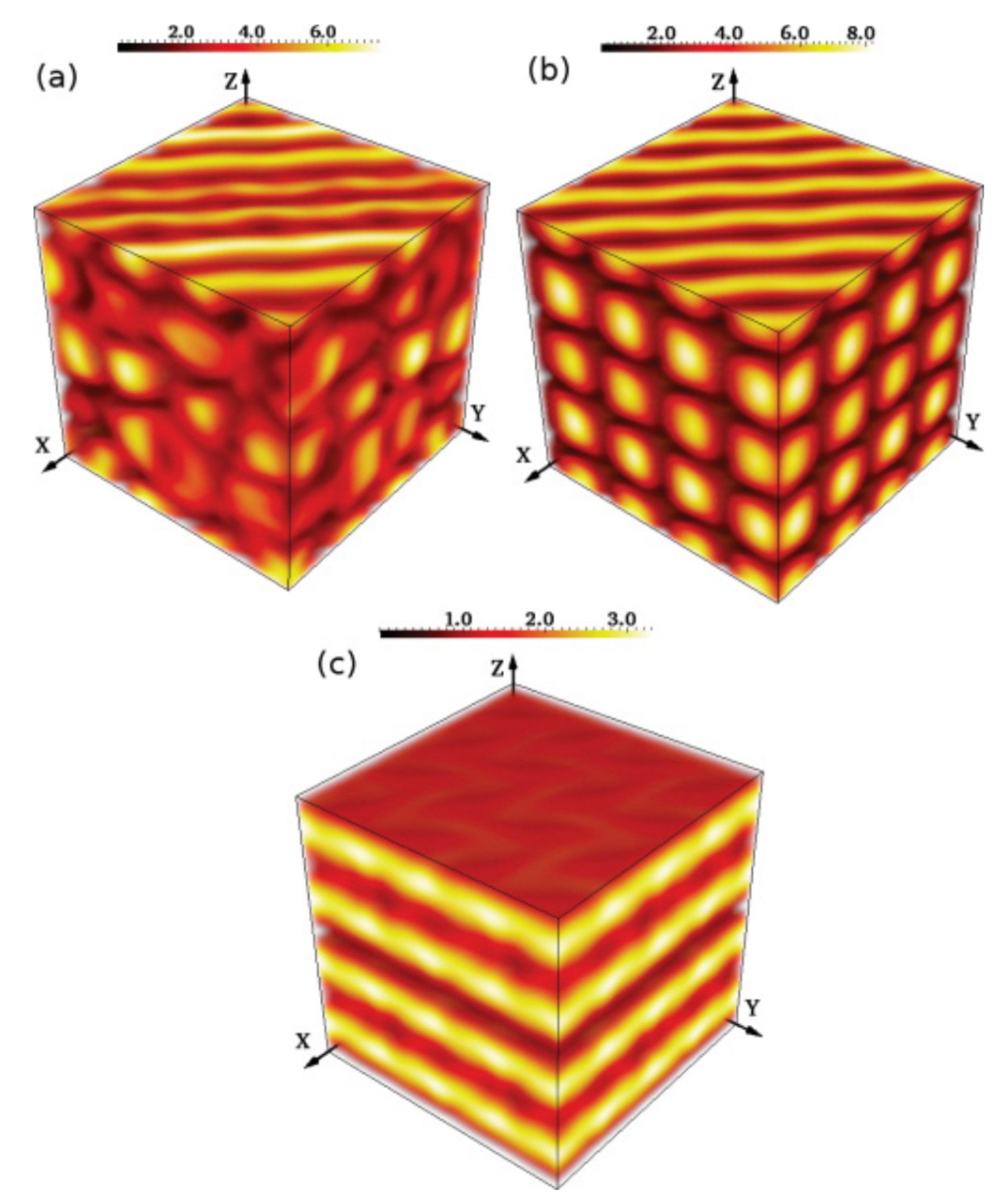}
	\caption{For the Taylor-Green dynamo with the forcing amplitude $ F_0 = 15.2 $, (a) 3D plot of the spatially chaotic velocity field for a no-dynamo state; (b) ordered velocity field for a dynamo state arising due to the suppression of chaos in the presence of a finite mean magnetic field; (c) ordered magnetic field.   From Yadav \etal \cite{Yadav:PRE2012}. Reprinted with the permission of APS.}
	\label{fig:dynamo_yadav1}
\end{figure}
In Fig.~\ref{fig:dynamo_yadav1} we exhibit the intensities of the magnitudes of the velocity and magnetic fields for the  forcing amplitude $ F_0 = 15.2 $. Before the dynamo transition, the velocity field is quite turbulent, as shown in Fig.~\ref{fig:dynamo_yadav1}(a). However,  after the dynamo transition or emergence of magnetic field, both the velocity and magnetic fields, shown in Fig.~\ref{fig:dynamo_yadav1}(b,c), become more ordered compared to the pure HD state of Fig.~\ref{fig:dynamo_yadav1}(a). Yadav \etal observed similar features at several other $ F_0 $'s. For example, at $ F_0 = 15.8$, after the emergence of magnetic field,  the velocity fluctuations are suppressed, and  the velocity and magnetic fields become quite coherent (see  Fig.~\ref{fig:dynamo_yadav2}).  The emergence of  ordered velocity field is akin to an enhancement of the mean velocity in a pipe  flow with polymers.  

\begin{figure}%[tbhp]
	\centering
	\includegraphics[width=0.5\linewidth]{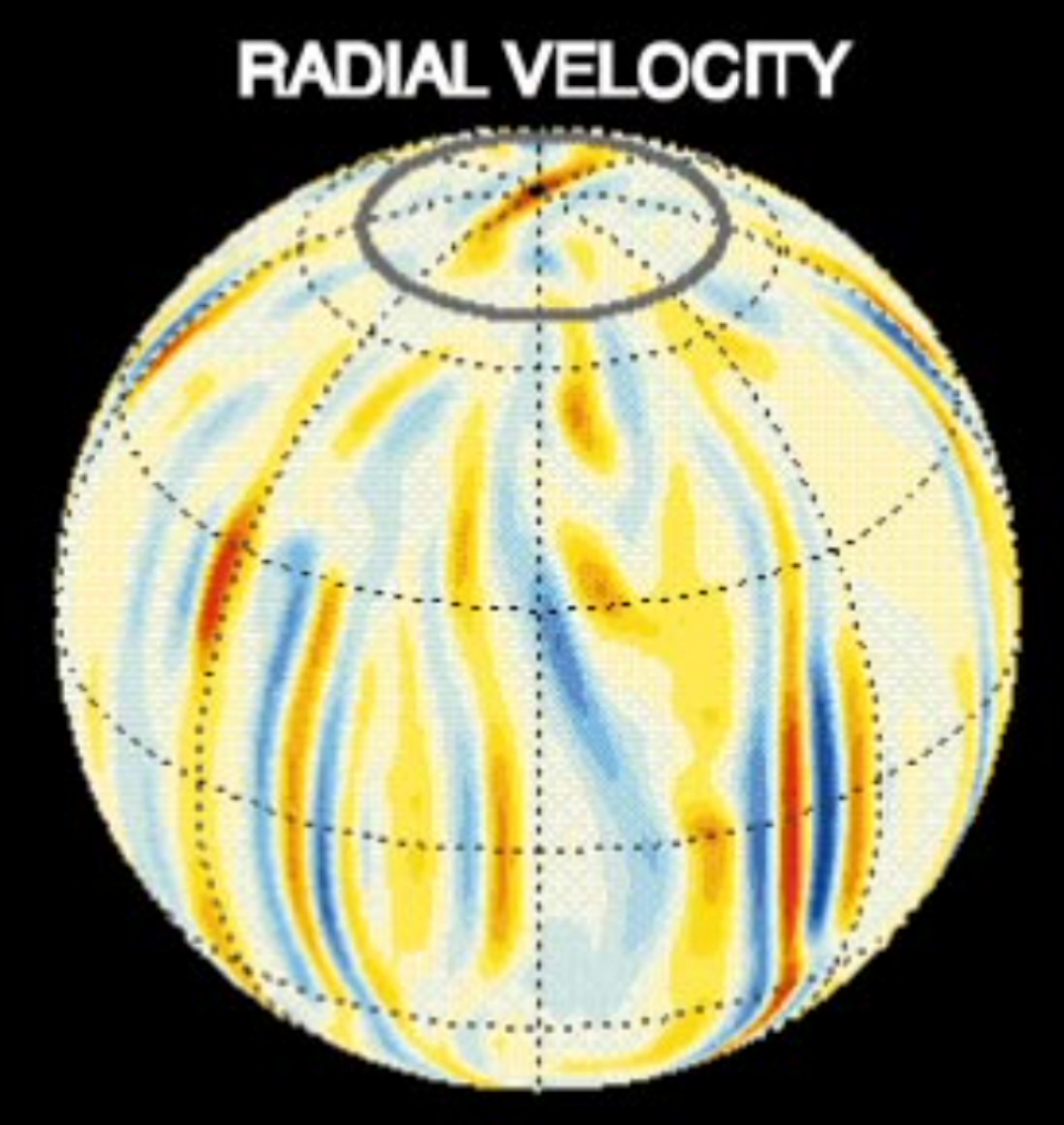}
	\caption{The radial component of the  velocity field in a numerical simulation of geodynamo by Olson et al.~\cite{Olson:JGR1999}. From Olson et al.~\cite{Olson:JGR1999}. Reproduced with permission from  John Wiley \& Sons. }
	\label{fig:geodynamo}
\end{figure}

The aforementioned simulation of Yadav \etal \cite{Yadav:PRE2012} is somewhat idealized in comparison to spherical geo- and solar dynamos with rotation and thermal convection at  extreme parameters. Interestingly, spherical dynamos share certain common features with Taylor-Green dynamo.  As shown in Fig.~\ref{fig:geodynamo}, the velocity field of spherical dynamo~\cite{Olson:JGR1999} is organized in vertical columns, which is also a feature of rotating turbulence~\cite{Davidson:book:TurbulenceRotating,Sharma:PF2018}. It is possible that thermal convection and magnetic field too contribute to the structural organization of the flow; this feature however needs a careful examination.

Even though $ \la \lvert {\bf u \cdot \nabla u} \rvert \ra$ and the energy fluxes for dynamos have been studied widely (e.g., \cite{Roberts:RMP2000,Verma:PR2004,Kumar:EPL2014}), TDR in dynamos has not been analyzed in detail.  It is hoped that a systematic study of TDR in dynamos would be performed in future.

In the next section, we describe TDR in QSMHD turbulence.

\section{TDR in QSMHD turbulence via energy flux}
\label{sec:QSMHD}

Liquid metals have small magnetic Prandtl number (Pm), and they are described using QSMHD equations, which are  a limiting case of MHD equations~\cite{Moreau:book:MHD,	Knaepen:ARFM2008,Verma:ROPP2017}.    The  equations for QSMHD with a strong external magnetic field $ {\bf B}_0 $ are \cite{Moreau:book:MHD,	Knaepen:ARFM2008,Verma:ROPP2017}
\bea
\frac{\partial{\bf u}}{\partial t} + ({\bf u}\cdot\nabla){\bf u}
& = & -\nabla({p}/{\rho}) - \frac{\sigma}{\rho} \Delta^{-1} [({\bf B}_0 \cdot \nabla)^2 {\bf u} ]  +  \nu\nabla^2 {\bf u}   + {\bf F}_\mathrm{ext},  \label{eq:U_QSMHD}  \\
\nabla \cdot {\bf u}  & = & 0, \label{eq:incompress_QSMHD}
\eea
where $\sigma $  is the electrical conductivity, and $ \Delta^{-1} $  is the inverse Laplacian operator.  In Fourier space, a nondimensionalized version of QSMHD equations    is
\bea
\frac{d}{dt} {\bf u(k)} &=&- i \sum_{\bf p} \{ {\bf k \cdot u(q)}  \} {\bf u(p)}  - i{\bf k} p({\bf k} )/\rho  -N (\cos^2\theta) {\bf u(k)} \nonumber \\
&& -\nu k^2 {\bf u(k)} + {\bf F}_\mathrm{ext}({\bf k}), \label{eq:uk_qsmhd} \\
{\bf k \cdot u(k)} & = & 0,
\eea
where  $N$ is the  {\em interaction parameter}, and $ \theta $ is the angle between the wavenumber $ {\bf k} $ and $ {\bf B}_0 $. The interaction parameter $ N $ is the ratio of the Lorentz force and nonlinear term ${\bf (u \cdot  \nabla) u}$, or
\be
N = \frac{\sigma B_0^2 L}{\rho U}.
\ee
Using Eq.~(\ref{eq:uk_qsmhd}), we derive an  equation for the modal energy as
\bea
\frac{d}{dt} E_u(\mathbf{k})  & = & T_{u}({\bf k})  - 2 N   E_u(k) \cos^2 \theta + \mathcal{F}_\mathrm{ext}({\bf k})-D_u(\mathbf{k}),
\label{eq:Eu_dot_Fext_qsmhd} 
\eea
where $ T_{u}({\bf k}) $ is defined in Eq.~(\ref{eq:Tuk}), and  the dissipation induced by Lorentz term is \cite{Knaepen:ARFM2008,Verma:ROPP2017}
\bea
\mathcal{F}_u({\bf k}) & = &   - 2 N  E_u({\bf k})  \cos^2 \theta<0 .
\label{eq:QSMHD_diss_Fu}
\eea
Hence, the magnetic field induces additional dissipation in QSMHD turbulence.

Equation~(\ref{eq:QSMHD_diss_Fu}) represents   the energy transfers from the velocity field  to the magnetic field at a wavenumber $ {\bf k} $. A sum of $ \mathcal{F}_u({\bf k}) $ over a wavenumber sphere of radius $ K $ yields the following expression for the energy flux  $ \Pi_B(K)$: 
\be
\Pi_B(K)=-\sum_{k\le K} \mathcal{F}_u({\bf k}) = \sum_{k\le K}  2 N   E_u({\bf k})  \cos^2 \theta>0.
\ee
Thus, the    Lorentz force transfers the kinetic energy to the magnetic energy, which is immediately dissipated by the Joule dissipation; this feature is due to $ \mathrm{Pm} =0 $.  As a consequence, for an injection rate $ \epsilon_\mathrm{inj} $,  $ \Pi_u(K) $ of a QSMHD run is suppressed compared to $ \Pi_u(K) $  of the corresponding HD run. Hence, in the inertial range,
\be
\Pi_u < \epsilon_\mathrm{inj}.
\ee
Therefore, following the same line arguments as in earlier sections, we deduce that  turbulent drag is suppressed in QSMHD turbulence.   In addition,  the velocity fields of the MHD runs are less random (or more ordered) compared to the corresponding HD runs, thus suppressing $\la \lvert ({\bf u}\cdot \nabla) {\bf u} \rvert  \ra$.  Therefore, we expect the turbulent drag  in QSMHD turbulence to be smaller than the corresponding HD counterpart.  In the following discussion, we will describe numerical results that are consistent with the above predictions.

\begin{figure}%[tbhp]
	\centering
	\includegraphics[width=0.7\linewidth]{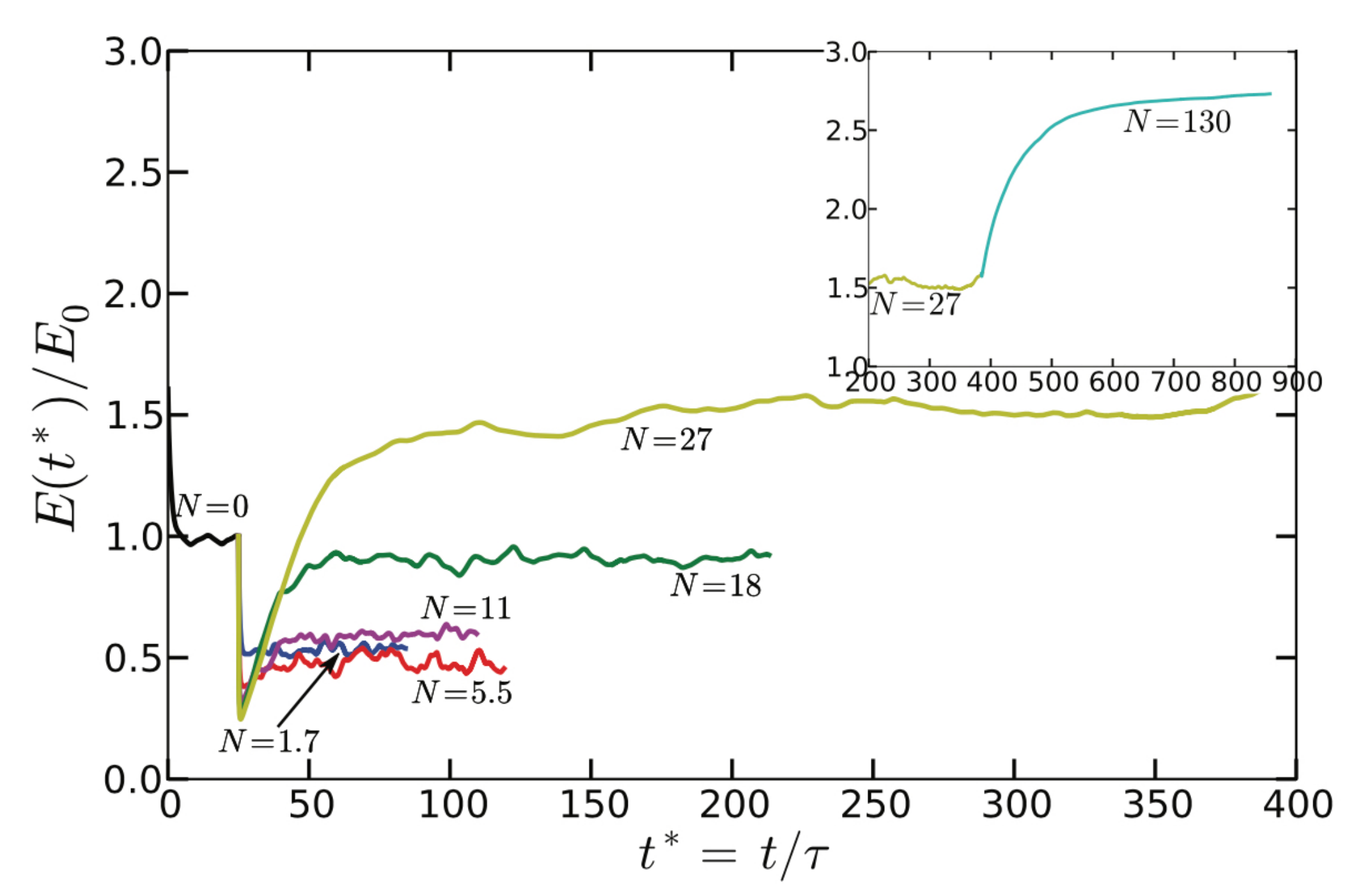}
	\caption{From the numerical simulation of QSMHD turbulence by Reddy and Verma~\cite{Reddy:PF2014}, the time series of the normalised KE, $ E(t)/E_0$, for $ N=5.5, 11, 18, 27, 130 $,  where $ E_0 $ is the energy at the final state of $  N = 0 $ simulation. For each $ N $, after an application of external magnetic field, the KE drops suddenly, and then  it increases and reaches a statistically steady value. The asymptotic KE for all the runs with $ N > 18 $ are larger than $ E_0 $. From Reddy and Verma~\cite{Reddy:PF2014}.  Reproduced with permission from AIP.}
	\label{fig:Reddy1}
\end{figure}

Reddy and Verma~\cite{Reddy:PF2014} simulated QSMHD turbulence in a periodic box for  $N$ ranging from 1.7 to 220. They employed  a constant  KE injection rate of 0.1 (in nondimensional units). In fact, the magnetic field $ {\bf B}_0 $ was switched on after the initial HD run was fully developed. After an introduction of $ {\bf B}_0 $,  KE first decreases abruptly due to Joule dissipation, and then it increases  due to reorganization of the flow.  As shown in Fig.~\ref{fig:Reddy1}, for $ N > 18 $, the total KE  is larger  than its HD counterpart ($ N=0 $).   In this range of $ N $, the flow becomes quasi two-dimensional with larger $ U $ and suppressed turbulent drag.  This is counter-intuitive because we expect the KE to decrease with the increase of Joule dissipation. However,  reorganization of the flow leads to  enhancement of $ U $ and TDR in the flow.

\begin{figure}%[tbhp]
	\centering
	\includegraphics[width=1\linewidth]{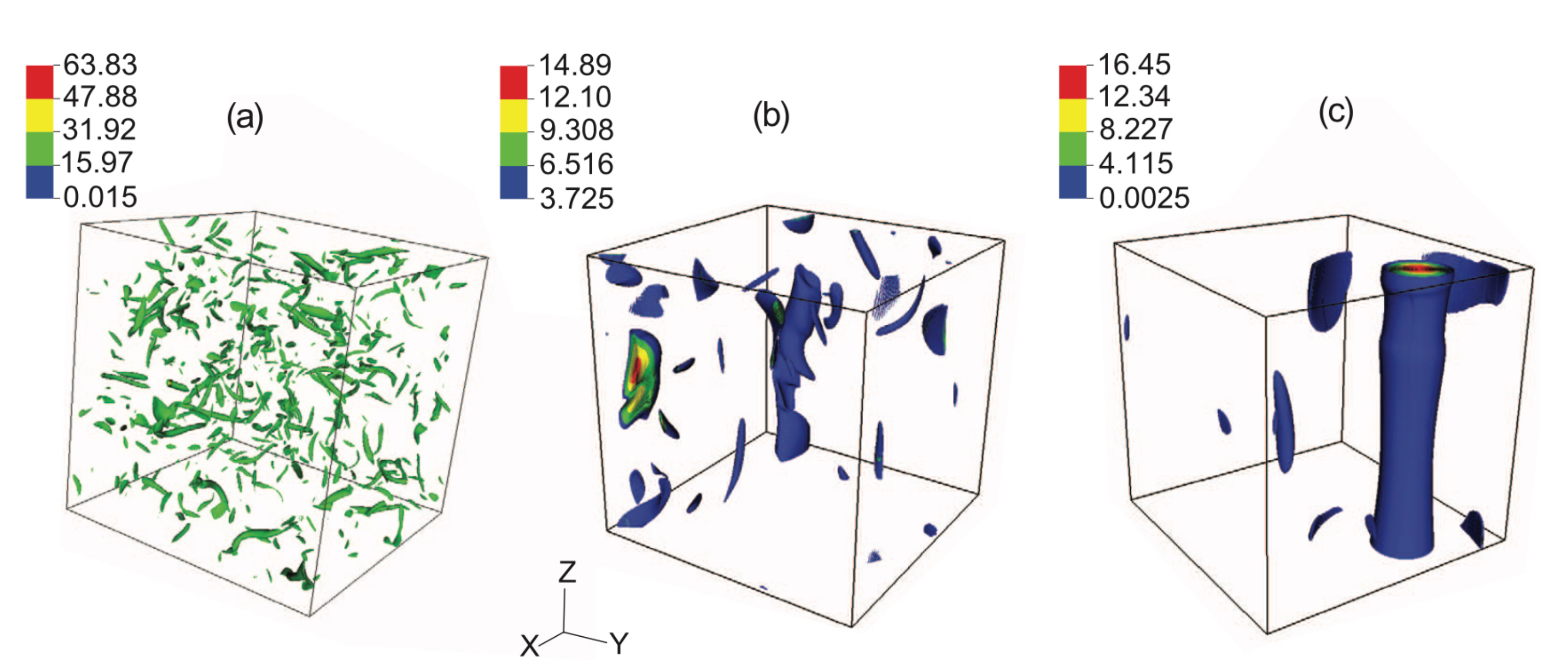}
	\caption{From the numerical simulation of QSMHD turbulence by Reddy and Verma~\cite{Reddy:PF2014}, the vorticity isosurfaces  for (a) N = 0, (b) N = 5.5, and (c) N = 18. The flow field becomes anisotropic and ordered with the increase of $ N $. We observe a vortex tube for $ N=18 $.  From Reddy and Verma~\cite{Reddy:PF2014}.  Reproduced with permission from AIP. }
	\label{fig:Reddy2}
\end{figure}

\begin{table}[h]
	\begin{center}
		\begin{minipage}{174pt}
			\caption{In numerical simulations of QSMHD turbulence by Verma and Reddy~\cite{Verma:PF2015QSMHD},  rms velocity  ($U$) for various  $N$'s.  Clearly, $ U $ increase with $N$. } \label{tab:QSMHD}%
			\vspace{10pt}%
			\begin{tabular}{@{}llllllll@{}}
				\toprule
				$N$ & 1.7 & 18 & 27 & 220  \\ 
				\midrule
				$U$ & 0.39 & 0.51 & 0.65 & 0.87	 \\ 
				\botrule
			\end{tabular}
		\end{minipage}
	\end{center}
\end{table}
In Table~\ref{tab:QSMHD}, we list  the rms velocity $U$ as a function of $N$.  Clearly,  $U$ increases monotonically with $N$ because $\la \lvert ({\bf u}\cdot \nabla) {\bf u} \rvert  \ra$  and turbulent drag decrease with the increase of $ N $.  In Fig.~\ref{fig:Reddy2}  we exhibit the vorticity isosurfaces for $ N=0, 5.5, $ and 18.  As is evident in the figure, the flow becomes quasi-2D and more orderly with the increase of $ N $.

The above results again indicate that a large $U$ does not necessarily imply  large  $\la \lvert ({\bf u}\cdot \nabla) {\bf u} \rvert  \ra$  because the nonlinear term  depends on $U$ and  the phase relations between the velocity modes.  In QSMHD turbulence, two-dimensionalization  leads to a reduction in $\la \lvert ({\bf u}\cdot \nabla) {\bf u} \rvert  \ra$  even with large $U$.  Note, however, that for a definitive demonstration of drag reduction in QSMHD turbulence, we still need to perform a comparative study of  $ \Pi_u $ and $\la \lvert ({\bf u}\cdot \nabla) {\bf u} \rvert  \ra$  for HD and QSMHD turbulence.  

Reduced turbulent flux is an important ingredient for drag reduction.   Note  that such a reduction does not occur in laminar QSMHD; here, the Lorentz force damps the flow further.   We illustrate this claim for a  channel flow. In  a HD channel flow, the maximum velocity at the centre of the pipe is (see Fig.~\ref{fig:pipe}) \cite{Kundu:book,Verma:book:Mechanics}
\be
U_\mathrm{HD} =- \frac{d^2}{2 \nu \rho} \left( \frac{dp}{dx} \right),
\ee
where $d$ is half-width of the channel (see Fig.~\ref{fig:pipe}). However, in a laminar QSMHD flow, the corresponding velocity is~\citep{Muller:book,Moreau:book:MHD, Verma:ROPP2017}
\be
U_\mathrm{QSMHD} =- \frac{1}{\sigma B_0^2} \left( \frac{\partial p}{\partial x} \right).
\ee
 The ratio of the two velocities is
\be
\frac{U_\mathrm{QSMHD}}{U_\mathrm{HD} } = \frac{2 \nu \rho}{\sigma B_0^2 d^2} = \frac{1}{\mathrm{Ha}^2},
\ee
where $\mathrm{Ha}$ is the Hartmann number, which is much larger than unity for a QSMHD flow.  Hence, the velocity in laminar QSMHD is much smaller than that in the HD channel.  In comparison,    $ U $  increases with $ N$ in QSMHD turbulence. Hence,   drag reduction is a nonlinear phenomena, which is a visible in a turbulent flow.   

In the next section, we will cover several more examples of  TDR.

\section{TDR in Miscellaneous Systems}
\label{sec:misc}

In this section, we briefly describe TDR in stably stratified turbulence,  over smooth surfaces, and in turbulent convection.

\subsection{TDR in stably stratified turbulence}

Many natural and laboratory flows are stably stratified with lighter fluid  above heavier fluid and gravity acting downwards. The governing equations for stably-stratified flows under Boussinesq approximation are~ \cite{Tritton:book,Kundu:book,Davidson:book:TurbulenceRotating,Verma:book:BDF} 
\bea
\frac{\partial{\mathbf{u}}}{\partial{t}}+ (\mathbf{u}\cdot \nabla)\mathbf{u} & = &  - \nabla p - \Omega \rho \hat{\bf z} + \nu \nabla^{2}\mathbf{u} +{\bf F}_\mathrm{LS},  
\label{eq:u_SS} \\
\frac{\partial{\rho}}{\partial{t}}+(\mathbf{u}\cdot\nabla)\rho& = & \Omega  u_{z} + \kappa \nabla^{2} \rho, \label{eq:b_SS} \\
{\bf \nabla \cdot u} & = & 0,
\eea 
where  $p$ is the pressure,    $\rho$ is the density fluctuation in velocity units,  $ - \Omega \rho \hat{\bf z} $ is  buoyancy, and $ \Omega $ is the {\em Brunt-V\"{a}is\"{a}l\"{a} frequency}, which is defined as \cite{Tritton:book, Davidson:book:TurbulenceRotating}
\be
\Omega = \sqrt{\frac{g}{\rho_m} \lvert \frac{d\bar{\rho}}{dz} \rvert}.
\ee
Here $\rho_m$ is the  mean density of the whole fluid, $ d\bar{\rho}/dz $ is the average density gradient, and $g$ is the acceleration due to gravity.   We convert the density in velocity units using the transformation, $ \rho \rightarrow \rho g/ (\Omega \rho_m)$. The ratio $ \nu/\kappa $ is called \textit{Schmidt number}, which is denoted by Sc.  {\em Richardson number}, Ri, which is a nondimensional number, is employed to quantify  the ratio of  buoyancy and the nonlinear term $ (\mathbf{u}\cdot \nabla)\mathbf{u} $.

For  periodic or vanishing boundary condition and in the absence of dissipative terms, the total energy,
\bea
E_u + E_\rho = \int d{\bf r} \frac{1}{2} u^2  +   \int d{\bf r} \frac{1}{2} \rho^2,   
\label{eq:buoyant:energy_conserv}
\eea
is conserved \cite{Tritton:book,Lindborg:JFM2006,	Davidson:book:TurbulenceRotating,Verma:JPA2022}.  Here, $E_\rho$ can be interpreted as the \textit{total potential energy}.  It has been shown that in the inertial range, the associated energy fluxes obey the following conservation law~\cite{Verma:PS2019,Verma:JPA2022}:
\be
\Pi_u + \Pi_\rho = \mathrm{const} = \epsilon_\mathrm{inj},
 \ee
 where $ \Pi_\rho $ is the potential energy flux, and $ \epsilon_\mathrm{inj} $ is the KE injection rate. Note that under steady state, $ \Pi_\rho $ equals the energy transfer rate from the velocity field to the density field. Using the stable nature of the flow, we can argue that $ \Pi_\rho > 0 $ \cite{Davidson:book:TurbulenceRotating,  Verma:PS2019,Verma:JPA2022,Verma:book:BDF}. 
 
\begin{figure}%[tbhp]
 	\centering
 	\includegraphics[width=0.7\linewidth]{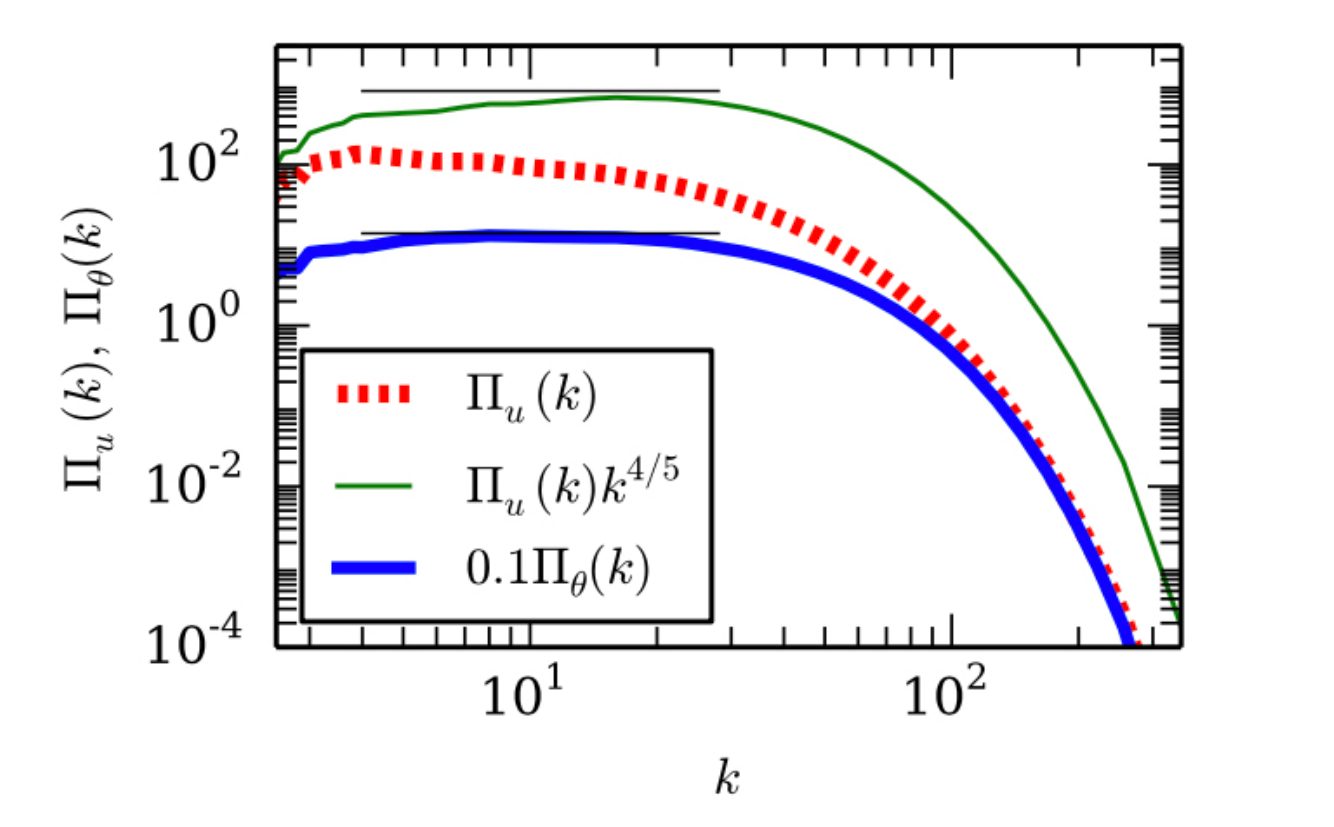}
 	\caption{Stably stratified simulation with Sc = 1 and Ri = 0.01: plots of KE flux $ \Pi_u(k) $, normalized KE flux $ \Pi_u(k)k^{4/5} $, and potential energy flux $ \Pi_\rho(k) $ (presented as $ \Pi_\theta(k) $ in the figure).  From Kumar \etal \cite{Kumar:PRE2014}. Reproduced with permission from APS.}
 	\label{fig:SST}
 \end{figure}

Nature of the stably stratified turbulence depends quite critically on the density gradient or Richardson number.  For moderate density gradient ($ \mathrm{Ri} \approx 1 $), Bolgiano~\cite{Bolgiano:JGR1959} and Obukhov~\cite{Obukhov:DANS1959} argued that $  \Pi_\rho  $ is positive and constant, whereas $ \Pi_u(k) \sim k^{-4/5} $. For small Richardson numbers, the scaling is closer to passive scalar turbulence \cite{Yeung:PF2005}, but the flow becomes quasi-2D for large Richardson numbers \cite{Davidson:book:TurbulenceRotating,Verma:book:BDF}. Here, we present only one numerical result.  Kumar \etal \cite{Kumar:PRE2014} simulated stably stratified turbulence  for Sc = 1 and Ri = 0.01, and observed that in the inertial range, $\Pi_\rho(k) = \mathrm{const}$ ($ > 0$) and  $ \Pi_u(k)  \sim k^{-4/5}  $. See Fig.~\ref{fig:SST} for an illustration. Researchers have observed that $ \Pi_\rho >0 $ for small and large Ri's as well~\cite{Yeung:PF2005,Davidson:book:TurbulenceRotating,Lindborg:JFM2006}.

Using the fact that $\Pi_\rho(k) > 0$, following the arguments described in Section~\ref{sec:drag}, we  argue that the turbulent drag will be reduced in stably stratified turbulence.  That is, for the same KE injection rate $ \epsilon_\mathrm{inj} $, $ \Pi_u(k) $ and $ \la {\bf u \cdot \nabla u} \ra $ for  stably stratified turbulence will be smaller than those for HD turbulence.  We remark that the flux-based arguments presented above are consistent with the observations of Narasimha and Sreenivasan~\cite{Narasimha:AAM1979} who argued that   stably stratified turbulence is relaminarized. 
 
In the next subsection, we will discuss TDR experienced by smooth bluff bodies.

%%%%
\subsection{TDR over smooth bluff bodies}

As discussed in Section~\ref{sec:Hydro}, bluff bodies experience turbulent drag at large Reynolds numbers. Models, experiments, and numerical simulations reveal that the turbulent drag on aerodynamic objects is a combination of the \textit{viscous drag} and \textit{adverse pressure gradient}~\cite{Kundu:book,Anderson:book:Aero,Anderson:book:History_aero}. Engineers have devised ingenious techniques to reduce this drag, which are  beyond the scope of this article.  

Equation~(\ref{eq:drag_foce_tot}) illustrates that the turbulent drag experienced by a bluff body is a combination of the inertial and viscous forces, and the adverse pressure gradient. However, for  bluff bodies like aerofoils and automobiles, the dominant contributions  come from the viscous drag and adverse pressure gradient~\cite{Anderson:book:History_aero,Anderson:book:Aero}.  Note, however, that the bulk flow  above the smooth surface is anisotropic, and it contains signatures of the surface properties. Hence, the nonlinear term $ \la \lvert {\bf u \cdot \nabla u} \rvert \ra$ and the drag coefficient $ \bar{C}_{d2} $ could yield interesting insights into TDR over bluff bodies. Narasimha and Sreenivasan \cite{Narasimha:AAM1979}  performed such analysis for a variety of flows.  In the following subsection,  we will use the above idea to explain TDR in turbulent thermal convection.

\subsection{TDR in turbulent thermal convection}

Turbulent convection exhibits interesting properties  related to TDR.  In this subsection, we consider Rayleigh-B\'{e}nard convection (RBC), which is an idealized  setup consisting of a thin fluid layer confined between two thermally conducting plates separated by a distance $d$.  The temperatures of the bottom and top plates are $T_b$ and $T_t$ respectively, with $ T_b > T_t $.

The equations for thermal convection under Boussinesq approximation are~ \cite{Chandrasekhar:book:Instability}
\bea
\frac{\partial {\bf u}}{\partial t} + ({\bf u} \cdot \nabla) {\bf u} & = &  -\frac{1}{\rho} \nabla p   + \alpha g T \hat{\bf z} + \nu \nabla^2 {\bf u}, \label{eq:buoyant:RBC1}
\\
\frac{\partial T}{\partial t} +  ({\bf u} \cdot \nabla)  T & = &   \kappa \nabla^2 T, \label{eq:buoyant:RBC2} \\
\nabla \cdot {\bf u} & = & 0, \label{eq:buoyant:RBC3}
\eea 
where $ T $ is the temperature field; $\alpha, \kappa$ are respectively  the thermal expansion coefficient and thermal diffusivity  of the fluid;  and $g$ is the acceleration due to gravity.   The two important parameters of turbulent thermal convection are  thermal  Prandtl number, $\mathrm{Pr}=\nu/\kappa$, and  Rayleigh number, 
\bea
\mathrm{Ra}  =  \frac{\alpha g d^3 (T_b-T_t) }{\nu \kappa}.
\eea

In turbulent thermal convection, the velocity field receives energy from the temperature field via buoyancy. Note that thermal plumes drive thermal convection. This feature  is opposite to what happens in polymeric,  MHD, and stably stratified turbulence, where the velocity field loses energy to the secondary field.  Yet, there are signatures of TDR in turbulent convection, which is due to the smooth thermal plates. Hence, the mechanism of TDR in turbulent thermal convection differs from that in polymeric,  MHD, and stably stratified turbulence.

\begin{figure}%[tbhp]
	\centering
	\includegraphics[width=0.7\linewidth]{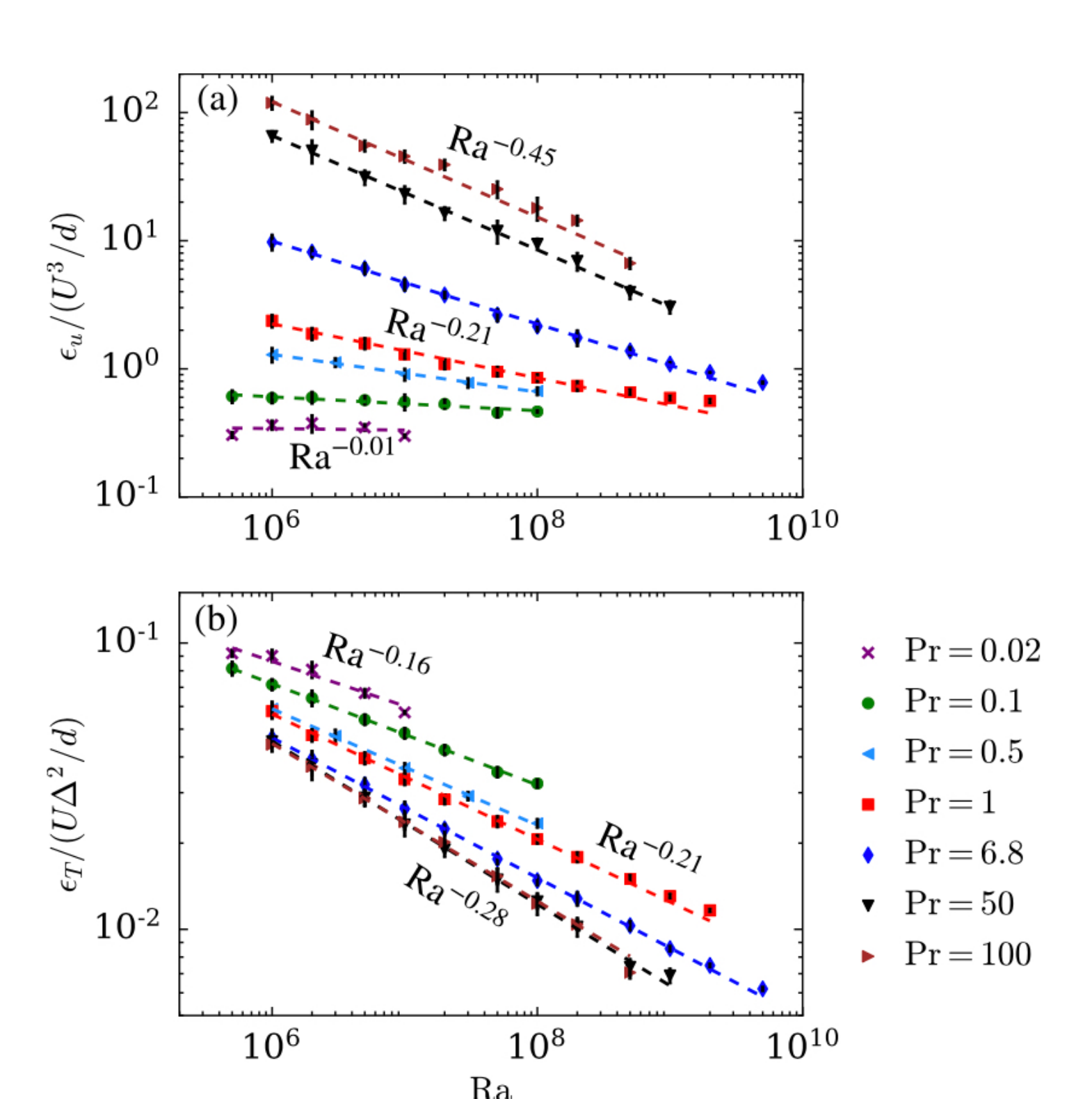}
	\caption{Plots exhibiting the Ra and Pr dependence of the viscous and thermal dissipation rates. For moderate Pr, $ \epsilon_u, \epsilon_T \sim \mathrm{Ra}^{-0.20} $. 	From Bhattacharya et al.~\cite{Bhattacharya:PF2021}. Reproduced with permission from AIP. }
	\label{fig:RBC_diss}
\end{figure}

In the following, we list some of the results related to TDR in thermal convection.
\begin{enumerate}
	\item Kraichnan~\cite{Kraichnan:PF1962Convection}  argued that turbulent thermal convection would become fully turbulent or reach \textit{ultimate regime} at very large Rayleigh number. In this asymptotic state, the effects of walls are expected to vanish, similar to the vanishing of boundary effects in the bulk of HD turbulence~\cite{Lesieur:book:Turbulence,Frisch:book,Verma:Pramana2005S2S}.  Kraichnan~\cite{Kraichnan:PF1962Convection} predicted that $ \mathrm{Nu} \propto \mathrm{Ra}^{1/2} $ in the ultimate regime.   However, experimental observations and numerical simulations reveal that for $ \mathrm{Ra} \lessapprox 10^{13} $, $ \mathrm{Nu} \sim \mathrm{Ra}^\beta  $ with  $ \beta $ ranging from 0.29 to 0.33 \cite{Grossmann:JFM2000,Niemela:Nature2000,Verma:book:BDF}. This reduction in the Nu exponent from 1/2 to approximately 0.30 is attributed to the suppression of heat flux due to the smooth thermal plates, boundary layers, and other complex properties \cite{Shraiman:PRA1990,Siggia:ARFM1994,Grossmann:JFM2000,Niemela:Nature2000,Verma:book:BDF}. 
	
	\item Pandey \etal \cite{Pandey:PF2016} performed numerical simulations of RBC for Pr = 1 and Ra ranging from $ 10^6 $ to $ 5\times 10^8 $, and  showed that  
	\be
	\frac{\mathrm{Nonlinear~term}}{\mathrm{Viscous~term}} = \frac{\lvert {\bf u \cdot \nabla u} \rvert}{\lvert \nu \nabla^2 {\bf u}\rvert} \sim \mathrm{Re}{\mathrm{Ra}^{-0.14}} .
	\ee
	Note that the above ratio is Re for HD turbulence.	Thus, nonlinearity ($ \la \lvert {\bf u \cdot \nabla u} \rvert \ra$) is suppressed in turbulent thermal convection at large Ra.

 \item Pandey \etal \cite{Pandey:PF2016}  and Bhattacharya \etal \cite{Bhattacharya:PF2021,Bhattacharya:PRF2021} showed that the viscous dissipation rate ($ \epsilon_u $) and thermal dissipation rate ($ \epsilon_T $)  depend on Rayleigh and Prandtl numbers, and that $ \epsilon_u $ and $ \epsilon_T $ are suppressed  compared to HD turbulence.  For moderate Pr and large Ra,
\bea
\epsilon_u  & \sim & \frac{U^3}{d} \mathrm{Ra}^{-0.2}, \\
\epsilon_T & \sim & \frac{U (T_b-T_t)^2}{d} \mathrm{Ra}^{-0.2} .
\eea
Interestingly, for small Prandtl numbers, $ \epsilon_u \sim U^3/d $  with very small Ra-dependent correction \cite{Bhattacharya:PF2021,Bhattacharya:PRF2021}.  See Fig.~\ref{fig:RBC_diss} for an illustration.  
\end{enumerate}

\begin{figure}%[tbhp]
	\centering
	\includegraphics[width=0.5\linewidth]{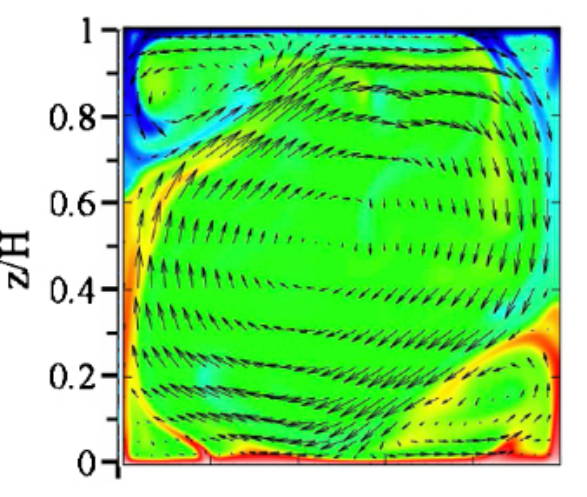}
	\caption{A LSC observed in 2D RBC by Sugiyama \etal \cite{Sugiyama:PRL2010}.   The arrows represent the velocity field, whereas the colors represent the temperature of the fluid, with red as hot and blue as cold fluid. 	From Sugiyama \etal \cite{Sugiyama:PRL2010}. Reproduced with permission from APS.}
	\label{fig:RBC_roll}
\end{figure}

 It is well known that a large-scale circulation (LSC) is present in turbulence convection (see Fig.~\ref{fig:RBC_roll})~\cite{Sreenivasan:PRE2002,Sugiyama:PRL2010,Chandra:PRE2011,Chandra:PRL2013,Verma:PF2015Reversal}. As we show below, the suppression of nonlinearity ($ \la \lvert {\bf u \cdot \nabla u} \rvert \ra$) and turbulent drag in RBC is related to this LSC and the smooth walls.

\begin{figure}%[tbhp]
	\centering
	\includegraphics[width=0.7\linewidth]{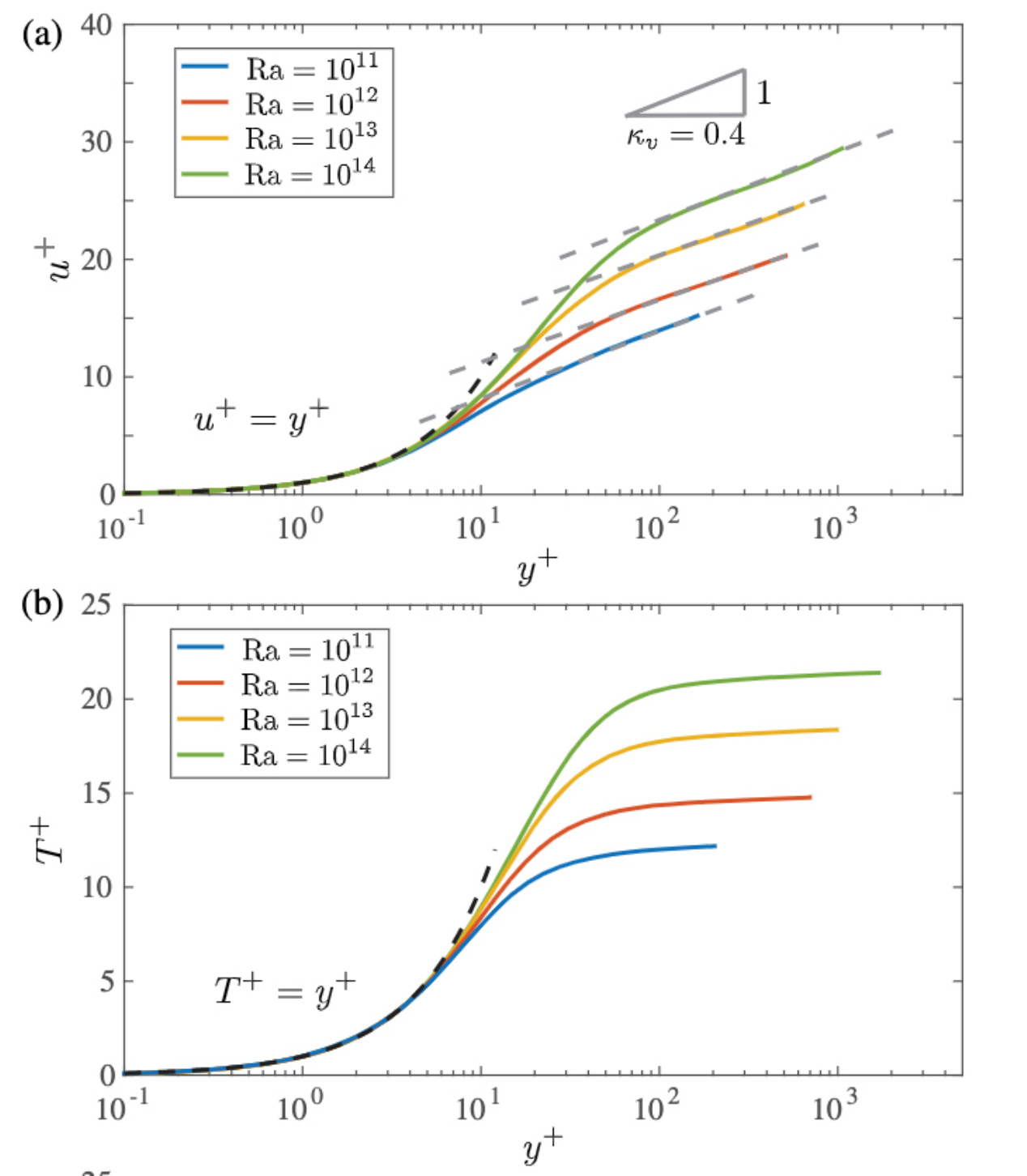}
	\caption{In numerical simulation of Zhu \etal \cite{Zhu:PRL2018}, the velocity (a) and temperature (b) profiles in wall units for various Ra's. The dashed lines illustrate the viscous sublayer  and the log-layer. A log layer is observed for the velocity field, but not for the temperature field.  
	From Zhu \etal \cite{Zhu:PRL2018}. Reproduced with permission from APS.}
	\label{fig:RBC_BL}
\end{figure}

As shown in Fig.~\ref{fig:RBC_roll},  the flows near the top and bottom plates have similarities with those near a flat plate. The LSC traverses vertically along the vertical walls, but moves horizontally along the thermal plates.  However, for a typical RBC flow, the horizontal extent of LSC is  shorter than  that in the flow past a flat plate.   Researchers have argued that for large Rayleigh numbers ($ \mathrm{Ra} \gtrapprox 10^{13} $), the boundary layers exhibit a transition to a log layer, which is a signature of transition from viscous to turbulent boundary layer, as in flow past a flat plate \cite{Landau:book:Fluid,Kundu:book,Grossmann:JFM2000,Verma:book:BDF,Anderson:book:Aero}. For example, Zhu \etal \cite{Zhu:PRL2018} simulated 2D RBC and showed that above the viscous layer, the normalized velocity field varies logarithmically with the normalized vertical distance.  In particular, Zhu \etal \cite{Zhu:PRL2018}  observed that $ u^+ \propto \log(y^+) $ for $ y^+ \gtrapprox 10 $ (see Fig.~\ref{fig:RBC_BL}). Note, however, that the thermal boundary layers do not show transition to log layer \cite{Zhu:PRL2018}. Several other experiments exhibit similar behaviour~\cite{He:PRL2014}. 

Since the boundary layers of turbulent thermal convection have similar properties  as those over a flat plate, we can argue that  the nonlinearity $ \la {\bf u \cdot \nabla u} \ra $ is suppressed in turbulent convection.  This is the reason why the dissipation rates and turbulent drag in turbulent convection are smaller than the corresponding quantities in HD turbulence.  Verma \etal \cite{Verma:PRE2012} studied the correlation $ \la u_z \theta \ra$, where $ \theta $ is the temperature fluctuation, and showed that for moderate Pr,
\be
\la u_z \theta \ra = \sqrt{\la u_z^{2} \ra} \sqrt{\la \theta^{2} \ra} 
(\mathrm{Pr Ra})^{-0.22} .
\ee
Note that $ \sqrt{\la u_z^{2} \ra} \approx \mathrm{Ra}^{1/2} $ and $ \sqrt{\la \theta^{2} \ra}  \approx (\Delta T)$.  Therefore, the correction $ (\mathrm{Pr Ra})^{-0.22}  $ of the above equation leads to 
 $ \la u_z \theta \ra \sim \mathrm{Ra}^{0.28}  $ or
$ \mathrm{Nu} \sim \mathrm{Ra}^{0.28} $. Verma \etal \cite{Verma:PRE2012,Verma:book:BDF}  argued that at very large Ra, the corrections would disappear and the flow will approach the ultimate regime with $ \la u_z \theta \ra  \sim \mathrm{Ra}^{1/2} $ or  $ \mathrm{Nu} \sim \mathrm{Ra}^{1/2} $.

 Note, however, that no experiment and numerical simulation has been able to achieve the ultimate regime, thus the ultimate regime remains a conjecture at present~\cite{Niemela:Nature2000,He:PRL2012,Zhu:PRL2018,Iyer:PNAS2020}, even though several experiments and numerical simulation report  a transition to the ultimate regime with the Nu exponent reaching up to 0.38 (but lower than 1/2)~\cite{Zhu:PRL2018,He:PRL2012}, while some others argue against the transition to the ultimate regime  \cite{Niemela:Nature2000,Iyer:PNAS2020}. It is interesting to note that for rough thermal plates, the heat transport is enhanced because of increase in turbulence due to the roughness \cite{Roche:NJP2010}.

RBC with periodic boundary condition exhibits $ \mathrm{Nu \propto Ra}^{1/2} $ due to the absence of boundary layers~\cite{Lohse:PRL2003,Verma:PRE2012}.  In addition, RBC with small Prandtl numbers too exhibit properties similar to those of periodic boundary condition. This is because the temperature gradient is linear in the bulk in both these systems~\cite{Mishra:EPL2010,	Bhattacharya:PRF2021}.
  
  In summary, turbulent thermal convection exhibits suppression of nonlinearity ($ \la \lvert {\bf u \cdot \nabla u} \rvert \ra$) and KE flux compared to HD turbulence.  This suppression, which occurs essentially due to the smooth walls,  leads to TDR in thermal convection.

\section{Discussions and conclusions}
\label{sec:conclusions}

Experiments and numerical simulations show that turbulent flows with dilute polymers exhibit TDR. Many factors--boundary layers, polymer properties, bulk properties of the flow--are responsible for this phenomena \cite{Lumley:ARFM1969,Tabor:EPL1986,deGennes:book:Intro,deGennes:book:Polymer,Sreenivasan:JFM2000,Lvov:PRL2004,Benzi:PRE2003,White:ARFM2008,Benzi:PD2010,Benzi:ARCMP2018,Verma:PP2020}. There are many interesting works in this field, however, in this review, we focus on the role of bulk turbulence on TDR. The KE flux, $ \Pi_u(k) $, is suppressed in the presence of polymers.  This reduction in $ \Pi_u(k) $ leads to suppression of nonlinearity $ \la {\bf u \cdot \nabla u} \ra $  and  turbulent drag.

MHD turbulence exhibits very similar behaviour as the polymeric turbulence \cite{Verma:PP2020}.  Here too, $ \Pi_u(k) $ is suppressed because a major fraction of the injected KE is transferred to the magnetic field. Consequently, $ \la {\bf u \cdot \nabla u} \ra $ and the turbulent drag  are suppressed in MHD turbulence. For the same KE injection rate at large scales, $ \Pi_u(k) $  and $ \la {\bf u \cdot \nabla u} \ra $ for MHD turbulence are smaller than the respective quantities of HD turbulence. These properties are borne out in  DNS and shell models.

The KE flux $ \Pi_u(k)  $ of stably stratified turbulence too is suppressed compared to  HD turbulence. Hence, we expect TDR in stably stratified turbulence. Narasimha and Sreenivasan \cite{Narasimha:AAM1979} made a similar observation.  We need detailed numerical simulations to verify the above statement.  An interesting point to note that for the above three flows,
\be
\Pi_u(k) + \Pi_B(k) = \mathrm{const}  = \epsilon_\mathrm{inj},
\label{eq:Pi_sum_const}
\ee
where $ \Pi_B(k)  $ represents the energy flux associated with the secondary field $ B $, which could be polymer, magnetic field, or density. The constancy of the sum of fluxes in Eq.~(\ref{eq:Pi_sum_const}) arises due to the stable nature of system \cite{Davidson:book:TurbulenceRotating,	Verma:PS2019,Verma:JPA2022}. The above constancy also represents a redistribution of the injected kinetic energy at large scales to (a)  the velocity field in the intermediate scales, and to (b) the secondary field. Positive $ \Pi_B $  implies that $  \Pi_u(k) < \epsilon_\mathrm{inj} $ which leads to TDR in the flow. Thus, TDR is intimately related to the  conservation law of Eq.~(\ref{eq:Pi_sum_const}).

Another important feature of TDR is that the mean flow or large scale velocity ($ U $) is enhanced in the presence of polymers or magnetic field. This is because the velocity field gets more ordered under TDR.  Suppression of  $ \Pi_u(k) $  and $ \la {\bf u \cdot \nabla u} \ra $ even with strong $ U $ is due to the correlations in the velocity field.  An emergence of ordered $ U $ is also observed in dynamo and QSMHD turbulence.  Unfortunately, DNS of MHD turbulence with magnetic Prandtl number Pm = 1/3, 1, and 10/3 do not show enhancement in $ U $ compared to the respective HD turbulence.    Based on the findings of QSMHD turbulence ($ \mathrm{Pm} \approx 0 $) and dynamo, we conjecture that $ U $ of MHD turbulence  with very small Pm  will be larger than that of  corresponding HD turbulence.

TDR  is also observed in turbulent thermal convection. This observation is based on the  suppression of viscous and thermal dissipation rates, and that of nonlinearity $ \la {\bf u \cdot \nabla u} \ra $ \cite{Pandey:PF2016,Bhattacharya:PF2021,Verma:book:ET}. Note, however, that unlike MHD, polymeric, and stably-stratified turbulence, $ \Pi_u(k) $ for turbulent thermal convection is not suppressed due to the unstable nature of thermal convection~\cite{Verma:PS2019,Verma:JPA2022}.  Therefore, the mechanism for TDR in turbulent thermal convection differs from that for TDR in MHD, polymeric, and stably-stratified turbulence. In this review, we argue that TDR in turbulent thermal convection occurs due to the smooth thermal plates.  Near the thermal plates, the large-scale circulation (LSC) are akin to the flow past a flat plate.  This feature has important consequences on the possible transition to the ultimate regime in thermal convection.
\begin{figure}[tbhp]
	\centering
	\includegraphics[scale = 0.4]{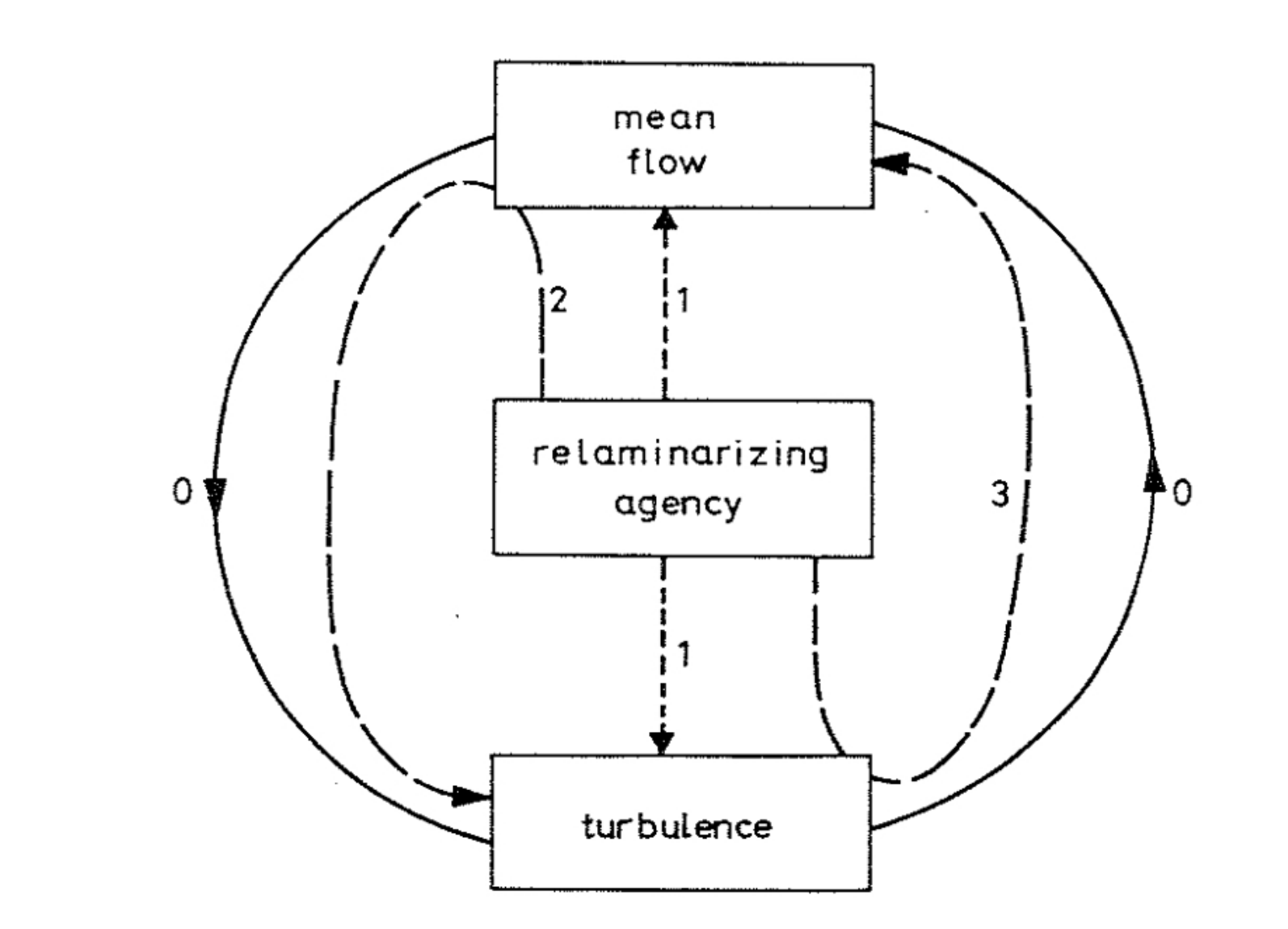}
	\caption{Interactions between the mean flow and turbulence via relaminarizing agency. The interaction channels 1,2,3 relaminarize the flow in comparison  to the HD turbulence where interactions occurs via channel 0.  From Narasimha and Sreenivasan \cite{Narasimha:AAM1979}. Reproduced with permission from K. R. Sreenivasan. }
	\label{fig:Sreeni-relam}
\end{figure} 

The enhancement of $ U $ under TDR is similar to the increase in the mean flow during relaminarization.  Narasimha and Sreenivasan \cite{Narasimha:AAM1979}  showed  reversion of flows from random to smooth profiles by relaminarizing agencies, which could be stably stratification, rotation,  thermal convection, etc.   Figure~\ref{fig:Sreeni-relam} illustrates  interactions between the mean flow and turbulence via a relaminarizing agency. In this figure, the channels 1, 2, and 3 represent complex interactions between the mean flow and fluctuations during relaminarization, whereas channel 0 represents these interactions in the HD turbulence. The arguments of Verma \etal \cite{Verma:PP2020} have certain similarities with those of Narasimha and Sreenivasan \cite{Narasimha:AAM1979}. 

In summary, this review  discusses a general framework based on KE flux to explain TDR in a wide range of phenomena---polymeric, MHD, QSMHD, and stably stratified turbulence;   dynamo; and turbulent thermal convection. This kind of study is relatively new, and it is hoped that it will be explored further in future.  We also expect TDR to emerge in other systems, such as drift-wave turbulence, astrophysical MHD, rotating turbulence, etc.  Such a study has an added benefit that TDR has practical applications in engineering flows, liquid metals, polymeric flows, etc.

\bmhead{Acknowledgments} The authors thank Abhishek Kumar and Shashwat Bhattacharya for useful discussions. This project was supported by Indo-French project 6104-1 from CEFIPRA.  S. Chatterjee is supported by INSPIRE fellowship (No. IF180094) of the Department of Science \& Technology, India. 

\section*{Declarations}

\bmhead{Conflict of interest statement} The authors have no actual or potential conflicts of interest to declare in relation to this article.

%\bibliography{/Users/mkv/Dropbox/docs-pub/bib/journal,/Users/mkv/Dropbox/docs-pub/bib/book,/Users/mkv/Dropbox/docs-pub/bib/book_chapter,/Users/mkv/Dropbox/docs-pub/bib/thesis,/Users/mkv/Dropbox/docs-pub/bib/conf} 

%\bibliography{bib/journal,bib/book,bib/book_chapter, bib/conf, bib/report, bib/mkv_journals, bib/thesis}% common bib file
%\bibliography{bib/book}

\begin{thebibliography}{106}
\ifx \bisbn   \undefined \def \bisbn  #1{ISBN #1}\fi
\ifx \binits  \undefined \def \binits#1{#1}\fi
\ifx \bauthor  \undefined \def \bauthor#1{#1}\fi
\ifx \batitle  \undefined \def \batitle#1{#1}\fi
\ifx \bjtitle  \undefined \def \bjtitle#1{#1}\fi
\ifx \bvolume  \undefined \def \bvolume#1{\textbf{#1}}\fi
\ifx \byear  \undefined \def \byear#1{#1}\fi
\ifx \bissue  \undefined \def \bissue#1{#1}\fi
\ifx \bfpage  \undefined \def \bfpage#1{#1}\fi
\ifx \blpage  \undefined \def \blpage #1{#1}\fi
\ifx \burl  \undefined \def \burl#1{\textsf{#1}}\fi
\ifx \doiurl  \undefined \def \doiurl#1{\url{https://doi.org/#1}}\fi
\ifx \betal  \undefined \def \betal{\textit{et al.}}\fi
\ifx \binstitute  \undefined \def \binstitute#1{#1}\fi
\ifx \binstitutionaled  \undefined \def \binstitutionaled#1{#1}\fi
\ifx \bctitle  \undefined \def \bctitle#1{#1}\fi
\ifx \beditor  \undefined \def \beditor#1{#1}\fi
\ifx \bpublisher  \undefined \def \bpublisher#1{#1}\fi
\ifx \bbtitle  \undefined \def \bbtitle#1{#1}\fi
\ifx \bedition  \undefined \def \bedition#1{#1}\fi
\ifx \bseriesno  \undefined \def \bseriesno#1{#1}\fi
\ifx \blocation  \undefined \def \blocation#1{#1}\fi
\ifx \bsertitle  \undefined \def \bsertitle#1{#1}\fi
\ifx \bsnm \undefined \def \bsnm#1{#1}\fi
\ifx \bsuffix \undefined \def \bsuffix#1{#1}\fi
\ifx \bparticle \undefined \def \bparticle#1{#1}\fi
\ifx \barticle \undefined \def \barticle#1{#1}\fi
\bibcommenthead
\ifx \bconfdate \undefined \def \bconfdate #1{#1}\fi
\ifx \botherref \undefined \def \botherref #1{#1}\fi
\ifx \url \undefined \def \url#1{\textsf{#1}}\fi
\ifx \bchapter \undefined \def \bchapter#1{#1}\fi
\ifx \bbook \undefined \def \bbook#1{#1}\fi
\ifx \bcomment \undefined \def \bcomment#1{#1}\fi
\ifx \oauthor \undefined \def \oauthor#1{#1}\fi
\ifx \citeauthoryear \undefined \def \citeauthoryear#1{#1}\fi
\ifx \endbibitem  \undefined \def \endbibitem {}\fi
\ifx \bconflocation  \undefined \def \bconflocation#1{#1}\fi
\ifx \arxivurl  \undefined \def \arxivurl#1{\textsf{#1}}\fi
\csname PreBibitemsHook\endcsname

%%% 1
\bibitem{Lumley:ARFM1969}
\begin{barticle}
\bauthor{\bsnm{Lumley}, \binits{J.}},
\bauthor{\bsnm{Blossey}, \binits{P.}}:
\batitle{{Drag reduction by additives}}.
\bjtitle{Annu. Rev. Fluid Mech.}
\bvolume{1}(\bissue{1}),
\bfpage{367}--\blpage{384}
(\byear{1969})
\end{barticle}
\endbibitem

%%% 2
\bibitem{Tabor:EPL1986}
\begin{barticle}
\bauthor{\bsnm{Tabor}, \binits{M.}},
\bauthor{\bparticle{de} \bsnm{Gennes}, \binits{P.G.}}:
\batitle{{A cascade theory of drag reduction}}.
\bjtitle{EPL}
\bvolume{2}(\bissue{7}),
\bfpage{519}--\blpage{522}
(\byear{1986})
\end{barticle}
\endbibitem

%%% 3
\bibitem{deGennes:book:Intro}
\begin{bbook}
\bauthor{\bparticle{de} \bsnm{Gennes}, \binits{P.G.}}:
\bbtitle{{Introduction to Polymer Dynamics}}.
\bpublisher{Cambridge University Press},
\blocation{Cambridge}
(\byear{1990})
\end{bbook}
\endbibitem

%%% 4
\bibitem{deGennes:book:Polymer}
\begin{bbook}
\bauthor{\bparticle{de} \bsnm{Gennes}, \binits{P.G.}}:
\bbtitle{{Scaling Concepts in Polymer Physics}}.
\bpublisher{Cornell University Press},
\blocation{Ithaca}
(\byear{1979})
\end{bbook}
\endbibitem

%%% 5
\bibitem{Sreenivasan:JFM2000}
\begin{barticle}
\bauthor{\bsnm{Sreenivasan}, \binits{K.R.}},
\bauthor{\bsnm{White}, \binits{C.M.}}:
\batitle{{The onset of drag reduction by dilute polymer additives, and the
  maximum drag reduction asymptote}}.
\bjtitle{J. Fluid Mech.}
\bvolume{409},
\bfpage{149}--\blpage{164}
(\byear{2000})
\end{barticle}
\endbibitem

%%% 6
\bibitem{Lvov:PRL2004}
\begin{barticle}
\bauthor{\bsnm{L'vov}, \binits{V.S.}},
\bauthor{\bsnm{Pomyalov}, \binits{A.}},
\bauthor{\bsnm{Procaccia}, \binits{I.}},
\bauthor{\bsnm{Tiberkevich}, \binits{V.}}:
\batitle{{Drag reduction by polymers in wall bounded turbulence}}.
\bjtitle{Phys. Rev. Lett.}
\bvolume{92}(\bissue{24}),
\bfpage{244503}
(\byear{2004})
\end{barticle}
\endbibitem

%%% 7
\bibitem{Benzi:PRE2003}
\begin{barticle}
\bauthor{\bsnm{Benzi}, \binits{R.}},
\bauthor{\bsnm{De~Angelis}, \binits{E.}},
\bauthor{\bsnm{Govindarajan}, \binits{R.}},
\bauthor{\bsnm{Procaccia}, \binits{I.}}:
\batitle{{Shell model for drag reduction with polymer additives in homogeneous
  turbulence}}.
\bjtitle{Phys. Rev. E}
\bvolume{68}(\bissue{1}),
\bfpage{016308}
(\byear{2003})
\end{barticle}
\endbibitem

%%% 8
\bibitem{White:ARFM2008}
\begin{barticle}
\bauthor{\bsnm{White}, \binits{C.M.}},
\bauthor{\bsnm{Mungal}, \binits{M.G.}}:
\batitle{Mechanics and prediction of turbulent drag reduction with polymer
  additives}.
\bjtitle{Annu. Rev. Fluid Mech.}
\bvolume{40},
\bfpage{235}--\blpage{256}
(\byear{2008})
\end{barticle}
\endbibitem

%%% 9
\bibitem{Benzi:PD2010}
\begin{barticle}
\bauthor{\bsnm{Benzi}, \binits{R.}}:
\batitle{{A short review on drag reduction by polymers in wall bounded
  turbulence}}.
\bjtitle{Physica D}
\bvolume{239}(\bissue{14}),
\bfpage{1338}--\blpage{1345}
(\byear{2010})
\end{barticle}
\endbibitem

%%% 10
\bibitem{Benzi:ARCMP2018}
\begin{barticle}
\bauthor{\bsnm{Benzi}, \binits{R.}},
\bauthor{\bsnm{Ching}, \binits{E.S.C.}}:
\batitle{{Polymers in Fluid Flows}}.
\bjtitle{Annu. Rev. Condens. Matter Phys.}
\bvolume{9}(\bissue{1}),
\bfpage{163}--\blpage{181}
(\byear{2018})
\end{barticle}
\endbibitem

%%% 11
\bibitem{Verma:PP2020}
\begin{barticle}
\bauthor{\bsnm{Verma}, \binits{M.K.}},
\bauthor{\bsnm{Alam}, \binits{S.}},
\bauthor{\bsnm{Chatterjee}, \binits{S.}}:
\batitle{{Turbulent drag reduction in magnetohydrodynamic and quasi-static
  magnetohydrodynamic turbulence}}.
\bjtitle{Phys. Plasmas}
\bvolume{27},
\bfpage{052301}
(\byear{2020})
\end{barticle}
\endbibitem

%%% 12
\bibitem{Landau:book:Fluid}
\begin{bbook}
\bauthor{\bsnm{Landau}, \binits{L.D.}},
\bauthor{\bsnm{Lifshitz}, \binits{E.M.}}:
\bbtitle{{Fluid Mechanics}},
\bedition{2nd} edn.
\bsertitle{Course of Theoretical Physics}.
\bpublisher{Elsevier},
\blocation{Oxford}
(\byear{1987})
\end{bbook}
\endbibitem

%%% 13
\bibitem{Kundu:book}
\begin{bbook}
\bauthor{\bsnm{Kundu}, \binits{P.K.}},
\bauthor{\bsnm{Cohen}, \binits{I.M.}},
\bauthor{\bsnm{Dowling}, \binits{D.R.}}:
\bbtitle{{Fluid Mechanics}},
\bedition{6th} edn.
\bpublisher{Academic Press},
\blocation{San Diego}
(\byear{2015})
\end{bbook}
\endbibitem

%%% 14
\bibitem{Anderson:book:Aero}
\begin{bbook}
\bauthor{\bsnm{Anderson}, \binits{J.D.}}:
\bbtitle{{Fundamentals of Aerodynamics}}.
\bpublisher{McGraw-Hill Education},
\blocation{New York}
(\byear{2017})
\end{bbook}
\endbibitem

%%% 15
\bibitem{Anderson:book:History_aero}
\begin{bbook}
\bauthor{\bsnm{Anderson~Jr.}, \binits{J.D.}}:
\bbtitle{{A History of Aerodynamics}}.
\bpublisher{Cambridge University Press},
\blocation{New York}
(\byear{1998})
\end{bbook}
\endbibitem

%%% 16
\bibitem{Dar:PD2001}
\begin{barticle}
\bauthor{\bsnm{Dar}, \binits{G.}},
\bauthor{\bsnm{Verma}, \binits{M.K.}},
\bauthor{\bsnm{Eswaran}, \binits{V.}}:
\batitle{{Energy transfer in two-dimensional magnetohydrodynamic turbulence:
  formalism and numerical results}}.
\bjtitle{Physica D}
\bvolume{157}(\bissue{3}),
\bfpage{207}--\blpage{225}
(\byear{2001})
\end{barticle}
\endbibitem

%%% 17
\bibitem{Mininni:ApJ2005}
\begin{barticle}
\bauthor{\bsnm{Mininni}, \binits{P.D.}},
\bauthor{\bsnm{Ponty}, \binits{Y.}},
\bauthor{\bsnm{Montgomery}, \binits{D.C.}},
\bauthor{\bsnm{Pinton}, \binits{J.-F.}},
\bauthor{\bsnm{Politano}, \binits{H.}},
\bauthor{\bsnm{Pouquet}, \binits{A.G.}}:
\batitle{{Dynamo regimes with a nonhelical forcing}}.
\bjtitle{ApJ}
\bvolume{626},
\bfpage{853}--\blpage{863}
(\byear{2005})
\end{barticle}
\endbibitem

%%% 18
\bibitem{Valente:JFM2014}
\begin{barticle}
\bauthor{\bsnm{Valente}, \binits{P.C.}},
\bauthor{\bparticle{da} \bsnm{Silva}, \binits{C.B.}},
\bauthor{\bsnm{Pinho}, \binits{F.T.}}:
\batitle{{The effect of viscoelasticity on the turbulent kinetic energy
  cascade}}.
\bjtitle{J. Fluid Mech.}
\bvolume{760},
\bfpage{39}--\blpage{62}
(\byear{2014})
\end{barticle}
\endbibitem

%%% 19
\bibitem{Valente:PF2016}
\begin{barticle}
\bauthor{\bsnm{Valente}, \binits{P.C.}},
\bauthor{\bparticle{da} \bsnm{Silva}, \binits{C.B.}},
\bauthor{\bsnm{Pinho}, \binits{F.T.}}:
\batitle{{Energy spectra in elasto-inertial turbulence}}.
\bjtitle{Phys. Fluids}
\bvolume{28}(\bissue{7}),
\bfpage{075108}--\blpage{17}
(\byear{2016})
\end{barticle}
\endbibitem

%%% 20
\bibitem{Moreau:book:MHD}
\begin{bbook}
\bauthor{\bsnm{Moreau}, \binits{R.J.}}:
\bbtitle{{Magnetohydrodynamics}}.
\bpublisher{Springer},
\blocation{Berlin}
(\byear{1990})
\end{bbook}
\endbibitem

%%% 21
\bibitem{Knaepen:ARFM2008}
\begin{barticle}
\bauthor{\bsnm{Knaepen}, \binits{B.}},
\bauthor{\bsnm{Moreau}, \binits{R.}}:
\batitle{{Magnetohydrodynamic turbulence at low magnetic reynolds number}}.
\bjtitle{Annu. Rev. Fluid Mech.}
\bvolume{40},
\bfpage{25}--\blpage{45}
(\byear{2008})
\end{barticle}
\endbibitem

%%% 22
\bibitem{Reddy:PF2014}
\begin{barticle}
\bauthor{\bsnm{Reddy}, \binits{K.S.}},
\bauthor{\bsnm{Verma}, \binits{M.K.}}:
\batitle{{Strong anisotropy in quasi-static magnetohydrodynamic turbulence for
  high interaction parameters}}.
\bjtitle{Phys. Fluids}
\bvolume{26},
\bfpage{025109}
(\byear{2014})
\end{barticle}
\endbibitem

%%% 23
\bibitem{Reddy:PP2014}
\begin{barticle}
\bauthor{\bsnm{Reddy}, \binits{K.S.}},
\bauthor{\bsnm{Kumar}, \binits{R.}},
\bauthor{\bsnm{Verma}, \binits{M.K.}}:
\batitle{{Anisotropic energy transfers in quasi-static magnetohydrodynamic
  turbulence}}.
\bjtitle{Phys. Plasmas}
\bvolume{21}(\bissue{10}),
\bfpage{102310}
(\byear{2014})
\end{barticle}
\endbibitem

%%% 24
\bibitem{Moffatt:book}
\begin{bbook}
\bauthor{\bsnm{Moffatt}, \binits{H.K.}}:
\bbtitle{{Magnetic Field Generation in Electrically Conducting Fluids}}.
\bpublisher{Cambridge University Press},
\blocation{Cambridge}
(\byear{1978})
\end{bbook}
\endbibitem

%%% 25
\bibitem{Roberts:RMP2000}
\begin{barticle}
\bauthor{\bsnm{Roberts}, \binits{P.H.}},
\bauthor{\bsnm{Glatzmaier}, \binits{G.A.}}:
\batitle{{Geodynamo theory and simulations}}.
\bjtitle{Rev. Mod. Phys.}
\bvolume{72},
\bfpage{1081}--\blpage{1123}
(\byear{2000})
\end{barticle}
\endbibitem

%%% 26
\bibitem{Brandenburg:PR2005}
\begin{barticle}
\bauthor{\bsnm{Brandenburg}, \binits{A.}},
\bauthor{\bsnm{Subramanian}, \binits{K.}}:
\batitle{{Astrophysical magnetic fields and nonlinear dynamo theory}}.
\bjtitle{Phys. Rep.}
\bvolume{417}(\bissue{1-4}),
\bfpage{1}--\blpage{209}
(\byear{2005})
\end{barticle}
\endbibitem

%%% 27
\bibitem{Yadav:PRE2012}
\begin{barticle}
\bauthor{\bsnm{Yadav}, \binits{R.K.}},
\bauthor{\bsnm{Verma}, \binits{M.K.}},
\bauthor{\bsnm{Wahi}, \binits{P.}}:
\batitle{{Bistability and chaos in the Taylor-Green dynamo}}.
\bjtitle{Phys. Rev. E}
\bvolume{85}(\bissue{3}),
\bfpage{036301}
(\byear{2012})
\end{barticle}
\endbibitem

%%% 28
\bibitem{Olson:JGR1999}
\begin{barticle}
\bauthor{\bsnm{Olson}, \binits{P.L.}},
\bauthor{\bsnm{Christensen}, \binits{U.R.}},
\bauthor{\bsnm{Glatzmaier}, \binits{G.A.}}:
\batitle{{Numerical modeling of the geodynamo: mechanisms of field generation
  and equilibration}}.
\bjtitle{J. Geophys. Res. B: Solid Earth}
\bvolume{104}(\bissue{B5}),
\bfpage{10383}--\blpage{10404}
(\byear{1999})
\end{barticle}
\endbibitem

%%% 29
\bibitem{Davidson:book:TurbulenceRotating}
\begin{bbook}
\bauthor{\bsnm{Davidson}, \binits{P.A.}}:
\bbtitle{{Turbulence in Rotating, Stratified and Electrically Conducting
  Fluids}}.
\bpublisher{Cambridge University Press},
\blocation{Cambridge}
(\byear{2013})
\end{bbook}
\endbibitem

%%% 30
\bibitem{Verma:book:BDF}
\begin{bbook}
\bauthor{\bsnm{Verma}, \binits{M.K.}}:
\bbtitle{Physics of Buoyant Flows: From Instabilities to Turbulence}.
\bpublisher{World Scientific},
\blocation{Singapore}
(\byear{2018})
\end{bbook}
\endbibitem

%%% 31
\bibitem{Pandey:PF2016}
\begin{barticle}
\bauthor{\bsnm{Pandey}, \binits{A.}},
\bauthor{\bsnm{Verma}, \binits{M.K.}}:
\batitle{{Scaling of large-scale quantities in Rayleigh-B{\'e}nard
  convection}}.
\bjtitle{Phys. Fluids}
\bvolume{28}(\bissue{9}),
\bfpage{095105}
(\byear{2016})
\end{barticle}
\endbibitem

%%% 32
\bibitem{Bhattacharya:PF2021}
\begin{barticle}
\bauthor{\bsnm{Bhattacharya}, \binits{S.}},
\bauthor{\bsnm{Verma}, \binits{M.K.}},
\bauthor{\bsnm{Samtaney}, \binits{R.}}:
\batitle{{Revisiting Reynolds and Nusselt numbers in turbulent thermal
  convection}}.
\bjtitle{Phys. Fluids}
\bvolume{33},
\bfpage{015113}
(\byear{2021})
\end{barticle}
\endbibitem

%%% 33
\bibitem{Kolmogorov:DANS1941Dissipation}
\begin{barticle}
\bauthor{\bsnm{Kolmogorov}, \binits{A.N.}}:
\batitle{{Dissipation of Energy in Locally Isotropic Turbulence}}.
\bjtitle{Dokl Acad Nauk SSSR}
\bvolume{32},
\bfpage{16}--\blpage{18}
(\byear{1941})
\end{barticle}
\endbibitem

%%% 34
\bibitem{Kolmogorov:DANS1941Structure}
\begin{barticle}
\bauthor{\bsnm{Kolmogorov}, \binits{A.N.}}:
\batitle{{The local structure of turbulence in incompressible viscous fluid for
  very large Reynolds numbers}}.
\bjtitle{Dokl Acad Nauk SSSR}
\bvolume{30},
\bfpage{301}--\blpage{305}
(\byear{1941})
\end{barticle}
\endbibitem

%%% 35
\bibitem{Lesieur:book:Turbulence}
\begin{bbook}
\bauthor{\bsnm{Lesieur}, \binits{M.}}:
\bbtitle{{Turbulence in Fluids}}.
\bpublisher{Springer},
\blocation{Dordrecht}
(\byear{2008})
\end{bbook}
\endbibitem

%%% 36
\bibitem{Frisch:book}
\begin{bbook}
\bauthor{\bsnm{Frisch}, \binits{U.}}:
\bbtitle{{Turbulence: The Legacy of A. N. Kolmogorov}}.
\bpublisher{Cambridge University Press},
\blocation{Cambridge}
(\byear{1995})
\end{bbook}
\endbibitem

%%% 37
\bibitem{Verma:book:ET}
\begin{bbook}
\bauthor{\bsnm{Verma}, \binits{M.K.}}:
\bbtitle{Energy Transfers in Fluid Flows: Multiscale and Spectral
  Perspectives}.
\bpublisher{Cambridge University Press},
\blocation{Cambridge}
(\byear{2019})
\end{bbook}
\endbibitem

%%% 38
\bibitem{Narasimha:AAM1979}
\begin{barticle}
\bauthor{\bsnm{Narasimha}, \binits{R.}},
\bauthor{\bsnm{Sreenivasan}, \binits{K.R.}}:
\batitle{{Relaminarization of fluid flows}}.
\bjtitle{Adv. Appl. Mech.}
\bvolume{19},
\bfpage{221}--\blpage{309}
(\byear{1979})
\end{barticle}
\endbibitem

%%% 39
\bibitem{Verma:book:Mechanics}
\begin{bbook}
\bauthor{\bsnm{Verma}, \binits{M.K.}}:
\bbtitle{{Introduction to Mechanics}},
\bedition{2nd} edn.
\bpublisher{Universities Press},
\blocation{Hyderabad}
(\byear{2016})
\end{bbook}
\endbibitem

%%% 40
\bibitem{Verma:JPA2022}
\begin{barticle}
\bauthor{\bsnm{Verma}, \binits{M.K.}}:
\batitle{{Variable energy flux in turbulence}}.
\bjtitle{Journal of Physics A: Mathematical and Theoretical}
\bvolume{55}(\bissue{1}),
\bfpage{013002}
(\byear{2022})
{\href{https://arxiv.org/abs/2011.07291}{{2011.07291}}}.
\doiurl{10.1088/1751-8121/ac354e}
\end{barticle}
\endbibitem

%%% 41
\bibitem{Kraichnan:JFM1959}
\begin{barticle}
\bauthor{\bsnm{Kraichnan}, \binits{R.H.}}:
\batitle{{The structure of isotropic turbulence at very high Reynolds
  numbers}}.
\bjtitle{J. Fluid Mech.}
\bvolume{5},
\bfpage{497}--\blpage{543}
(\byear{1959})
\end{barticle}
\endbibitem

%%% 42
\bibitem{Verma:PR2004}
\begin{barticle}
\bauthor{\bsnm{Verma}, \binits{M.K.}}:
\batitle{{Statistical theory of magnetohydrodynamic turbulence: recent
  results}}.
\bjtitle{Phys. Rep.}
\bvolume{401}(\bissue{5}),
\bfpage{229}--\blpage{380}
(\byear{2004})
\end{barticle}
\endbibitem

%%% 43
\bibitem{Sreenivasan:PF1998}
\begin{barticle}
\bauthor{\bsnm{Sreenivasan}, \binits{K.R.}}:
\batitle{{An update on the energy dissipation rate in isotropic turbulence}}.
\bjtitle{Phys. Fluids}
\bvolume{10}(\bissue{2}),
\bfpage{528}--\blpage{529}
(\byear{1998})
\end{barticle}
\endbibitem

%%% 44
\bibitem{Onsagar:Nouvo1949_SH}
\begin{barticle}
\bauthor{\bsnm{Onsager}, \binits{L.}}:
\batitle{{Statistical hydrodynamics}}.
\bjtitle{Il Nuovo Cimento}
\bvolume{6}(\bissue{2}),
\bfpage{279}--\blpage{287}
(\byear{1949})
\end{barticle}
\endbibitem

%%% 45
\bibitem{Prandtl}
\begin{bchapter}
\bauthor{\bsnm{Prandtl}, \binits{L.}}:
\bctitle{{On the motion of fluids with very little friction}}.
In: \beditor{\bsnm{Krazer}, \binits{A.}} (ed.)
\bbtitle{Verhandlungen des Dritten Internationalen Mathematiker-Kongresses in
  Heidelberg}.
\bpublisher{Teubner},
\blocation{Leipzig}
(\byear{1905})
\end{bchapter}
\endbibitem

%%% 46
\bibitem{Fouxon:PF2003}
\begin{barticle}
\bauthor{\bsnm{Fouxon}, \binits{A.}},
\bauthor{\bsnm{Lebedev}, \binits{V.}}:
\batitle{{Spectra of turbulence in dilute polymer solutions}}.
\bjtitle{Phys. Fluids}
\bvolume{15}(\bissue{7}),
\bfpage{2060}--\blpage{2072}
(\byear{2003})
\end{barticle}
\endbibitem

%%% 47
\bibitem{Perlekar:PRL2006}
\begin{barticle}
\bauthor{\bsnm{Perlekar}, \binits{P.}},
\bauthor{\bsnm{Mitra}, \binits{D.}},
\bauthor{\bsnm{Pandit}, \binits{R.}}:
\batitle{{Manifestations of drag reduction by polymer additives in decaying,
  homogeneous, isotropic turbulence}}.
\bjtitle{Phys. Rev. Lett.}
\bvolume{97}(\bissue{26}),
\bfpage{264501}
(\byear{2006})
\end{barticle}
\endbibitem

%%% 48
\bibitem{Sagaut:book}
\begin{bbook}
\bauthor{\bsnm{Sagaut}, \binits{P.}},
\bauthor{\bsnm{Cambon}, \binits{C.}}:
\bbtitle{{Homogeneous Turbulence Dynamics}},
\bedition{2nd} edn.
\bpublisher{Cambridge University Press},
\blocation{Cambridge}
(\byear{2018})
\end{bbook}
\endbibitem

%%% 49
\bibitem{Ray:EPL2016}
\begin{barticle}
\bauthor{\bsnm{Ray}, \binits{S.S.}},
\bauthor{\bsnm{Vincenzi}, \binits{D.}}:
\batitle{{Elastic turbulence in a shell model of polymer solution}}.
\bjtitle{EPL}
\bvolume{114},
\bfpage{44001}
(\byear{2016})
\end{barticle}
\endbibitem

%%% 50
\bibitem{Thais:IJHFF2013}
\begin{barticle}
\bauthor{\bsnm{Thais}, \binits{L.}},
\bauthor{\bsnm{Gatski}, \binits{T.B.}},
\bauthor{\bsnm{Mompean}, \binits{G.}}:
\batitle{{Analysis of polymer drag reduction mechanisms from energy budgets}}.
\bjtitle{Int. J. Heat Mass Transfer}
\bvolume{43}(\bissue{C}),
\bfpage{52}--\blpage{61}
(\byear{2013})
\end{barticle}
\endbibitem

%%% 51
\bibitem{Nguyen:PRF2016}
\begin{barticle}
\bauthor{\bsnm{Nguyen}, \binits{M.Q.}},
\bauthor{\bsnm{Delache}, \binits{A.}},
\bauthor{\bsnm{Simo{\"e}ns}, \binits{S.}},
\bauthor{\bsnm{Bos}, \binits{W.J.T.}},
\bauthor{\bsnm{El~Hajem}, \binits{M.}}:
\batitle{{Small scale dynamics of isotropic viscoelastic turbulence}}.
\bjtitle{Phys. Rev. Fluids}
\bvolume{1}(\bissue{8}),
\bfpage{083301}
(\byear{2016})
\end{barticle}
\endbibitem

%%% 52
\bibitem{Lvov:PRL2005}
\begin{barticle}
\bauthor{\bsnm{L'vov}, \binits{V.S.}},
\bauthor{\bsnm{Pomyalov}, \binits{A.}},
\bauthor{\bsnm{Procaccia}, \binits{I.}},
\bauthor{\bsnm{Tiberkevich}, \binits{V.}}:
\batitle{{Drag Reduction by Microbubbles in Turbulent Flows: The Limit of
  Minute Bubbles}}.
\bjtitle{Phys. Rev. Lett.}
\bvolume{94}(\bissue{17}),
\bfpage{174502}
(\byear{2005})
\end{barticle}
\endbibitem

%%% 53
\bibitem{Cowling:book}
\begin{bbook}
\bauthor{\bsnm{Cowling}, \binits{T.G.}}:
\bbtitle{{Magnetohydrodynamics}}.
\bpublisher{Adam Hilger},
\blocation{London}
(\byear{1976})
\end{bbook}
\endbibitem

%%% 54
\bibitem{Priest:book}
\begin{bbook}
\bauthor{\bsnm{Priest}, \binits{E.R.}}:
\bbtitle{{Magnetohydrodynamics of the Sun}}.
\bpublisher{Cambridge University Press},
\blocation{Cambridge}
(\byear{2014})
\end{bbook}
\endbibitem

%%% 55
\bibitem{Goldstein:ARAA1995}
\begin{barticle}
\bauthor{\bsnm{Goldstein}, \binits{M.L.}},
\bauthor{\bsnm{Roberts}, \binits{D.A.}}:
\batitle{{Magnetohydrodynamic turbulence in the solar wind}}.
\bjtitle{Annu. Rev. Astron. Astrophys.}
\bvolume{33},
\bfpage{283}--\blpage{325}
(\byear{1995})
\end{barticle}
\endbibitem

%%% 56
\bibitem{Davidson:book:Turbulence}
\begin{bbook}
\bauthor{\bsnm{Davidson}, \binits{P.A.}}:
\bbtitle{{Turbulence: An Introduction for Scientists and Engineers}}.
\bpublisher{Oxford University Press},
\blocation{Oxford}
(\byear{2004})
\end{bbook}
\endbibitem

%%% 57
\bibitem{Debliquy:PP2005}
\begin{barticle}
\bauthor{\bsnm{Debliquy}, \binits{O.}},
\bauthor{\bsnm{Verma}, \binits{M.K.}},
\bauthor{\bsnm{Carati}, \binits{D.}}:
\batitle{{Energy fluxes and shell-to-shell transfers in three-dimensional
  decaying magnetohydrodynamic turbulence}}.
\bjtitle{Phys. Plasmas}
\bvolume{12}(\bissue{4}),
\bfpage{042309}
(\byear{2005})
\end{barticle}
\endbibitem

%%% 58
\bibitem{Kumar:EPL2014}
\begin{barticle}
\bauthor{\bsnm{Kumar}, \binits{R.}},
\bauthor{\bsnm{Verma}, \binits{M.K.}},
\bauthor{\bsnm{Samtaney}, \binits{R.}}:
\batitle{{Energy transfers and magnetic energy growth in small-scale dynamo}}.
\bjtitle{EPL}
\bvolume{104}(\bissue{5}),
\bfpage{54001}
(\byear{2014})
\end{barticle}
\endbibitem

%%% 59
\bibitem{Kumar:JoT2015}
\begin{barticle}
\bauthor{\bsnm{Kumar}, \binits{R.}},
\bauthor{\bsnm{Verma}, \binits{M.K.}},
\bauthor{\bsnm{Samtaney}, \binits{R.}}:
\batitle{{Energy transfers in dynamos with small magnetic Prandtl numbers}}.
\bjtitle{J. of Turbulence}
\bvolume{16}(\bissue{11}),
\bfpage{1114}--\blpage{1134}
(\byear{2015})
\end{barticle}
\endbibitem

%%% 60
\bibitem{Verma:PRE2001}
\begin{barticle}
\bauthor{\bsnm{Verma}, \binits{M.K.}}:
\batitle{{Field theoretic calculation of renormalized viscosity, renormalized
  resistivity, and energy fluxes of magnetohydrodynamic turbulence}}.
\bjtitle{Phys. Rev. E}
\bvolume{64}(\bissue{2}),
\bfpage{026305}
(\byear{2001})
\end{barticle}
\endbibitem

%%% 61
\bibitem{Boyd:book:Plasma}
\begin{bbook}
\bauthor{\bsnm{Boyd}, \binits{T.J.M.}},
\bauthor{\bsnm{Sanderson}, \binits{J.J.}}:
\bbtitle{{The Physics of Plasmas}}.
\bpublisher{Cambridge University Press},
\blocation{Cambridge}
(\byear{2003})
\end{bbook}
\endbibitem

%%% 62
\bibitem{Verma:Pramana2013tarang}
\begin{barticle}
\bauthor{\bsnm{Verma}, \binits{M.K.}},
\bauthor{\bsnm{Chatterjee}, \binits{A.G.}},
\bauthor{\bsnm{Yadav}, \binits{R.K.}},
\bauthor{\bsnm{Paul}, \binits{S.}},
\bauthor{\bsnm{Chandra}, \binits{M.}},
\bauthor{\bsnm{Samtaney}, \binits{R.}}:
\batitle{{Benchmarking and scaling studies of pseudospectral code Tarang for
  turbulence simulations}}.
\bjtitle{Pramana-J. Phys.}
\bvolume{81}(\bissue{4}),
\bfpage{617}--\blpage{629}
(\byear{2013})
\end{barticle}
\endbibitem

%%% 63
\bibitem{Chatterjee:JPDC2018}
\begin{barticle}
\bauthor{\bsnm{Chatterjee}, \binits{A.G.}},
\bauthor{\bsnm{Verma}, \binits{M.K.}},
\bauthor{\bsnm{Kumar}, \binits{A.}},
\bauthor{\bsnm{Samtaney}, \binits{R.}},
\bauthor{\bsnm{Hadri}, \binits{B.}},
\bauthor{\bsnm{Khurram}, \binits{R.}}:
\batitle{{Scaling of a Fast Fourier Transform and a pseudo-spectral fluid
  solver up to 196608 cores}}.
\bjtitle{J. Parallel Distrib. Comput.}
\bvolume{113},
\bfpage{77}--\blpage{91}
(\byear{2018})
\end{barticle}
\endbibitem

%%% 64
\bibitem{Craya:thesis}
\begin{botherref}
\oauthor{\bsnm{Craya}, \binits{A.}}:
{Contribution {\`a} l'analyse de la turbulence associ{\'e}e {\`a} des vitesses
  moyennes}.
PhD thesis,
Universit{\'e} de Granoble
(1958)
\end{botherref}
\endbibitem

%%% 65
\bibitem{Herring:PF1974}
\begin{barticle}
\bauthor{\bsnm{Herring}, \binits{J.R.}}:
\batitle{{Approach of axisymmetric turbulence to isotropy}}.
\bjtitle{Phys. Fluids}
\bvolume{17}(\bissue{5}),
\bfpage{859}--\blpage{872}
(\byear{1974})
\end{barticle}
\endbibitem

%%% 66
\bibitem{Sadhukhan:PRF2019}
\begin{barticle}
\bauthor{\bsnm{Sadhukhan}, \binits{S.}},
\bauthor{\bsnm{Verma}, \binits{M.K.}},
\bauthor{\bsnm{Stepanov}, \binits{R.}},
\bauthor{\bsnm{Plunian}, \binits{F.}},
\bauthor{\bsnm{Samtaney}, \binits{R.}}:
\batitle{{Kinetic helicity and enstrophy transfers in helical hydrodynamic
  turbulence}}.
\bjtitle{Phys. Rev. Fluids}
\bvolume{4},
\bfpage{84607}
(\byear{2019})
\end{barticle}
\endbibitem

%%% 67
\bibitem{Gledzer:DANS1973}
\begin{barticle}
\bauthor{\bsnm{Gledzer}, \binits{E.B.}}:
\batitle{{System of hydrodynamic type allowing 2 quadratic integrals of motion
  }}.
\bjtitle{Dokl Acad Nauk SSSR}
\bvolume{209},
\bfpage{1046}--\blpage{1048}
(\byear{1973})
\end{barticle}
\endbibitem

%%% 68
\bibitem{Yamada:PRE1998}
\begin{barticle}
\bauthor{\bsnm{Yamada}, \binits{M.}},
\bauthor{\bsnm{Ohkitani}, \binits{K.}}:
\batitle{{Asymptotic formulas for the lyapunov spectrum of fully developed
  shell model turbulence}}.
\bjtitle{Phys. Rev. E}
\bvolume{57}(\bissue{6}),
\bfpage{6257}
(\byear{1998})
\end{barticle}
\endbibitem

%%% 69
\bibitem{Ditlevsen:book}
\begin{bbook}
\bauthor{\bsnm{Ditlevsen}, \binits{P.D.}}:
\bbtitle{{Turbulence and Shell Models}}.
\bpublisher{Cambridge University Press},
\blocation{Cambridge}
(\byear{2010})
\end{bbook}
\endbibitem

%%% 70
\bibitem{Frick:PRE1998}
\begin{barticle}
\bauthor{\bsnm{Frick}, \binits{P.}},
\bauthor{\bsnm{Sokoloff}, \binits{D.D.}}:
\batitle{{Cascade and dynamo action in a shell model of magnetohydrodynamic
  turbulence}}.
\bjtitle{Phys. Rev. E}
\bvolume{57}(\bissue{4}),
\bfpage{4155}--\blpage{4164}
(\byear{1998})
\end{barticle}
\endbibitem

%%% 71
\bibitem{Stepanov:ApJ2008}
\begin{barticle}
\bauthor{\bsnm{Stepanov}, \binits{R.}},
\bauthor{\bsnm{Plunian}, \binits{F.}}:
\batitle{{Phenomenology of turbulent dynamo growth and saturation}}.
\bjtitle{ApJ}
\bvolume{680}(\bissue{1}),
\bfpage{809}--\blpage{815}
(\byear{2008})
\end{barticle}
\endbibitem

%%% 72
\bibitem{Plunian:PR2012}
\begin{barticle}
\bauthor{\bsnm{Plunian}, \binits{F.}},
\bauthor{\bsnm{Stepanov}, \binits{R.}},
\bauthor{\bsnm{Frick}, \binits{P.}}:
\batitle{{Shell models of magnetohydrodynamic turbulence}}.
\bjtitle{Phys. Rep.}
\bvolume{523},
\bfpage{1}--\blpage{60}
(\byear{2012})
\end{barticle}
\endbibitem

%%% 73
\bibitem{Verma:JoT2016}
\begin{barticle}
\bauthor{\bsnm{Verma}, \binits{M.K.}},
\bauthor{\bsnm{Kumar}, \binits{R.}}:
\batitle{{Dynamos at extreme magnetic Prandtl numbers: insights from shell
  models}}.
\bjtitle{J. Turbul.}
\bvolume{17}(\bissue{12}),
\bfpage{1112}--\blpage{1141}
(\byear{2016})
\end{barticle}
\endbibitem

%%% 74
\bibitem{Stepanov:JoT2006}
\begin{barticle}
\bauthor{\bsnm{Stepanov}, \binits{R.}},
\bauthor{\bsnm{Plunian}, \binits{F.}}:
\batitle{{Fully developed turbulent dynamo at low magnetic Prandtl numbers}}.
\bjtitle{J. Turbul.}
\bvolume{7}(\bissue{39}),
\bfpage{1}--\blpage{15}
(\byear{2006})
\end{barticle}
\endbibitem

%%% 75
\bibitem{Sharma:PF2018}
\begin{barticle}
\bauthor{\bsnm{Sharma}, \binits{M.K.}},
\bauthor{\bsnm{Kumar}, \binits{A.}},
\bauthor{\bsnm{Verma}, \binits{M.K.}},
\bauthor{\bsnm{Chakraborty}, \binits{S.}}:
\batitle{{Statistical features of rapidly rotating decaying turbulence:
  Enstrophy and energy spectra and coherent structures}}.
\bjtitle{Phys. Fluids}
\bvolume{30}(\bissue{4}),
\bfpage{045103}
(\byear{2018})
\end{barticle}
\endbibitem

%%% 76
\bibitem{Verma:ROPP2017}
\begin{barticle}
\bauthor{\bsnm{Verma}, \binits{M.K.}}:
\batitle{{Anisotropy in Quasi-Static Magnetohydrodynamic Turbulence}}.
\bjtitle{Rep. Prog. Phys.}
\bvolume{80}(\bissue{8}),
\bfpage{087001}
(\byear{2017})
\end{barticle}
\endbibitem

%%% 77
\bibitem{Verma:PF2015QSMHD}
\begin{barticle}
\bauthor{\bsnm{Verma}, \binits{M.K.}},
\bauthor{\bsnm{Reddy}, \binits{K.S.}}:
\batitle{{Modeling quasi-static magnetohydrodynamic turbulence with variable
  energy flux}}.
\bjtitle{Phys. Fluids}
\bvolume{27}(\bissue{2}),
\bfpage{025114}
(\byear{2015})
\end{barticle}
\endbibitem

%%% 78
\bibitem{Muller:book}
\begin{bbook}
\bauthor{\bsnm{M{\"u}ller}, \binits{U.}},
\bauthor{\bsnm{B{\"u}hler}, \binits{L.}}:
\bbtitle{{Magnetofluiddynamics in Channels and Containers}}.
\bpublisher{Springer},
\blocation{Berlin Heidelberg}
(\byear{2001})
\end{bbook}
\endbibitem

%%% 79
\bibitem{Tritton:book}
\begin{bbook}
\bauthor{\bsnm{Tritton}, \binits{D.J.}}:
\bbtitle{{Physical Fluid Dynamics}}.
\bpublisher{Clarendon Press},
\blocation{Oxord}
(\byear{1988})
\end{bbook}
\endbibitem

%%% 80
\bibitem{Lindborg:JFM2006}
\begin{barticle}
\bauthor{\bsnm{Lindborg}, \binits{E.}}:
\batitle{{The energy cascade in a strongly stratified fluid}}.
\bjtitle{J. Fluid Mech.}
\bvolume{550},
\bfpage{207}--\blpage{242}
(\byear{2006})
\end{barticle}
\endbibitem

%%% 81
\bibitem{Verma:PS2019}
\begin{barticle}
\bauthor{\bsnm{Verma}, \binits{M.K.}}:
\batitle{{Contrasting turbulence in stably stratified flows and thermal
  convection}}.
\bjtitle{Phys. Scr.}
\bvolume{94}(\bissue{6}),
\bfpage{064003}
(\byear{2019})
\end{barticle}
\endbibitem

%%% 82
\bibitem{Bolgiano:JGR1959}
\begin{barticle}
\bauthor{\bsnm{Bolgiano}, \binits{R.}}:
\batitle{{Turbulent spectra in a stably stratified atmosphere}}.
\bjtitle{J. Geophys. Res.}
\bvolume{64}(\bissue{12}),
\bfpage{2226}--\blpage{2229}
(\byear{1959})
\end{barticle}
\endbibitem

%%% 83
\bibitem{Obukhov:DANS1959}
\begin{barticle}
\bauthor{\bsnm{Obukhov}, \binits{A.M.}}:
\batitle{{On influence of buoyancy forces on the structure of temperature field
  in a turbulent flow}}.
\bjtitle{Dokl Acad Nauk SSSR}
\bvolume{125},
\bfpage{1246}
(\byear{1959})
\end{barticle}
\endbibitem

%%% 84
\bibitem{Yeung:PF2005}
\begin{barticle}
\bauthor{\bsnm{Yeung}, \binits{P.K.}},
\bauthor{\bsnm{Donzis}, \binits{D.A.}},
\bauthor{\bsnm{Sreenivasan}, \binits{K.R.}}:
\batitle{{High-Reynolds-number simulation of turbulent mixing}}.
\bjtitle{Phys. Fluids}
\bvolume{17}(\bissue{8}),
\bfpage{081703}
(\byear{2005})
\end{barticle}
\endbibitem

%%% 85
\bibitem{Kumar:PRE2014}
\begin{barticle}
\bauthor{\bsnm{Kumar}, \binits{A.}},
\bauthor{\bsnm{Chatterjee}, \binits{A.G.}},
\bauthor{\bsnm{Verma}, \binits{M.K.}}:
\batitle{{Energy spectrum of buoyancy-driven turbulence}}.
\bjtitle{Phys. Rev. E}
\bvolume{90}(\bissue{2}),
\bfpage{023016}
(\byear{2014})
\end{barticle}
\endbibitem

%%% 86
\bibitem{Chandrasekhar:book:Instability}
\begin{bbook}
\bauthor{\bsnm{Chandrasekhar}, \binits{S.}}:
\bbtitle{{Hydrodynamic and Hydromagnetic Stability}}.
\bpublisher{Oxford University Press},
\blocation{Clarendon}
(\byear{1961})
\end{bbook}
\endbibitem

%%% 87
\bibitem{Kraichnan:PF1962Convection}
\begin{barticle}
\bauthor{\bsnm{Kraichnan}, \binits{R.H.}}:
\batitle{{Turbulent thermal convection at arbitrary prandtl number}}.
\bjtitle{Phys. Fluids}
\bvolume{5}(\bissue{11}),
\bfpage{1374}--\blpage{1389}
(\byear{1962})
\end{barticle}
\endbibitem

%%% 88
\bibitem{Verma:Pramana2005S2S}
\begin{barticle}
\bauthor{\bsnm{Verma}, \binits{M.K.}},
\bauthor{\bsnm{Ayyer}, \binits{A.}},
\bauthor{\bsnm{Debliquy}, \binits{O.}},
\bauthor{\bsnm{Kumar}, \binits{S.}},
\bauthor{\bsnm{Chandra}, \binits{A.V.}}:
\batitle{{Local shell-to-shell energy transfer via nonlocal interactions in
  fluid turbulence}}.
\bjtitle{Pramana-J. Phys.}
\bvolume{65}(\bissue{2}),
\bfpage{297}--\blpage{310}
(\byear{2005})
\end{barticle}
\endbibitem

%%% 89
\bibitem{Grossmann:JFM2000}
\begin{barticle}
\bauthor{\bsnm{Grossmann}, \binits{S.}},
\bauthor{\bsnm{Lohse}, \binits{D.}}:
\batitle{{Scaling in thermal convection: a unifying theory}}.
\bjtitle{J. Fluid Mech.}
\bvolume{407},
\bfpage{27}--\blpage{56}
(\byear{2000})
\end{barticle}
\endbibitem

%%% 90
\bibitem{Niemela:Nature2000}
\begin{barticle}
\bauthor{\bsnm{Niemela}, \binits{J.J.}},
\bauthor{\bsnm{Skrbek}, \binits{L.}},
\bauthor{\bsnm{Sreenivasan}, \binits{K.R.}},
\bauthor{\bsnm{Donnelly}, \binits{R.J.}}:
\batitle{{Turbulent convection at very high Rayleigh numbers}}.
\bjtitle{Nature}
\bvolume{404},
\bfpage{837}--\blpage{840}
(\byear{2000})
\end{barticle}
\endbibitem

%%% 91
\bibitem{Shraiman:PRA1990}
\begin{barticle}
\bauthor{\bsnm{Shraiman}, \binits{B.I.}},
\bauthor{\bsnm{Siggia}, \binits{E.D.}}:
\batitle{{Heat transport in high-Rayleigh-number convection}}.
\bjtitle{Phys. Rev. A}
\bvolume{42}(\bissue{6}),
\bfpage{3650}--\blpage{3653}
(\byear{1990})
\end{barticle}
\endbibitem

%%% 92
\bibitem{Siggia:ARFM1994}
\begin{barticle}
\bauthor{\bsnm{Siggia}, \binits{E.D.}}:
\batitle{{High Rayleigh number convection}}.
\bjtitle{Annu. Rev. Fluid Mech.}
\bvolume{26}(\bissue{1}),
\bfpage{137}--\blpage{168}
(\byear{1994})
\end{barticle}
\endbibitem

%%% 93
\bibitem{Bhattacharya:PRF2021}
\begin{barticle}
\bauthor{\bsnm{Bhattacharya}, \binits{S.}},
\bauthor{\bsnm{Verma}, \binits{M.K.}},
\bauthor{\bsnm{Samtaney}, \binits{R.}}:
\batitle{{Prandtl number dependence of the small-scale properties in turbulent
  Rayleigh-B{\'e}nard convection}}.
\bjtitle{Phys. Rev. Fluids}
\bvolume{6},
\bfpage{063501}
(\byear{2021})
\end{barticle}
\endbibitem

%%% 94
\bibitem{Sreenivasan:PRE2002}
\begin{barticle}
\bauthor{\bsnm{Sreenivasan}, \binits{K.R.}},
\bauthor{\bsnm{Bershadskii}, \binits{A.}},
\bauthor{\bsnm{Niemela}, \binits{J.J.}}:
\batitle{{Mean wind and its reversal in thermal convection}}.
\bjtitle{Phys. Rev. E}
\bvolume{65}(\bissue{5}),
\bfpage{056306}
(\byear{2002})
\end{barticle}
\endbibitem

%%% 95
\bibitem{Sugiyama:PRL2010}
\begin{barticle}
\bauthor{\bsnm{Sugiyama}, \binits{K.}},
\bauthor{\bsnm{Ni}, \binits{R.}},
\bauthor{\bsnm{Stevens}, \binits{R.J.A.M.}},
\bauthor{\bsnm{Chan}, \binits{T.-S.}},
\bauthor{\bsnm{Zhou}, \binits{S.-Q.}},
\bauthor{\bsnm{Xi}, \binits{H.-D.}},
\bauthor{\bsnm{Sun}, \binits{C.}},
\bauthor{\bsnm{Grossmann}, \binits{S.}},
\bauthor{\bsnm{Xia}, \binits{K.-Q.}},
\bauthor{\bsnm{Lohse}, \binits{D.}}:
\batitle{{Flow reversals in thermally driven turbulence}}.
\bjtitle{Phys. Rev. Lett.}
\bvolume{105}(\bissue{3}),
\bfpage{034503}
(\byear{2010})
\end{barticle}
\endbibitem

%%% 96
\bibitem{Chandra:PRE2011}
\begin{barticle}
\bauthor{\bsnm{Chandra}, \binits{M.}},
\bauthor{\bsnm{Verma}, \binits{M.K.}}:
\batitle{{Dynamics and symmetries of flow reversals in turbulent convection}}.
\bjtitle{Phys. Rev. E}
\bvolume{83},
\bfpage{067303}
(\byear{2011})
\end{barticle}
\endbibitem

%%% 97
\bibitem{Chandra:PRL2013}
\begin{barticle}
\bauthor{\bsnm{Chandra}, \binits{M.}},
\bauthor{\bsnm{Verma}, \binits{M.K.}}:
\batitle{{Flow Reversals in Turbulent Convection via Vortex Reconnections}}.
\bjtitle{Phys. Rev. Lett.}
\bvolume{110}(\bissue{11}),
\bfpage{114503}
(\byear{2013})
\end{barticle}
\endbibitem

%%% 98
\bibitem{Verma:PF2015Reversal}
\begin{barticle}
\bauthor{\bsnm{Verma}, \binits{M.K.}},
\bauthor{\bsnm{Ambhire}, \binits{S.C.}},
\bauthor{\bsnm{Pandey}, \binits{A.}}:
\batitle{{Flow reversals in turbulent convection with free-slip walls}}.
\bjtitle{Phys. Fluids}
\bvolume{27}(\bissue{4}),
\bfpage{047102}
(\byear{2015})
\end{barticle}
\endbibitem

%%% 99
\bibitem{Zhu:PRL2018}
\begin{barticle}
\bauthor{\bsnm{Zhu}, \binits{X.}},
\bauthor{\bsnm{Mathai}, \binits{V.}},
\bauthor{\bsnm{Stevens}, \binits{R.J.A.M.}},
\bauthor{\bsnm{Verzicco}, \binits{R.}},
\bauthor{\bsnm{Lohse}, \binits{D.}}:
\batitle{{Transition to the Ultimate Regime in Two-Dimensional
  Rayleigh-B{\'e}nard Convection}}.
\bjtitle{Phys. Rev. Lett.}
\bvolume{120}(\bissue{14}),
\bfpage{144502}
(\byear{2018})
\end{barticle}
\endbibitem

%%% 100
\bibitem{He:PRL2014}
\begin{barticle}
\bauthor{\bsnm{He}, \binits{X.}},
\bauthor{\bparticle{van} \bsnm{Gils}, \binits{D.P.M.}},
\bauthor{\bsnm{Bodenschatz}, \binits{E.}},
\bauthor{\bsnm{Ahlers}, \binits{G.}}:
\batitle{{Logarithmic Spatial Variations and Universal Power Spectra of
  Temperature Fluctuations in Turbulent Rayleigh-B{\'e}nard Convection}}.
\bjtitle{Phys. Rev. Lett.}
\bvolume{112}(\bissue{17}),
\bfpage{174501}
(\byear{2014})
\end{barticle}
\endbibitem

%%% 101
\bibitem{Verma:PRE2012}
\begin{barticle}
\bauthor{\bsnm{Verma}, \binits{M.K.}},
\bauthor{\bsnm{Mishra}, \binits{P.K.}},
\bauthor{\bsnm{Pandey}, \binits{A.}},
\bauthor{\bsnm{Paul}, \binits{S.}}:
\batitle{{Scalings of field correlations and heat transport in turbulent
  convection}}.
\bjtitle{Phys. Rev. E}
\bvolume{85}(\bissue{1}),
\bfpage{016310}
(\byear{2012})
\end{barticle}
\endbibitem

%%% 102
\bibitem{He:PRL2012}
\begin{barticle}
\bauthor{\bsnm{He}, \binits{X.}},
\bauthor{\bsnm{Funfschilling}, \binits{D.}},
\bauthor{\bsnm{Nobach}, \binits{H.}},
\bauthor{\bsnm{Bodenschatz}, \binits{E.}},
\bauthor{\bsnm{Ahlers}, \binits{G.}}:
\batitle{{Transition to the Ultimate State of Turbulent Rayleigh-B{\'e}nard
  Convection}}.
\bjtitle{Phys. Rev. Lett.}
\bvolume{108}(\bissue{2}),
\bfpage{024502}
(\byear{2012})
\end{barticle}
\endbibitem

%%% 103
\bibitem{Iyer:PNAS2020}
\begin{barticle}
\bauthor{\bsnm{Iyer}, \binits{K.P.}},
\bauthor{\bsnm{Scheel}, \binits{J.D.}},
\bauthor{\bsnm{Schumacher}, \binits{J.}},
\bauthor{\bsnm{Sreenivasan}, \binits{K.R.}}:
\batitle{Classical 1/3 scaling of convection holds up to $ra= 10^{15}$}.
\bjtitle{PNAS}
\bvolume{117}(\bissue{14}),
\bfpage{7594}--\blpage{7598}
(\byear{2020})
\end{barticle}
\endbibitem

%%% 104
\bibitem{Roche:NJP2010}
\begin{barticle}
\bauthor{\bsnm{Roche}, \binits{P.-E.}},
\bauthor{\bsnm{Gauthier}, \binits{F.}},
\bauthor{\bsnm{Kaiser}, \binits{R.}},
\bauthor{\bsnm{Salort}, \binits{J.}}:
\batitle{{On the triggering of the ultimate regime of convection}}.
\bjtitle{New J. Phys.}
\bvolume{12}(\bissue{8}),
\bfpage{085014}
(\byear{2010})
\end{barticle}
\endbibitem

%%% 105
\bibitem{Lohse:PRL2003}
\begin{barticle}
\bauthor{\bsnm{Lohse}, \binits{D.}},
\bauthor{\bsnm{Toschi}, \binits{F.}}:
\batitle{{Ultimate state of thermal convection}}.
\bjtitle{Phys. Rev. Lett.}
\bvolume{90}(\bissue{3}),
\bfpage{034502}
(\byear{2003})
\end{barticle}
\endbibitem

%%% 106
\bibitem{Mishra:EPL2010}
\begin{barticle}
\bauthor{\bsnm{Mishra}, \binits{P.K.}},
\bauthor{\bsnm{Wahi}, \binits{P.}},
\bauthor{\bsnm{Verma}, \binits{M.K.}}:
\batitle{{Patterns and bifurcations in low{\textendash}Prandtl-number
  Rayleigh{\textendash}B{\'e}nard convection}}.
\bjtitle{EPL}
\bvolume{89}(\bissue{4}),
\bfpage{44003}
(\byear{2010})
\end{barticle}
\endbibitem

\end{thebibliography}


\end{document}